\documentclass[floatfix,twocolumn,prl,amsmath]{revtex4-2}
\usepackage{graphicx}
\usepackage{bm}
\usepackage{amssymb}
\usepackage{dcolumn}
\usepackage{subfigure}
\usepackage{threeparttable}
\usepackage{multirow}
\usepackage{mathrsfs}
\DeclareMathAlphabet{\mathscrbf}{OMS}{mdugm}{b}{n}
\usepackage{color}

\begin{document}
\newcommand{\intqspa}{\int\!\!\frac{\rmd^d q}{(2\pi)^d}}
\newcommand{\intqspathr}{\int\!\!\frac{\rmd^3 q}{(2\pi)^3}}
\newcommand{\intqspatwo}{\int\!\!\frac{\rmd^2 q}{(2\pi)^2}}
\newcommand{\intkspatwo}{\int\!\!\frac{\rmd^2 k}{(2\pi)^2}}
\newcommand{\intkspa}{\int\!\!\frac{\rmd^d k}{(2\pi)^d}}
\newcommand{\intkspapri}{\int\!\!\frac{\rmd^d k'}{(2\pi)^d}}
\newcommand{\vn}[1]{{\boldsymbol{#1}}}
\newcommand{\vht}[1]{{\boldsymbol{#1}}}
\newcommand{\matn}[1]{{\bf{#1}}}
\newcommand{\matnht}[1]{{\boldsymbol{#1}}}
\newcommand{\bege}{\begin{equation}}
\newcommand{\gretke}{G_{\vn{k} }^{\rm R}(\mathcal{E})}
\newcommand{\gret}{G^{\rm R}}
\newcommand{\gadv}{G^{\rm A}}
\newcommand{\gmat}{G^{\rm M}}
\newcommand{\gles}{G^{<}}
\newcommand{\ghat}{\hat{G}}
\newcommand{\sigmahat}{\hat{\Sigma}}
\newcommand{\glesone}{G^{<,{\rm I}}}
\newcommand{\glestwo}{G^{<,{\rm II}}}
\newcommand{\gspec}{G^{\rm S}}
\newcommand{\glesthree}{G^{<,{\rm III}}}
\newcommand{\magdir}{\hat{\vn{n}}}
\newcommand{\sigmaret}{\Sigma^{\rm R}}
\newcommand{\sigmales}{\Sigma^{<}}
\newcommand{\sigmalesone}{\Sigma^{<,{\rm I}}}
\newcommand{\sigmalestwo}{\Sigma^{<,{\rm II}}}
\newcommand{\sigmalesthree}{\Sigma^{<,{\rm III}}}
\newcommand{\sigmaadv}{\Sigma^{A}}
\newcommand{\ee}{\end{equation}}
\newcommand{\bal}{\begin{aligned}}
\newcommand{\defbar}{\overline}
\newcommand{\SM}{\scriptstyle}
\newcommand{\rmd}{{\rm d}}
\newcommand{\rme}{{\rm e}}
\newcommand{\eal}{\end{aligned}}
\newcommand{\crea}[1]{{c_{#1}^{\dagger}}}
\newcommand{\annihi}[1]{{c_{#1}^{\phantom{\dagger}}}}
\newcommand{\udot}{\overset{.}{u}}
\newcommand{\exponential}[1]{{\exp(#1)}}
\newcommand{\phandot}[1]{\overset{\phantom{.}}{#1}}
\newcommand{\phandag}{\phantom{\dagger}}
\newcommand{\Trace}{\text{Tr}}
\setcounter{secnumdepth}{2}
%\title{Anisotropy and temperature-dependence of spin-orbit torques: Role of magnons}
\title{Effect of magnons on the temperature dependence and anisotropy of spin-orbit torque}
\author{Frank Freimuth$^{1,2}$}
\email[Corresp.~author:~]{f.freimuth@fz-juelich.de}
\author{Stefan Bl\"ugel$^{1}$}
\author{Yuriy Mokrousov$^{1,2}$}
\affiliation{$^1$Peter Gr\"unberg Institut and Institute for Advanced Simulation,
Forschungszentrum J\"ulich and JARA, 52425 J\"ulich, Germany}
\affiliation{$^2$ Institute of Physics, Johannes Gutenberg University Mainz, 55099 Mainz, Germany
}
\begin{abstract}
We investigate the influence of magnons on the temperature-dependence and the
anisotropy of the spin-orbit torque (SOT).
For this purpose we use 3rd order perturbation theory in the framework of the
Keldysh formalism in order to derive suitable equations to compute the
magnonic SOT. We find several contributions to the magnonic SOT,
which depend differently on the 
spin-wave stiffness $\mathcal{A}$ 
and on the temperature $T$, with the dominating contribution scaling 
like $T^{2}/\mathcal{A}^{2}$.
Based on this formalism we compute the magnonic SOT 
in the ferromagnetic Rashba model.
For large Rashba parameters the magnonic SOT is strongly anisotropic
and for small quasiparticle broadening it may become larger than the
non-magnonic SOT.
\end{abstract}
\maketitle
\section{Introduction}
Spin-orbit torques (SOTs) allow us to excite magnetization
dynamics and to switch the magnetization in
magnetic bits, which may be used for MRAM applications~\cite{mat_today_mram}.
Therefore, 
they have become a cornerstone in 
spintronics research (see Ref.~\cite{rmp_sot} for a recent review).
A magnetic bilayer, such as Co/Pt, is composed of a magnetic layer (Co) on 
a heavy metal layer (Pt). When an electric current is applied in-plane along
the $x$ direction in the magnetic bilayer, 
the torque exerted on the magnetization $\hat{\vn{M}}$ 
due to the SOT consists of the field-like 
torque $\propto \hat{\vn{M}}\times \hat{\vn{e}}_y$    
and the antidamping 
torque $\propto \hat{\vn{M}}\times[\hat{\vn{M}}\times \hat{\vn{e}}_y]$.
Additional contributions, which depend neither like $\hat{\vn{M}}\times \hat{\vn{e}}_y$ 
nor like $\hat{\vn{M}}\times[\hat{\vn{M}}\times \hat{\vn{e}}_y]$ on 
the magnetization direction $\hat{\vn{M}}$ have been found
both experimentally~\cite{symmetry_spin_orbit_torques}
and theoretically~\cite{anisotropic_sot_hanke,PhysRevMaterials.3.011401}. 
They are referred to as the
anisotropy of the SOT.

Electronic structure calculations describe
many properties of the SOTs
measured in experiments correctly~\cite{current_induced_torques_haney,ibcsoit,sot_NiMnSb}.
However, the SOT measured in
Ta/CoFeB/MgO
nanowires exhibits a strong
increase of the field-like 
component
with increasing temperature
suggesting thermally induced excitation
processes to be at play, which have not yet
been considered in microscopic calculations of
the SOT~\cite{angular_and_temperature_dependence_Ta_CoFeB_MgO,PhysRevB.89.174424,PhysRevB.94.140414}.
The same observation is made in
Pt/Co/C~\cite{temperature_effects_PtCoC},
Pt/Hf/FeCoB/MgO and
W/Hf/FeCoB/MgO~\cite{PhysRevB.94.140414}.
A second phenomenon that calls for extensions of the
existing theoretical models is the strong
anisotropy found in experiments~\cite{symmetry_spin_orbit_torques},
which contrasts the often roughly isotropic or only weakly 
anisotropic SOT obtained in
first principles electronic structure calculations~\cite{PhysRevB.97.224426}.

In this work we extend our formalism for 
calculations of the SOT~\cite{ibcsoit}
by including magnons. 
Theoretical approaches to compute the effect of magnons
on the electrical conductivity in models use rate 
equations~\cite{resistance_ferromagnetic_kasuya},
the Boltzmann equation~\cite{goodings},
and diagrammatic perturbation theory~\cite{PhysRevB.79.140408,PhysRevB.90.024405}.
First-principles methods are based on the
disordered-local-moment approach~\cite{dlm_kudrnovsky_PhysRevB.86.144423}
or on the Kubo-Landauer formalism applied to a large supercell with 
spin disorder~\cite{spin_disorder_landauer}.
In this paper we make use of the torque operator $\vn{\mathcal{T}}$
to include the effect of magnons.
In previous works we showed that the torque operator
may be used to compute the response to 
magnetization dynamics~\cite{invsot}
and to calculate the effects of magnetic 
texture~\cite{mothedmisot,phase_space_berry}.
Here, we use perturbations of the 
form $\mathcal{\vn{T}}_{x}\sin(\vn{q}\cdot\vn{r}-\omega_{\rm mag}t)$
to take into account the wave-vectors $\vn{q}$ and the 
frequencies $\omega_{\rm mag}$ of magnons. We employ the
Keldysh nonequilibrium formalism to assess the SOT in the presence
of these perturbations by magnons.

This paper is structured as follows.
In Sec.~\ref{sec_formalism_sot} we develop the
equations suitable to compute the magnonic SOT.
Several contributions to the magnonic SOT are detailed in Appendix~\ref{sec_appendix}. 
The necessary integrals of the magnon dispersion are dealt with in
Sec.~\ref{sec_magnonic_integrals}.
In Sec.~\ref{sec_scaling} we explain how the magnonic torque
scales with temperature
and spin-wave stiffness.
In
Sec.~\ref{sec_formalism_anisotropy}
we generalize the formalism for general magnetization directions,
which is necessary for the calculation of the anisotropy of the SOT.
In Sec.~\ref{sec_results}
we present our results on the
magnonic torque in the ferromagnetic Rashba model.
Additional results for various different parameters are
presented in Appendix~\ref{sec_appendix_plots}.
This paper ends with a summary in
Sec.~\ref{sec_conclusions}.

\section{Formalism}
\subsection{SOT in the presence of magnons}
\label{sec_formalism_sot}
The one-magnon state is described by the normalized
magnetization
\bege\label{eq_mag_magnon}
\hat{\vn{M}}(\vn{r},t)=
\begin{pmatrix}
\eta \cos(\vn{q}\cdot \vn{r}-\omega_{\rm mag}(\vn{q}) t)\\
\eta \sin(\vn{q}\cdot \vn{r}-\omega_{\rm mag}(\vn{q}) t)\\
1-\eta^2/2\\
\end{pmatrix},
\ee
where
$\eta$ determines the cone-angle of the magnon, 
$\omega_{\rm mag}(\vn{q})$ is the dispersion,
and $\vn{q}$ is the magnon wave-vector.
From the solution of the Heisenberg model it is known 
that $M_z$ is reduced in the 1-magnon state by
the factor $1-1/N_{\rm mag}$, 
where $N_{\rm mag}$ is the number of sites.
Consequently, $1-\eta^2/2=1-1/N_{\rm mag}$ and therefore
\bege
\label{eq_relate_eta_N}
\eta=\sqrt{2/N_{\rm mag}}.
\ee
We consider the ferromagnetic ground state
with magnetization in $z$ direction as the 
unperturbed reference
state (in Sec.~\ref{sec_formalism_anisotropy} we will generalize
the formalism to general magnetization directions)
and add the perturbation term
\bege\label{eq_perturbation_by_magnon}
\begin{aligned}
\delta H_{\rm mag}(t)=&
\mu_{\rm B}\Omega^{\rm xc}(\vn{r})
\eta\sigma_{x}
\cos(\vn{q}\cdot\vn{r}-\omega_{\rm mag}(\vn{q})t)\\
+&
\mu_{\rm B}\Omega^{\rm xc}(\vn{r})
\eta\sigma_{y}
\sin(\vn{q}\cdot\vn{r}-\omega_{\rm mag}(\vn{q})t)\\
=&
\eta\mathcal{T}_{y}
\cos(\vn{q}\cdot\vn{r}-\omega_{\rm mag}(\vn{q})t)\\
-&
\eta\mathcal{T}_{x}
\sin(\vn{q}\cdot\vn{r}-\omega_{\rm mag}(\vn{q})t)\\
\end{aligned}
\ee
to the Hamiltonian in order to compute the
electronic states in the presence of the
magnon, Eq.~\eqref{eq_mag_magnon}.
Here, $\Omega^{\rm xc}(\vn{r})=\frac{1}{2\mu_{\rm B}}\left[
V^{\rm eff}_{\rm minority}(\vn{r})-V^{\rm eff}_{\rm majority}(\vn{r})
\right]$ is the exchange field, i.e., the difference
between the effective potentials of minority and
majority electrons, $\mu_{\rm B}$ is the
Bohr magneton, and $\vn{\mathcal{T}}$
is the torque operator~\cite{ibcsoit}.
We include the effect of this 
perturbation, Eq.~\eqref{eq_perturbation_by_magnon},
on the electronic states with the help of the Keldysh 
nonequilibrium formalism. A single perturbation by
Eq.~\eqref{eq_perturbation_by_magnon}
leads to a response that oscillates spatially proportional
to the $\cos$ and $\sin$ so that its spatial average is
zero. We therefore consider the quadratic response to the
perturbation Eq.~\eqref{eq_perturbation_by_magnon}.
A sampling over the magnon distribution is performed in
the course of the derivation.

The perturbation by the applied electric field is given by
\bege\label{eq_pert_elefield}
\delta H_{\rm em}(t)=e\vn{v}\cdot{\vn{A}}(t),
\ee
where
\bege\label{eq_vector_poten}
\vn{A}(t)=
\frac{1}{2}
\left[
\frac{\vn{E}_{0} e^{-i\omega t}}{i\omega}
-
\frac{\vn{E}_{0} e^{i \omega t}}{i\omega}
\right]
=
-\frac{\vn{E}_{0}\sin(\omega t)}{\omega}
\ee
is the vector potential, $\vn{v}$ is the velocity operator,
and $e$ is the elementary positive charge. We will
take the limit $\omega\rightarrow 0$ at the end of 
the calculation in order to extract the dc response to
the applied electric field.

Since we need the response quadratic in $\delta H_{\rm mag}$
and linear in $\delta H_{\rm em}$, we take the
3rd order perturbation
from the Dyson equation~\cite{rammer_smith}:
\bege\label{eq_dy_third}
\begin{aligned}
&G^{<}_{3}=\gret_{\rm eq}
\frac{\delta H_{\rm tot}(t_1)}{\hbar}
\gret_{\rm eq}
\frac{\delta H_{\rm tot}(t_2)}{\hbar}
\gret_{\rm eq}
\frac{\delta H_{\rm tot}(t_3)}{\hbar}
\gles_{\rm eq}+\\
&+\gret_{\rm eq}
\frac{\delta H_{\rm tot}(t_1)}{\hbar}
\gret_{\rm eq}
\frac{\delta H_{\rm tot}(t_2)}{\hbar}
\gles_{\rm eq}
\frac{\delta H_{\rm tot}(t_3)}{\hbar}
\gadv_{\rm eq}+\\
&+\gret_{\rm eq}
\frac{\delta H_{\rm tot}(t_1)}{\hbar}
\gles_{\rm eq}
\frac{\delta H_{\rm tot}(t_2)}{\hbar}
\gadv_{\rm eq}
\frac{\delta H_{\rm tot}(t_3)}{\hbar}
\gadv_{\rm eq}+\\
&+\gles_{\rm eq}
\frac{\delta H_{\rm tot}(t_1)}{\hbar}
\gadv_{\rm eq}
\frac{\delta H_{\rm tot}(t_2)}{\hbar}
\gadv_{\rm eq}
\frac{\delta H_{\rm tot}(t_3)}{\hbar}\gadv_{\rm eq},
\end{aligned}
\ee
where 
$\gret_{\rm eq}$, $\gadv_{\rm eq}$ and $\gles_{\rm eq}$
are the retarded, advanced, and lesser Green's functions of the
unperturbed system, respectively,
and
\bege
\delta H_{\rm tot}(t)=\delta H_{\rm mag}(t)+\delta H_{\rm em}(t).
\ee
In Eq.~\eqref{eq_dy_third} we suppressed the two time arguments 
that each Green's function has for notational convenience. Additionally,
we suppressed the time-integrations over the intermediate times $t_1$,
$t_2$ and $t_3$ for notational brevity. How these time-integrals
are performed is clarified in the following Eq.~\eqref{eq_prod_four_green}.
The time-integration of 
the product of four Green's functions is given by
\bege\label{eq_prod_four_green}
\begin{aligned}
&\int\!\!\!
\rmd t_1
\rmd t_2
\rmd t_3
e^{-i\Omega_1 t_1}
e^{-i\Omega_2 t_2}
e^{-i\Omega_3 t_3} \\
&\times
G_{\rm eq}^{\alpha}(t,t_1)
G_{\rm eq}^{\alpha'}(t_1,t_2)
G_{\rm eq}^{\alpha''}(t_2,t_3)
G_{\rm eq}^{\alpha'''}(t_3,t)
=\\
&=
\frac{\rme^{-i[\Omega_1+\Omega_2+\Omega_3]t}}{2\pi}
\!\int\!\!
\rmd \Omega
G_{\rm eq}^{\alpha}(\Omega)
G_{\rm eq}^{\alpha'}(\Omega-\Omega_1)\\
&\times G_{\rm eq}^{\alpha''}(\Omega-\Omega_1-\Omega_2)
G_{\rm eq}^{\alpha'''}(\Omega-\Omega_1-\Omega_2-\Omega_3)
,\\
\end{aligned}
\ee
where $\alpha={\rm R}, {\rm A},<$ and $\Omega_i$ may take the
values $\pm\omega$ and $\pm\omega_{\rm mag}$ ($i$=1,2,3).
The following frequency combinations may contribute to the
magnonic SOT:
Case 1: $\Omega_1=\pm\omega$ and $\Omega_2=-\Omega_3=\pm\omega_{\rm mag}$. 
Case 2: $\Omega_2=\pm\omega$ and $\Omega_1=-\Omega_3=\pm\omega_{\rm mag}$. 
Case 3: $\Omega_3=\pm\omega$ and $\Omega_1=-\Omega_2=\pm\omega_{\rm mag}$. 

In order to make the equations more compact, we introduce the Keldysh Green's
function
\bege
\ghat_{\rm eq}(\Omega)=\begin{pmatrix}
\gret_{\rm eq}(\Omega) &\gles_{\rm eq}(\Omega)\\
0 &\gadv_{\rm eq}(\Omega)
\end{pmatrix}.
\ee
In case $j$ we obtain ($j=1,2,3$):
\bege
\ghat_{3,j}=\lim_{\omega\rightarrow 0}\frac{1}{8i\omega}
\sum_{u=\pm}\sum_{u'=\pm}u\ghat_{3,j}(u,u'),
\ee
where 
\bege\label{eq_case1}
\begin{aligned}
&\ghat_{3,1}(u,u')=\frac{\eta^2 e
\vn{E}_{0}}{2\pi\hbar^3}\cdot\!\!\!\!
\sum_{\gamma=x,y}\!\int\!\! \rmd\Omega \ghat_{\rm eq}(\Omega)
\vn{v} \ghat_{\rm eq}(\Omega-u\omega)\\
&\,\,\,\,\,\,\,\,\times \mathcal{T }_{\gamma}
\ghat_{{\rm eq},-u'\vn{q}}(\Omega-u\omega-u'\omega_{\rm mag})
\mathcal{T}_{\gamma}
\ghat_{\rm eq}(\Omega-u\omega)
\end{aligned}
\ee
in case 1,
\bege\label{eq_case2}
\begin{aligned}
&\ghat_{3,2}(u,u')=\frac{\eta^2 e
\vn{E}_{0}}{2\pi\hbar^3}\cdot\!\!\!\!
\sum_{\gamma=x,y}\!\int\!\! 
\rmd\Omega \ghat_{\rm eq}(\Omega)
\mathcal{T}_{\gamma}\\
&\times\ghat_{{\rm eq},-u'\vn{q}}(\Omega-u'\omega_{\rm mag})
\vn{v}_{-u'\vn{q}}\\
&\times\ghat_{{\rm eq},-u'\vn{q}}(\Omega-u\omega-u'\omega_{\rm mag})
\mathcal{T}_{\gamma}
\ghat_{\rm eq}(\Omega-u\omega)
\end{aligned}
\ee
in case 2, and
\bege\label{eq_case3}
\begin{aligned}
&\ghat_{3,3}(u,u')=\frac{\eta^2 e
\vn{E}_{0}}{2\pi\hbar^3}\cdot\!\!\!\!
\sum_{\gamma=x,y}\!\int\!\! 
\rmd\Omega \ghat_{\rm eq}(\Omega)
\mathcal{T}_{\gamma}\\
&\times\ghat_{{\rm eq},-u'\vn{q}}(\Omega-u'\omega_{\rm mag})
\mathcal{T}_{\gamma}
\ghat_{\rm eq}(\Omega)
\vn{v}
\ghat_{\rm eq}(\Omega-u\omega)
\end{aligned}
\ee
in case 3.
Green's functions and velocity operators that carry the momentum subscript $-u'\vn{q}$
are shifted in momentum space by $-u'\vn{q}$.

Summing up cases 1,2 and 3 we obtain 
\bege\label{eq_sum_up_cases}
\hat{G}_{3}=\sum_{j=1}^3
\hat{G}_{3,j}
=
\lim_{\omega \rightarrow 0}
\frac{1}{8i\omega}\sum_{u=\pm}\sum_{u' =\pm}
u \hat{G}_{3} (u,u'),
\ee
where
\bege\label{eq_g3_beginning}
\begin{aligned}
&\hat{G}_{3} (u,u')=\frac{\eta^2 e}{2\pi\hbar^3}
\int\rmd\Omega
\sum_{\gamma=x,y}\Bigl[\\
&\ghat_{\Omega} 
\vn{v}\cdot\vn{E}_{0} \ghat_{\Omega-u\omega} 
\mathcal{T}_{\gamma} 
\ghat_{\Omega-u\omega-u'\omega_{\rm mag},-u'\vn{q} }
\mathcal{T}_{\gamma} 
\ghat_{\Omega-u\omega }\\
&+
\ghat_{\Omega}
\mathcal{T}_{\gamma} 
\ghat_{\Omega-u'\omega_{\rm mag},-u'\vn{q} } 
\vn{v}_{-u'\vn{q}}\cdot\vn{E}_{0} \\
&\times\ghat_{\Omega-u\omega-u'\omega_{\rm mag},-u'\vn{q} } 
\mathcal{T}_{\gamma} 
\ghat_{\Omega-u\omega} \\
&+
\ghat_{\Omega}
\mathcal{T}_{\gamma} 
\ghat_{\Omega-u'\omega_{\rm mag},-u'\vn{q} } 
\mathcal{T}_{\gamma} 
\ghat_{\Omega} 
\vn{v}\cdot\vn{E}_{0} 
\ghat_{\Omega-u\omega}\Bigr].\\
\end{aligned}
\ee
Here, in order to save space we introduced the 
notation $\ghat_{\Omega}=\ghat_{\rm eq}(\Omega)$.
The SOT due to $\ghat_{3}$ is given by
\bege
\label{eq_torque_from_lesser}
\vn{T}_{\rm mag}=i{\rm Tr}
\left[
\vn{\mathcal{T}}
G^{<}_{3}
\right].
\ee

An important consistency check is that
Eq.~\eqref{eq_torque_from_lesser}
predicts a SOT of zero when there is no spin-orbit interaction.
This may be seen as follows: In the absence of spin-orbit interaction
the velocity operator is diagonal in spin-space and the Green
functions
$\hat{G}_{\Omega}$ are diagonal in spin-space as well. In contrast,
every torque operator causes a transition from spin-up to spin-down
or from spin-down to spin-up. Since the number of torque operators
in all summands is three, taking the trace in spin-space will yield
zero
when there is no spin-orbit coupling included in the calculation.

In the non-magnonic SOT the application
of an electric field generates a non-equilibrium spin density
perpendicular
to the magnetization, which exerts a torque on the magnetization.
The magnonic SOT described by Eq.~\eqref{eq_torque_from_lesser},
Eq.~\eqref{eq_sum_up_cases},
and Eq.~\eqref{eq_g3_beginning}
corresponds to processes where these non-equilibrium spins are
additionally flipped two times by a magnon. In the presence of 
spin-orbit interaction two consecutive spin-flips by a magnon
constitutes a non-trivial process because the non-equilibrium spins
may precess in the spin-orbit field in between the two spin flips.
This leads to a modification of the non-equilibrium spin density by
the magnons.

For the numerical evaluation of Eq.~\eqref{eq_g3_beginning}
it is convenient to perform a Taylor-expansion in $\vn{q}$ 
and $\omega_{\rm mag}$
as follows:
\bege\label{eq_taylor_gles}
\begin{aligned}
G^{<}_{3}(u,u')&=G^{<, (0,0)}_{3}(u,u')+G^{<, (1,0)}_{3}(u,u')+\\
&+G^{<, (2,0)}_{3}(u,u')+G^{<, (0,1)}_{3}(u,u')+\\
&+G^{<, (0,2)}_{3}(u,u')+\dots,
\end{aligned}
\ee
where
\bege\label{eq_gles_ij}
G^{<, (i,j)}_{3}(u,u')\propto
(u'\omega_{\rm mag})^{i} q^{j},
\ee
i.e., $G^{<, (i,j)}_{3}(u,u')$ is $i$-th order in $\omega_{\rm mag}$
and $j$-th order in $\vn{q}$ in the Taylor expansion of $G^{< }_{3}(u,u')$. 
A priori it is unclear whether all terms in the expansion
Eq.~\eqref{eq_taylor_gles} contribute to the magnonic SOT. Therefore,
we will evaluate them separately so that we can compare their magnitudes
later.

The contributions to $G^{<, (i,j)}_{3}(u,u')$ may be further
distinguished
according to the order of the derivative of the Fermi function that
they
contain. 
Derivatives of the Fermi function are produced when the derivatives
$\partial/\partial\omega$ or $\partial/\partial\omega_{\rm mag}$
act on the lesser Green's functions $\gles_{\rm eq}(\Omega-u\omega)$,
$\gles_{\rm eq}(\Omega-u'\omega_{\rm mag})$, or
$\gles_{\rm eq}(\Omega-u\omega-u'\omega_{\rm mag})$, because
$\gles_{\rm eq}(\Omega)=f(\Omega)[\gadv_{\rm eq}(\Omega)-\gret_{\rm
  eq}(\Omega)]$
contains the Fermi function $f(\Omega)$.
While we use the derivative $\partial/\partial\omega$ in order to take the
$\omega\rightarrow 0$ limit, the  $\partial/\partial\omega_{\rm mag}$
derivatives
are necessary for the Taylor expansion in $\omega_{\rm mag}$
according to Eq.~\eqref{eq_taylor_gles}
and Eq.~\eqref{eq_gles_ij}.
Following the standard notation used in linear response
theory
we label terms that contain $f$ with a superscript ${\rm II}$
(so-called 'lesser-two') and
terms that contain $f'$ with a superscript ${\rm I}$ (so-called 'lesser-one'). However, due to
the Taylor-expansion in $\omega_{\rm mag}$ we will encounter also higher
derivatives of $f$ that do not occur in standard linear response
theory.
We denote terms that involve the second derivative $f''$ with a
superscript ${\rm III}$ and terms that involve the third derivative
$f'''$
with a superscript ${\rm IV}$.

At zeroth order in $\omega_{\rm mag}$
and $\vn{q}$
the lesser-one contribution from
Eq.~\eqref{eq_g3_beginning}
is given by
\bege\label{eq_g3_zeroth_order}
\begin{aligned}
&G^{<,\rm I,  (0,0)}_{3} (u,u')=-u\omega e \int \rmd \Omega \sum_{\gamma=x,y} f'(\hbar\Omega)
\frac{\eta^2 \vn{E}_{0}}{2\pi\hbar^2}\cdot\Bigl[\\
&\gret_{\Omega} 
\vn{v}
\gspec_{\Omega}
\mathcal{T}_{\gamma} 
\gadv_{\Omega }
\mathcal{T}_{\gamma} 
\gadv_{\Omega }+\gret_{\Omega} 
\vn{v}  \gret_{\Omega} 
\mathcal{T}_{\gamma}  
\gspec_{\Omega}
\mathcal{T}_{\gamma} 
\gadv_{\Omega }\\
&+\gret_{\Omega} 
\vn{v}  \gret_{\Omega} 
\mathcal{T}_{\gamma}  
\gret_{\Omega }
\mathcal{T}_{\gamma} 
\gspec_{\Omega}
+
\gret_{\Omega}
\mathcal{T}_{\gamma} 
\gret_{\Omega} 
\vn{v}
\gspec_{\Omega}
\mathcal{T}_{\gamma} 
\gadv_{\Omega} \\
&+
\gret_{\Omega}
\mathcal{T}_{\gamma} 
\gret_{\Omega} 
\vn{v}
\gret_{\Omega} 
\mathcal{T}_{\gamma} 
\gspec_{\Omega}
+
\gret_{\Omega}
\mathcal{T}_{\gamma} 
\gret_{\Omega} 
\mathcal{T}_{\gamma} 
\gret_{\Omega} 
\vn{v}
\gspec_{\Omega}
\Bigr],\\
\end{aligned}
\ee
where $\gspec_{\Omega}=\gadv_{\Omega}-\gret_{\Omega}$.

$G^{<,\rm I,  (0,0)}_{3} (u,u')$ still needs to be summed over the
populated magnon modes. $G^{<,\rm I,  (0,0)}_{3} (u,u')$ itself
depends on the magnons only through $\eta$.
The effect of summing  $G^{<,(0,0)}_{3} (u,u')$ over the magnon
modes is therefore the multiplication by the number of magnons.
We Taylor-expand only the electronic lesser Green's function
in terms of $\omega_{\rm mag}$ and $\vn{q}$ and not the Bose-Einstein
distribution function. Therefore, we introduce the integral
\bege\label{eq_i00}
\begin{aligned}
I^{(0,0)}(T)&=\frac{1}{N_{\rm mag} A_{\rm mag}}\sum_{\vn{q}} F(\omega_{\rm mag}(\vn{q}),T)\\
&=\intqspatwo F(\omega_{\rm mag}(\vn{q}),T),
\end{aligned}
\ee
where $F(\omega_{\rm mag}(\vn{q}),T)$ is the Bose-Einstein distribution function
and $A_{\rm mag}$ is the area occupied by one magnetic site. For example,
in the case of Co/Pt magnetic bilayers, $A_{\rm mag}$ 
is the 
area of the unit cell.
This integral is evaluated below in section~\ref{sec_magnonic_integrals}.

Plugging Eq.~\eqref{eq_g3_zeroth_order} 
into Eq.~\eqref{eq_sum_up_cases}, summing over magnon-modes,
and using Eq.~\eqref{eq_torque_from_lesser} to evaluate the torque we obtain therefore
\bege\label{eq_g3_zeroth_order_magav}
\begin{aligned}
&
\vn{T}^{\rm I, (0,0)}_{\rm mag}=-
\frac{A_{\rm mag}  I^{(0,0)}(T)}{2\pi\hbar^2}
\int\rmd\Omega
\sum_{\gamma=x,y}
f'(\hbar\Omega)
{\rm Tr}\Bigl\{
\vn{\mathcal{T}}\\
&\times\gret_{\Omega} 
\Bigl[
\vn{v}
\gspec_{\Omega} 
\mathcal{T}_{\gamma} \gadv_{\Omega }
\mathcal{T}_{\gamma}
\gadv_{\Omega }+
\vn{v}  \gret_{\Omega} 
\mathcal{T}_{\gamma} 
\gspec_{\Omega }
\mathcal{T}_{\gamma}
\gadv_{\Omega }\\
&+
\vn{v} \gret_{\Omega} 
\mathcal{T}_{\gamma} \gret_{\Omega }
\mathcal{T}_{\gamma}
\gspec_{\Omega}+
\mathcal{T}_{\gamma}
\gret_{\Omega} 
\vn{v}
\gspec_{\Omega} 
\mathcal{T}_{\gamma}
\gadv_{\Omega} \\
&+
\mathcal{T}_{\gamma}
\gret_{\Omega} 
\vn{v}
\gret_{\Omega} 
\mathcal{T}_{\gamma}
\gspec_{\Omega} +
\mathcal{T}_{\gamma}
\gret_{\Omega} 
\mathcal{T}_{\gamma}
\gret_{\Omega} 
\vn{v} 
\gspec_{\Omega}\Bigr]
\cdot\vn{E}_{0}e
\Bigl\},\\
\end{aligned}
\ee
where we made use of $\eta^2=2/N_{\rm mag}$ (see Eq.~\eqref{eq_relate_eta_N}).

Similarly, we may extract the lesser-two contribution from
Eq.~\eqref{eq_g3_beginning}
and evaluate the corresponding torque, which is given in 
Eq.~\eqref{eq_app_t00_fermisea} in the Appendix.

The next contribution to the Taylor-expansion 
is $G^{<,\rm I, (1,0)}_{3}(u,u')$ (see Eq.~\eqref{eq_taylor_gles}).
According to Eq.~\eqref{eq_gles_ij}
we have $G^{<,\rm I, (1,0)}_{3}(u,u')\propto  u'\omega_{\rm mag}$.
Since we need to sum over $u'=\pm 1$, 
this does not contribute to the magnonic
SOT.
The following contribution $G^{<,\rm I, (2,0)}_{3}(u,u')$ (see Eq.~\eqref{eq_taylor_gles})
requires us to extract the terms quadratic in $\omega_{\rm mag}$ 
from Eq.~\eqref{eq_g3_beginning}. We obtain
\bege\label{eq_g3_2ndorderomegamag}
\begin{aligned}
&\hat{G}^{(2,0)}_{3} (u,u')=
-\frac{e u\omega
[\omega_{\rm mag}]^2}{4\pi}
\frac{\eta^2}{\hbar^3}\int\rmd\Omega\sum_{\gamma=x,y}
\ghat_{\Omega} 
\vn{E}_{0}\cdot\Bigl[
\\
&
\vn{v}
\frac{\partial
\ghat_{\Omega}}{\partial \Omega}
\mathcal{T}_{\gamma} 
\frac{\partial^2
\ghat_{\Omega}}
{\partial \Omega^2}
\mathcal{T}_{\gamma}
\ghat_{\Omega }+
\vn{v}\ghat_{\Omega} 
\mathcal{T}_{\gamma} 
\frac{\partial^3
\ghat_{\Omega }}
{\partial \Omega^3}
\mathcal{T}_{\gamma}
\ghat_{\Omega }\\
&
+
\vn{v} \ghat_{\Omega} 
\mathcal{T}_{\gamma} 
\frac{\partial^2
\ghat_{\Omega}}
{\partial \Omega^2}
\mathcal{T}_{\gamma}
\frac{\partial
\ghat_{\Omega }}{\partial \Omega}+
\mathcal{T}_{\gamma}
\frac{\partial^2
\ghat_{\Omega} 
}{\partial^2 \Omega}
\vn{v}
\frac{\partial
\ghat_{\Omega} 
}{\partial \Omega}
\mathcal{T}_{\gamma}
\ghat_{\Omega} \\
&+
\mathcal{T}_{\gamma}
\frac{\partial^2
\ghat_{\Omega} 
}{\partial \Omega^2}
\vn{v}
\ghat_{\Omega} 
\mathcal{T}_{\gamma}
\frac{\partial
\ghat_{\Omega} 
}{\partial \Omega}+
\mathcal{T}_{\gamma}
\ghat_{\Omega} 
\vn{v}
\frac{\partial^3
\ghat_{\Omega} 
}{\partial \Omega^3}
\mathcal{T}_{\gamma}
\ghat_{\Omega}\\
&+
\mathcal{T}_{\gamma}
\ghat_{\Omega} 
\vn{v}
\frac{\partial^2
\ghat_{\Omega} 
}{\partial \Omega^2}
\mathcal{T}_{\gamma}
\frac{\partial
\ghat_{\Omega} 
}{\partial \Omega}
+2
\mathcal{T}_{\gamma}
\frac{\partial
\ghat_{\Omega} 
}{\partial \Omega}
\vn{v}
\frac{\partial^2
\ghat_{\Omega} 
}{\partial \Omega^2}
\mathcal{T}_{\gamma}
\ghat_{\Omega} \\
&+\!2
\mathcal{T}_{\gamma}
\frac{\partial
\ghat_{\Omega} 
}{\partial \Omega}
\vn{v}
\frac{\partial
\ghat_{\Omega} 
}{\partial \Omega}
\mathcal{T}_{\gamma}
\frac{\partial
\ghat_{\Omega} 
}{\partial \Omega}
\!+\!
\mathcal{T}_{\gamma}
\frac{\partial^2
\ghat_{\Omega }}
{\partial \Omega^2}
\mathcal{T}_{\gamma}
\ghat_{\Omega} 
\vn{v}
\frac{\partial
\ghat_{\Omega}
}{\partial \Omega}\Bigr].
\\
\end{aligned}
\ee
From this we extract the lesser-one contribution
\bege\label{eq_glesone_20}
\begin{aligned}
&
G_{3}^{<,\rm I, (2,0)}(u,u')=
-
\int\rmd\Omega f'(\hbar\Omega) \sum_{\gamma=x,y}
\\
&
\gret_{\Omega}\Bigl[ 
\vn{v} 
\gspec_{\Omega}
\mathcal{T}_{\gamma} 
\frac{\partial^2
\gadv_{\Omega}}
{\partial \Omega^2}
\mathcal{T}_{\gamma}
\gadv_{\Omega }+2
\vn{v}
\frac{\partial
\gret_{\Omega}}{\partial \Omega}
\mathcal{T}_{\gamma} 
\frac{\partial
\gspec_{\Omega}}
{\partial \Omega}
\mathcal{T}_{\gamma}
\gadv_{\Omega }\\
&+3
\vn{v} \gret_{\Omega} 
\mathcal{T}_{\gamma} 
\frac{\partial^2
\gspec_{\Omega }}
{\partial {\Omega^2}}
\mathcal{T}_{\gamma}
\gadv_{\Omega }+2
\vn{v} \gret_{\Omega} 
\mathcal{T}_{\gamma} 
\frac{\partial
\gspec_{\Omega}}
{\partial \Omega}
\mathcal{T}_{\gamma}
\frac{\partial
\gadv_{\Omega }}{\partial \Omega}\\
&
+
\vn{v} \gret_{\Omega} 
\mathcal{T}_{\gamma} 
\frac{\partial^2
\gret_{\Omega}}
{\partial \Omega^2}
\mathcal{T}_{\gamma}
\gspec_{\Omega }+2
\mathcal{T}_{\gamma}
\frac{\partial
\gspec_{\Omega} 
}{\partial \Omega}
\vn{v}
\frac{\partial
\gadv_{\Omega} 
}{\partial \Omega}
\mathcal{T}_{\gamma}
\gadv_{\Omega} 
\\
&+
\mathcal{T}_{\gamma}
\frac{\partial^2
\gret_{\Omega} 
}{\partial^2 \Omega}
\vn{v}
\gspec_{\Omega} 
\mathcal{T}_{\gamma}
\gadv_{\Omega} 
+
2\mathcal{T}_{\gamma}
\frac{\partial
\gspec_{\Omega} 
}{\partial \Omega}
\vn{v}
\gadv_{\Omega} 
\mathcal{T}_{\gamma}
\frac{\partial
\gadv_{\Omega} 
}{\partial \Omega}
\\
&+
\mathcal{T}_{\gamma}
\frac{\partial^2
\gret_{\Omega} 
}{\partial^2 \Omega}
\vn{v}
\gret_{\Omega} 
\mathcal{T}_{\gamma}
\gspec_{\Omega} +3
\mathcal{T}_{\gamma}
\gret_{\Omega} 
\vn{v}
\frac{\partial^2
\gspec_{\Omega} 
}{\partial \Omega^2}
\mathcal{T}_{\gamma}
\gadv_{\Omega}\\
&+2
\mathcal{T}_{\gamma}
\gret_{\Omega} 
\vn{v}
\frac{\partial
\gspec_{\Omega} 
}{\partial \Omega}
\mathcal{T}_{\gamma}
\frac{\partial
\gadv_{\Omega} 
}{\partial \Omega}+
\mathcal{T}_{\gamma}
\gret_{\Omega} 
\vn{v}
\frac{\partial^2
\gret_{\Omega} 
}{\partial \Omega^2}
\mathcal{T}_{\gamma}
\gspec_{\Omega} 
\\
&+2
\mathcal{T}_{\gamma}
\gspec_{\Omega} 
\vn{v}
\frac{\partial^2
\gadv_{\Omega} 
}{\partial \Omega^2}
\mathcal{T}_{\gamma}
\gadv_{\Omega}+4
\mathcal{T}_{\gamma}
\frac{\partial
\gret_{\Omega} 
}{\partial \Omega}
\vn{v}
\frac{\partial
\gspec_{\Omega} 
}{\partial \Omega}
\mathcal{T}_{\gamma}
\gadv_{\Omega} \\
&+2
\mathcal{T}_{\gamma}
\gspec_{\Omega} 
\vn{v}
\frac{\partial
\gadv_{\Omega} 
}{\partial \Omega}
\mathcal{T}_{\gamma}
\frac{\partial
\gadv_{\Omega} 
}{\partial \Omega}
+2
\mathcal{T}_{\gamma}
\frac{\partial
\gret_{\Omega} 
}{\partial \Omega}
\vn{v}
\gspec_{\Omega} 
\mathcal{T}_{\gamma}
\frac{\partial
\gadv_{\Omega} 
}{\partial \Omega}
\\
&+2
\mathcal{T}_{\gamma}
\frac{\partial
\gret_{\Omega} 
}{\partial \Omega}
\vn{v}
\frac{\partial
\gret_{\Omega} 
}{\partial \Omega}
\mathcal{T}_{\gamma}
\gspec_{\Omega} 
+2
\mathcal{T}_{\gamma}
\frac{\partial
\gspec_{\Omega }}
{\partial \Omega}
\mathcal{T}_{\gamma}
\gadv_{\Omega} 
\vn{v}
\frac{\partial
\gadv_{\Omega}
}{\partial \Omega}
\\
&+
\mathcal{T}_{\gamma}
\frac{\partial^2
\gret_{\Omega }}
{\partial \Omega^2}
\mathcal{T}_{\gamma}
\gret_{\Omega} 
\vn{v}
\gspec_{\Omega}
\Bigr]\cdot\vn{E}_{0}\frac{e}{4\pi}
u\omega
[\omega_{\rm mag}]^2
\frac{\eta^2}{\hbar^2}.
\\
\end{aligned}
\ee

Eq.~\eqref{eq_glesone_20} depends on the magnons through
$\eta^2$ and through $\omega_{\rm mag}^2$. 
Consequently, in order to perform the sampling over magnon modes 
we introduce the integral
\bege\label{eq_i20}
I^{(2,0)}(T)=\intqspatwo 
[\hbar\omega_{\rm mag}(\vn{q})]^2
F(\vn{q},T),
\ee
which we discuss below in section~\ref{sec_magnonic_integrals}.
Thus, employing Eq.~\eqref{eq_sum_up_cases} and Eq.~\eqref{eq_torque_from_lesser}
yields the following contribution to the SOT
after summing over the magnon modes: 
\bege
\label{eq_torque_20_one}
\begin{aligned}
&
\vn{T}^{\rm I, (2,0)}_{\rm mag}=-
\frac{A_{\rm mag}  I^{(2,0)}(T)}{4\pi\hbar^4}
\int\rmd\Omega
\sum_{\gamma=x,y}
f'(\hbar\Omega)
{\rm Tr}\Bigl\{
\vn{\mathcal{T}}\\
&
\times\gret_{\Omega}\Bigl[ 
\vn{v} 
\gspec_{\Omega}
\mathcal{T}_{\gamma} 
\frac{\partial^2
\gadv_{\Omega}}
{\partial \Omega^2}
\mathcal{T}_{\gamma}
\gadv_{\Omega }+2
\vn{v}
\frac{\partial
\gret_{\Omega}}{\partial \Omega}
\mathcal{T}_{\gamma} 
\frac{\partial
\gspec_{\Omega}}
{\partial \Omega}
\mathcal{T}_{\gamma}
\gadv_{\Omega }\\
&+3
\vn{v} \gret_{\Omega} 
\mathcal{T}_{\gamma} 
\frac{\partial^2
\gspec_{\Omega }}
{\partial {\Omega^2}}
\mathcal{T}_{\gamma}
\gadv_{\Omega }+2
\vn{v} \gret_{\Omega} 
\mathcal{T}_{\gamma} 
\frac{\partial
\gspec_{\Omega}}
{\partial \Omega}
\mathcal{T}_{\gamma}
\frac{\partial
\gadv_{\Omega }}{\partial \Omega}\\
&
+
\vn{v} \gret_{\Omega} 
\mathcal{T}_{\gamma} 
\frac{\partial^2
\gret_{\Omega}}
{\partial \Omega^2}
\mathcal{T}_{\gamma}
\gspec_{\Omega }+2
\mathcal{T}_{\gamma}
\frac{\partial
\gspec_{\Omega} 
}{\partial \Omega}
\vn{v}
\frac{\partial
\gadv_{\Omega} 
}{\partial \Omega}
\mathcal{T}_{\gamma}
\gadv_{\Omega} 
\\
&+
\mathcal{T}_{\gamma}
\frac{\partial^2
\gret_{\Omega} 
}{\partial^2 \Omega}
\vn{v}
\gspec_{\Omega} 
\mathcal{T}_{\gamma}
\gadv_{\Omega} 
+
2\mathcal{T}_{\gamma}
\frac{\partial
\gspec_{\Omega} 
}{\partial \Omega}
\vn{v}
\gadv_{\Omega} 
\mathcal{T}_{\gamma}
\frac{\partial
\gadv_{\Omega} 
}{\partial \Omega}
\\
&+
\mathcal{T}_{\gamma}
\frac{\partial^2
\gret_{\Omega} 
}{\partial^2 \Omega}
\vn{v}
\gret_{\Omega} 
\mathcal{T}_{\gamma}
\gspec_{\Omega} +3
\mathcal{T}_{\gamma}
\gret_{\Omega} 
\vn{v}
\frac{\partial^2
\gspec_{\Omega} 
}{\partial \Omega^2}
\mathcal{T}_{\gamma}
\gadv_{\Omega}\\
&+2
\mathcal{T}_{\gamma}
\gret_{\Omega} 
\vn{v}
\frac{\partial
\gspec_{\Omega} 
}{\partial \Omega}
\mathcal{T}_{\gamma}
\frac{\partial
\gadv_{\Omega} 
}{\partial \Omega}+
\mathcal{T}_{\gamma}
\gret_{\Omega} 
\vn{v}
\frac{\partial^2
\gret_{\Omega} 
}{\partial \Omega^2}
\mathcal{T}_{\gamma}
\gspec_{\Omega} 
\\
&+2
\mathcal{T}_{\gamma}
\gspec_{\Omega} 
\vn{v}
\frac{\partial^2
\gadv_{\Omega} 
}{\partial \Omega^2}
\mathcal{T}_{\gamma}
\gadv_{\Omega}+4
\mathcal{T}_{\gamma}
\frac{\partial
\gret_{\Omega} 
}{\partial \Omega}
\vn{v}
\frac{\partial
\gspec_{\Omega} 
}{\partial \Omega}
\mathcal{T}_{\gamma}
\gadv_{\Omega} \\
&+2
\mathcal{T}_{\gamma}
\gspec_{\Omega} 
\vn{v}
\frac{\partial
\gadv_{\Omega} 
}{\partial \Omega}
\mathcal{T}_{\gamma}
\frac{\partial
\gadv_{\Omega} 
}{\partial \Omega}
+2
\mathcal{T}_{\gamma}
\frac{\partial
\gret_{\Omega} 
}{\partial \Omega}
\vn{v}
\gspec_{\Omega} 
\mathcal{T}_{\gamma}
\frac{\partial
\gadv_{\Omega} 
}{\partial \Omega}
\\
&+2
\mathcal{T}_{\gamma}
\frac{\partial
\gret_{\Omega} 
}{\partial \Omega}
\vn{v}
\frac{\partial
\gret_{\Omega} 
}{\partial \Omega}
\mathcal{T}_{\gamma}
\gspec_{\Omega} 
+2
\mathcal{T}_{\gamma}
\frac{\partial
\gspec_{\Omega }}
{\partial \Omega}
\mathcal{T}_{\gamma}
\gadv_{\Omega} 
\vn{v}
\frac{\partial
\gadv_{\Omega}
}{\partial \Omega}
\\
&+
\mathcal{T}_{\gamma}
\frac{\partial^2
\gret_{\Omega }}
{\partial \Omega^2}
\mathcal{T}_{\gamma}
\gret_{\Omega} 
\vn{v}
\gspec_{\Omega}
\Bigr]\Bigr\}\cdot\vn{E}_{0} e.
\\
\end{aligned}
\ee
Similarly, we obtain $\vn{T}^{\rm II, (2,0)}_{\rm mag}$
from the lesser-two contribution to the Green's function,
which is given in Eq.~\eqref{eq_torque_20_less2} in the Appendix.
Due to the derivatives with respect to $\omega_{\rm mag}$
there are additionally the contributions 
$\vn{T}^{\rm III, (2,0)}_{\rm mag}$
and 
$\vn{T}^{\rm IV, (2,0)}_{\rm mag}$
from the lesser-three and lesser-four Green's functions, respectively.
The explicit expressions are given in Eq.~\eqref{eq_torque_20_less3}
and Eq.~\eqref{eq_torque_20_less4}
 in the Appendix.

The next contribution to the Taylor-expansion is
$G^{<,\rm I, (0,1)}_{3}(u,u')$ (see Eq.~\eqref{eq_taylor_gles}).
Since it is linear in $\vn{q}$, the average
over magnon modes evaluates to zero for it.
The next non-zero contribution is therefore $G^{<,\rm I,(0,2)}_{3}(u,u')$. 
The Taylor-expansion of
Eq.~\eqref{eq_case1},
Eq.~\eqref{eq_case2}
and Eq.~\eqref{eq_case3} 
up to second order in $\vn{q}$
and up to zeroth order in $\omega_{\rm mag}$ yields
the lesser-one contributions 
\bege\label{eq_secondq_case1}
\begin{aligned}
&\ghat^{<,\rm I,(0,2)}_{3,1}(u,u')
=-u\omega\frac{1}{4\pi}
\frac{\eta^2}{\hbar^2}e\sum_{\gamma=x,y}
\sum_{ij}q_iq_j
\int\rmd\Omega  \\
&
\times f'(\hbar\Omega)\gret_{\Omega} \vn{E}_{0}\cdot\Bigl[
\vn{v}  
\gspec_{\Omega} 
\mathcal{T}_{\gamma} 
\frac{\partial^2 \gadv_{\Omega,\vn{q} }}{\partial q_i \partial q_j}
\mathcal{T}_{\gamma}
\gadv_{\Omega }
\\
&+
\vn{v} 
\gret_{\Omega} 
\mathcal{T}_{\gamma} 
\frac{\partial^2 \gspec_{\Omega,\vn{q} }}{\partial q_i \partial q_j}
\mathcal{T}_{\gamma}
\gadv_{\Omega }
+
\vn{v} 
\gret_{\Omega} 
\mathcal{T}_{\gamma} 
\frac{\partial^2 \gret_{\Omega,\vn{q} }}{\partial q_i \partial q_j}
\mathcal{T}_{\gamma}
\gspec_{\Omega }
\Bigr]
\\
\end{aligned}
\ee
in case 1, 
\bege\label{eq_secondq_case2}
\begin{aligned}
&
\ghat^{<,\rm I,(0,2)}_{3,2}(u,u')
=-u\omega
\sum_{ij}
\sum_{\gamma=x,y}
\int\rmd\Omega f'(\hbar\Omega)
\\
&\times\frac{1}{4\pi}q_iq_j\frac{\eta^2}{\hbar^2}e\vn{E}_{0}\cdot\gret_{\Omega}\mathcal{T}_{\gamma}\Bigl[
\frac{\partial^2
\gret_{\Omega,\vn{q}} 
}{\partial q_i \partial q_j }
\vn{v}
\gspec_{\Omega} 
\mathcal{T}_{\gamma}
\gadv_{\Omega}\\ 
&+
\frac{\partial^2
\gret_{\Omega,\vn{q}} 
}{\partial q_i \partial q_j }
\vn{v}
\gret_{\Omega} 
\mathcal{T}_{\gamma}
\gspec_{\Omega} 
+
\frac{\partial
\gret_{\Omega,\vn{q}} 
}{\partial q_i   }
\vn{v}
\frac{\partial
\gspec_{\Omega,\vn{q}} 
}{\partial q_j }
\mathcal{T}_{\gamma}
\gadv_{\Omega} \\
&+
\frac{\partial
\gret_{\Omega,\vn{q}} 
}{\partial q_i}
\vn{v}
\frac{\partial
\gret_{\Omega,\vn{q}} 
}{\partial q_j}
\mathcal{T}_{\gamma}
\gspec_{\Omega} 
+
\frac{\partial
\gret_{\Omega,\vn{q}} 
}{\partial q_j}
\vn{v}
\frac{\partial
\gspec_{\Omega,\vn{q}} 
}{\partial q_i  }
\mathcal{T}_{\gamma}
\gadv_{\Omega}\\
&+
\frac{\partial
\gret_{\Omega,\vn{q}} 
}{\partial q_j}
\vn{v}
\frac{\partial
\gret_{\Omega,\vn{q}} 
}{\partial q_i}
\mathcal{T}_{\gamma}
\gspec_{\Omega}+
\gret_{\Omega} 
\vn{v}
\frac{\partial^2
\gspec_{\Omega,\vn{q}} 
}{\partial q_i  \partial q_j  }
\mathcal{T}_{\gamma}
\gadv_{\Omega}\\
&+
\gret_{\Omega} 
\vn{v}
\frac{\partial^2
\gret_{\Omega,\vn{q}} 
}{\partial q_i  \partial q_j}
\mathcal{T}_{\gamma}
\gspec_{\Omega}\\
&+\frac{\partial
\gret_{\Omega,\vn{q}} 
}{\partial q_i  }
\frac{\partial
\vn{v}}{\partial q_j}
\gspec_{\Omega} 
\mathcal{T}_{\gamma}
\gadv_{\Omega}
+\frac{\partial
\gret_{\Omega,\vn{q}} 
}{\partial q_j  }
\frac{\partial
\vn{v}}{\partial q_i}
\gspec_{\Omega} 
\mathcal{T}_{\gamma}
\gadv_{\Omega}\\
&+
\gret_{\Omega} 
\frac{\partial
\vn{v}}{\partial q_j}
\frac{\partial
\gspec_{\Omega,\vn{q}} 
}{\partial q_i}
\mathcal{T}_{\gamma}
\gadv_{\Omega}
+
\gret_{\Omega} 
\frac{\partial
\vn{v}}{\partial q_i}
\frac{\partial
\gspec_{\Omega,\vn{q}} 
}{\partial q_j}
\mathcal{T}_{\gamma}
\gadv_{\Omega}
\\ 
&+\frac{\partial
\gret_{\Omega,\vn{q}} 
}{\partial q_i  }
\frac{\partial
\vn{v}}{\partial q_j}
\gret_{\Omega} 
\mathcal{T}_{\gamma}
\gspec_{\Omega}
+\frac{\partial
\gret_{\Omega,\vn{q}} 
}{\partial q_j  }
\frac{\partial
\vn{v}}{\partial q_i}
\gret_{\Omega} 
\mathcal{T}_{\gamma}
\gspec_{\Omega}\\
&+
\gret_{\Omega} 
\frac{\partial
\vn{v}}{\partial q_j}
\frac{\partial
\gret_{\Omega,\vn{q}} 
}{\partial q_i}
\mathcal{T}_{\gamma}
\gspec_{\Omega}
+
\gret_{\Omega} 
\frac{\partial
\vn{v}}{\partial q_i}
\frac{\partial
\gret_{\Omega,\vn{q}} 
}{\partial q_j}
\mathcal{T}_{\gamma}
\gspec_{\Omega}
\Bigr]
\end{aligned}
\ee
in case 2, and 
\bege\label{eq_secondq_case3}
\begin{aligned}
&
\ghat^{<,\rm I,(0,2)}_{3,3}(u,u')
=-u\omega e\vn{E}_{0}\cdot
\sum_{ij}q_iq_j
\sum_{\gamma=x,y}
\int\rmd\Omega \\
&\times f'(\hbar\Omega)\gret_{\Omega}
\mathcal{T}_{\gamma}
\frac{\partial^2
\gret_{\Omega,\vn{q}} 
}{\partial q_i  \partial q_j}
\mathcal{T}_{\gamma}
\gret_{\Omega} 
\vn{v}
\gspec_{\Omega}\frac{1}{4\pi} 
\frac{\eta^2}{\hbar^2}
\end{aligned}
\ee
in case 3.

These lesser-one Green's functions depend on the magnons 
through $\eta^2$ and through $q_{i}q_{j}$. Consequently,
we use the integral
\bege\label{eq_i02}
I^{(0,2)}_{ij}(T)=\intqspatwo q_{i}q_{j} F(\vn{q},T)
\ee
in order to average over the magnon distribution.
This integral is discussed below in section~\ref{sec_magnonic_integrals}. 
Using  Eq.~\eqref{eq_sum_up_cases} and Eq.~\eqref{eq_torque_from_lesser}
and summing over the magnon modes we obtain
the torque
$\vn{T}^{\rm I, (0,2)}_{\rm mag}$
from the lesser-one Green's functions in 
Eq.~\eqref{eq_secondq_case1},
Eq.~\eqref{eq_secondq_case2},
and Eq.~\eqref{eq_secondq_case3}.
The explicit expression is given in
Eq.~\eqref{eq_app_secondq_torque_lesserone}
in the Appendix.
Similarly, we obtain the torque Eq.~\eqref{eq_app_secondq_torque_lessertwo}
from the lesser-two
Green's function. 

We introduce the torkance tensors $t^{I,(J,K)}_{{\rm mag},ij}$~\cite{ibcsoit}
so that
\bege
\vn{T}^{ I,(J,K)}_{\rm mag}=\sum_{i=1}^3
\sum_{j=1}^2 \hat{\vn{e}}_{i}
t^{I,(J,K)}_{{\rm mag},ij} E_{0,j},
\ee
where $\vn{E}_{0}=(E_{0,x},E_{0,y},0)$ is the applied in-plane
electric field (applied in the plane of the magnetic bilayer, therefore
no $z$ component), $\hat{\vn{e}}_{i}$ is the unit vector along the
$i$-th Cartesian direction, $I={\rm I, II, III, IV}$, $J=0,2$ and $K=0,2$. 
When periodic boundary conditions are used, the Green's functions
depend on a $k$-point, which we suppress for notational simplicity.
In this case an additional $k$-integration is necessary, i.e.,
we use
\bege
\begin{aligned}
&t^{\rm tot}_{{\rm mag},ij}=
\intkspatwo 
[
t^{\rm I,(0,0)}_{{\rm mag},ij}
+
t^{\rm II,(0,0)}_{{\rm mag},ij}
+
t^{\rm I,(2,0)}_{{\rm mag},ij}
+
t^{\rm II,(2,0)}_{{\rm mag},ij}\\
&+
t^{\rm III,(2,0)}_{{\rm mag},ij}
+
t^{\rm IV,(2,0)}_{{\rm mag},ij}
+
t^{\rm I,(0,2)}_{{\rm mag},ij}
+
t^{\rm II,(0,2)}_{{\rm mag},ij}
]
\end{aligned}
\ee
in order to obtain the total torkance.

\subsection{Integrals over magnon modes}
\label{sec_magnonic_integrals}
In the previous subsection we introduced 
integrals over magnon modes in Eq.~\eqref{eq_i00},
Eq.~\eqref{eq_i20}, and Eq.~\eqref{eq_i02}.
In order to evaluate these
integrals, we assume that the magnon dispersion is given by
\bege\label{eq_magno_dispers_isotro}
\omega_{\rm mag}(\vn{q})=\mathcal{A}q^2+\mathcal{C},
\ee
where $\mathcal{A}$ is the spin-wave stiffness and
$\mathcal{C}$ is the spin-wave gap.
In principle, the $\vn{q}$ integrals should be restricted to the
first Brillouin-zone in $\vn{q}$-space, the volume of which is
reciprocal to $A_{\rm mag}$. However, for the examples considered
here one introduces only a small error by waiving the restriction 
to the first Brillouin-zone and integrating instead over the full
$\vn{q}$-space. Therefore, we integrate in the following
over the full $\vn{q}$-space, which has the advantage 
that the integrals are
given then by analytical expressions.

The first integral is
\bege
\begin{aligned}
I^{(0,0)}(T)&=\intqspatwo F(\omega_{\rm mag}(\vn{q}),T)=\\
&=\frac{1}{2\pi}\int \frac{q d\, q}{e^{\hbar \omega_{\rm mag }(q)/(k_{\rm B}T)}-1}=\\
&=\frac{1}{4\pi  \mathcal{A} }\int_{\mathcal{C}}^{\infty} \frac{ d\, \omega_{\rm mag }}{e^{\hbar \omega_{\rm mag }/(k_{\rm B}T)}-1}=\\
&=\frac{1}{4\pi  \mathcal{A} }
\frac{k_{\rm B}T}{\hbar}
\int^{\infty}_{\hbar\mathcal{C}/(k_{\rm B}T)} \frac{d\, \xi}{e^{\xi}-1}=\\
&=\frac{1}{4\pi  \mathcal{A} }
\frac{k_{\rm B}T}{\hbar}
\left[
\frac{\hbar\mathcal{C}}{k_{\rm B}T}-
\log\left(
e^{\frac{\hbar\mathcal{C}}{k_{\rm B}T}}-1
\right)
\right]
,
\\
\end{aligned}
\ee
which diverges when the magnon gap $\mathcal{C}$ goes to zero.
Here, $k_{\rm B}$ is the Boltzmann constant.

The second integral is (assuming $\mathcal{C}=0$)
\bege
\begin{aligned}
I^{(2,0)}(T)&=\intqspatwo 
[\hbar\omega_{\rm mag}(\vn{q})]^2
F(\omega_{\rm mag}(\vn{q}),T)=\\
%&=\frac{1}{2\pi}\int \frac{  [\hbar\omega_{\rm mag}(q)]^2   q d\, q}{e^{\hbar \omega_{\rm mag }(q)/(k_{\rm B}T)}-1}=\\
%&=\frac{\hbar^2}{4\pi  \mathcal{A} }\int \frac{ [\omega_{\rm mag }]^{2} d\, \omega_{\rm mag }}{e^{\hbar \omega_{\rm mag }/(k_{\rm B}T)}-1}=\\
%&=\frac{\hbar^2}{4\pi  \mathcal{A} }\left[\frac{k_{\rm B}T}{\hbar}\right]^{3}\int \frac{[\xi]^{2} d\, \xi}{e^{\xi}-1}=\\
&=\frac{\hbar^2}{2\pi  \mathcal{A} }
\left[
\frac{k_{\rm B}T}{\hbar}
\right]^{3}\zeta(3),
\end{aligned}
\ee
were  $\zeta$ denotes the Zeta function, i.e., 
$\zeta(3)\approx 1.202$.

For the isotropic dispersion of Eq.~\eqref{eq_magno_dispers_isotro}
the third integrals satisfy
$I^{(0,2)}_{xx}(T)=I^{(0,2)}_{yy}(T)=I^{(0,2)}(T)/2$,
with (assuming $\mathcal{C}=0$)
\bege
\begin{aligned}
I^{(0,2)}(T)&=\intqspatwo  q^2 F(\vn{q},T)\\
%&=\frac{1}{2\pi \mathcal{A}}\int \frac{ \omega_{\rm mag}(q)   q d\, q}{e^{\hbar \omega_{\rm mag }(q)/(k_{\rm B}T)}-1}=\\
%&=\frac{1}{4\pi  \mathcal{A}^{2} }\int \frac{ \omega_{\rm mag } d\, \omega_{\rm mag }}{e^{\hbar \omega_{\rm mag }/(k_{\rm B}T)}-1}=\\
%&=\frac{1}{4\pi  \mathcal{A}^{2} }\left[\frac{k_{\rm B}T}{\hbar}\right]^{2}\int \frac{\xi d\, \xi}{e^{\xi}-1}=\\
&=\frac{\pi}{24 \mathcal{A}^{2} }
\left[
\frac{k_{\rm B}T}{\hbar}
\right]^{2}.
\end{aligned}
\ee

In table~\ref{table_magnonic_integrals_twodim}
we list the values of these integrals for various ferromagnets. 
For the
spin-wave stiffnesses we took 
bulk values from
the literature~\cite{magnetic_excitations_iron,magnons_iron_cobalt_nickel_sandratskii,mook_spin_waves_Ni}.
$A_{\rm mag}$
is the in-plane area of the unit cell per magnetic atom.
These areas are
$A_{\rm mag}=4.109\,\mathring{{\rm A}}^{2}$
in the case of Fe,
$A_{\rm mag}=2.723\,\mathring{{\rm A}}^{2}$
in the case of Co,
and
$A_{\rm mag}=3.107\, \mathring{{\rm A}}^{2}$
in the case of Ni.
The data for Mn
correspond to a monolayer of Mn on W(001)~\cite{heide_dmi_mnw}
with $A_{\rm mag}= 10.018\, \mathring{{\rm A}}^{2}$.
For the first integral $I^{(0,0)}$ we used a magnon gap
of $0.1$~meV, which ensures convergence, while
the values of the second and third integrals are almost 
not affected by this small gap of $0.1$~meV and therefore
their values are almost identical to the analytical expressions
above with $\mathcal{C}=0$. 
Since the STM experiments on Mn/W(001) were performed at $T=$13~K
we set the temperature in the integrals to $T=$13~K (Mn-13K).
As the spin-wave stiffness of the Mn monolayer is much smaller than
the spin-wave stiffnesses of Fe, Co and Ni, the integrals $I^{(0,0)}$
and $I^{(0,2)}$ in Mn at $T=$13~K are similar in size to the ones of Fe, Co, and Ni
at $T=$300~K. 

\begin{threeparttable}
\caption{
Integrals  
$I^{(0,0)}(T)A_{\rm mag}$,
$I^{(2,0)}(T)A_{\rm mag}$,
and
$I^{(0,2)}(T)A_{\rm mag}$
at temperature $T=$300~K for various ferromagnets. In the case
of Mn-13K the temperature is $T=$13~K.
}
\label{table_magnonic_integrals_twodim}
\begin{ruledtabular}
\begin{tabular}{c|c|c|c|c}
&$\mathcal{A}$ 
&$I^{(0,0)}A_{\rm mag}$
&$I^{(2,0)}A_{\rm mag}$ 
&$I^{(0,2)}A_{\rm mag}$ 
\\
&[${\rm meV}\mathring{{\rm A}}^2$]
&
&[(eV)$^2$]
&[$\mathring{{\rm A}}^{-2}$]
\\
\hline
Fe
& 307 
& $0.153$
& $4.424\cdot 10^{-5}$
& $3.814\cdot 10^{-3}$
\\
\hline
Co
&539 
& $5.78\cdot 10^{-2}$
& $1.67\cdot 10^{-5}$
& $8.20\cdot 10^{-4}$
\\
\hline
Ni
&433
& $8.203\cdot 10^{-2}$
& $2.372\cdot 10^{-5}$
& $1.450\cdot 10^{-3}$
\\
\hline
Mn-13K
&56
& $3.923\cdot 10^{-2}$
& $4.808\cdot 10^{-8}$
& $5.246\cdot 10^{-4}$
\\
\hline
\end{tabular}
\end{ruledtabular}
\end{threeparttable}

\subsection{Dependence on temperature}
\label{sec_scaling}
Putting together the results from the previous two subsections,
we find that
the three magnonic contributions to the SOT exhibit the following
scaling behaviour with respect to temperature $T$ and spin-wave
stiffness $\mathcal{A}$: 
\bege
t^{(0,0)}_{{\rm mag},ij}\propto 
\frac{T}{\mathcal{A}}\left[
\frac{\hbar\mathcal{C}}{k_{\rm B}T}-
\log\left(
e^{\frac{\hbar\mathcal{C}}{k_{\rm B}T}}-1
\right)
\right],
\ee
\bege
t^{(2,0)}_{{\rm mag},ij}\propto 
\frac{T^{3} }{\mathcal{A}}, 
\ee
and 
\bege
t^{(0,2)}_{{\rm mag},ij}\propto
\frac{T^{2}}{\mathcal{A}^{2}}.
\ee

In the ferromagnetic Rashba model $t^{(0,2)}_{{\rm mag},ij}$ is the
dominant contribution. It depends quadratically on the temperature.
A scaling $\propto T^{d/2+1}$, where $d$ is the dimensionality
of the system, has also been found for the spin-wave-induced correction
to the conductivity of ferromagnets~\cite{PhysRevB.90.024405}.
This strong temperature dependence resembles the one
measured in experiments~\cite{angular_and_temperature_dependence_Ta_CoFeB_MgO,PhysRevB.89.174424,PhysRevB.94.140414,temperature_effects_PtCoC}.

Even though the relaxation time $\tau$ depends on temperature through
phonon and magnon scattering, we do not express the relaxation time
in terms of the temperature here, because
interfacial disorder is expected to provide major scattering channels
as well in
magnetic bilayers. Therefore, we treat temperature and 
relaxation time $\tau$ as independent parameters, because the latter 
can be controlled independently of temperature by tuning the disorder
in the system.

Spin disorder usually increases the electrical
resistivity~\cite{spin_disorder_landauer,dlm_kudrnovsky_PhysRevB.86.144423}
due to the additional scattering
channels, which may be
described effectively by a simple reduction of the relaxation time.
In contrast, the magnonic SOT discussed here 
cannot simply be accounted for by this reduction of the relaxation time.

\subsection{Generalizations of the formalism to treat the anisotropy of SOT }
\label{sec_formalism_anisotropy}
In Sec.~\ref{sec_formalism_sot} we assumed that the magnetization is oriented 
in $z$-direction. In order to compute the anisotropy of the SOT, it is necessary to
generalize this for general magnetization directions.
It is effective to express the magnetization direction in spherical 
coordinates:
\bege
\hat{\vn{M}}=
\begin{pmatrix}
\sin(\theta)\cos(\phi)\\
\sin(\theta)\sin(\phi)\\
\cos(\theta)\\
\end{pmatrix}.
\ee 
In order to discuss the anisotropy of the SOT it is convenient to
project the torques onto the unit 
vectors $\hat{\vn{e}}_{\theta}=\partial \hat{\vn{M}}/\partial \theta$
and $\hat{\vn{e}}_{\phi}=\partial \hat{\vn{M}}/\partial \phi/\sin(\theta)$
of the spherical coordinate system, because the torques are perpendicular
to the magnetization~\cite{symmetry_spin_orbit_torques}.

Eq.~\eqref{eq_g3_zeroth_order_magav},
Eq.~\eqref{eq_torque_20_one},
Eq.~\eqref{eq_app_t00_fermisea},
Eq.~\eqref{eq_torque_20_less2},
Eq.~\eqref{eq_torque_20_less3},
Eq.~\eqref{eq_torque_20_less4},
Eq.~\eqref{eq_app_secondq_torque_lesserone}, and
Eq.~\eqref{eq_app_secondq_torque_lessertwo}
become valid for general magnetization direction if the following
replacement is made:
\bege
\begin{aligned}
\sum_{\gamma=x,y}
\rightarrow
\sum_{\gamma=\theta,\phi},
\end{aligned}
\ee
where
\bege
\mathcal{T}_{\theta}=\sum_{\gamma=x,y,z}\hat{\vn{e}}_{\theta}\cdot \hat{\vn{e}}_{\gamma}\mathcal{T}_{\gamma}
\ee
and
\bege
\mathcal{T}_{\phi}=\sum_{\gamma=x,y,z}\hat{\vn{e}}_{\phi}\cdot \hat{\vn{e}}_{\gamma}\mathcal{T}_{\gamma}.
\ee

\section{Magnonic SOT in the ferromagnetic Rashba model}
\label{sec_results}

\begin{figure*}
\includegraphics[width=0.49\linewidth]{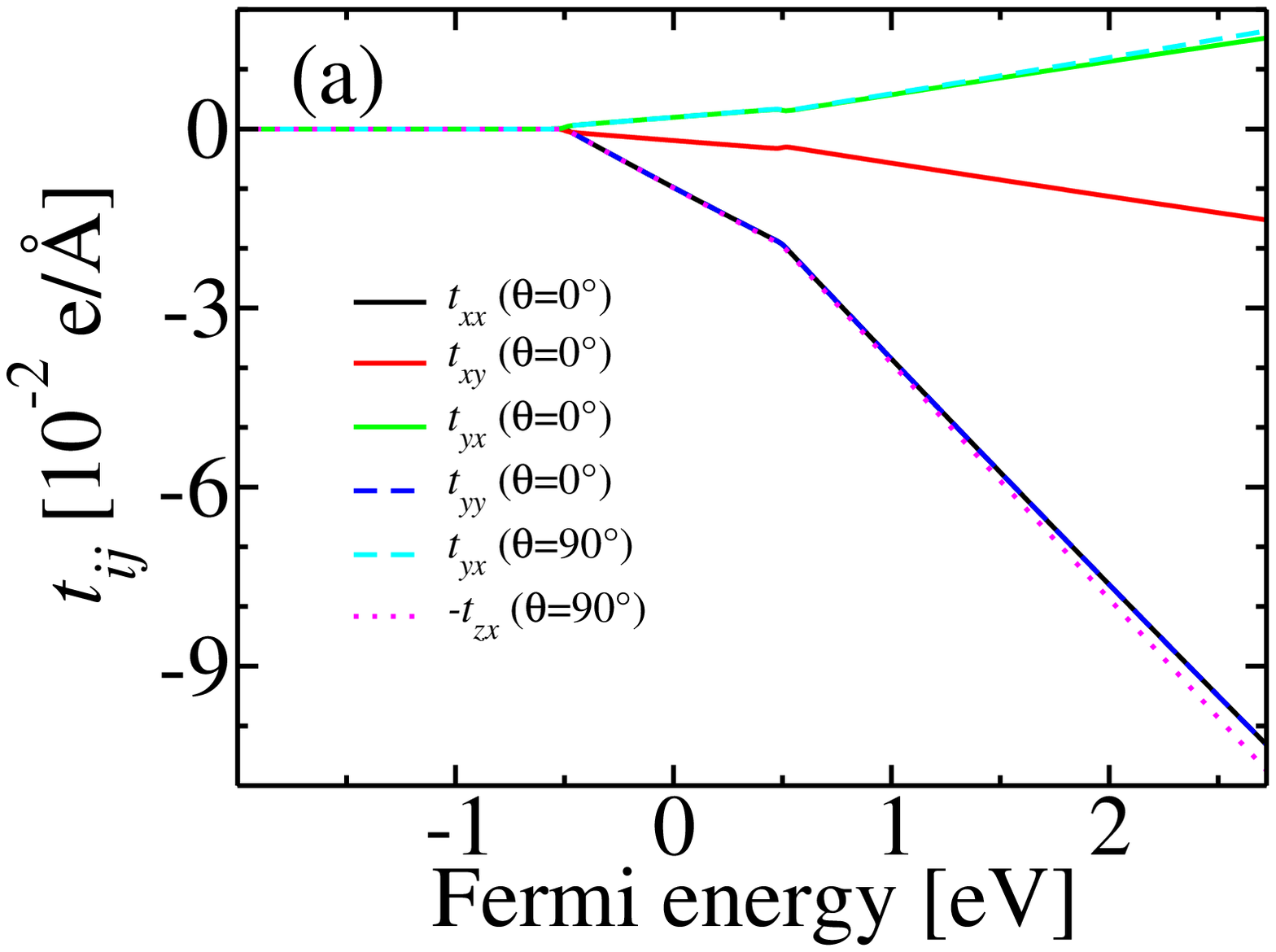}
\includegraphics[width=0.49\linewidth]{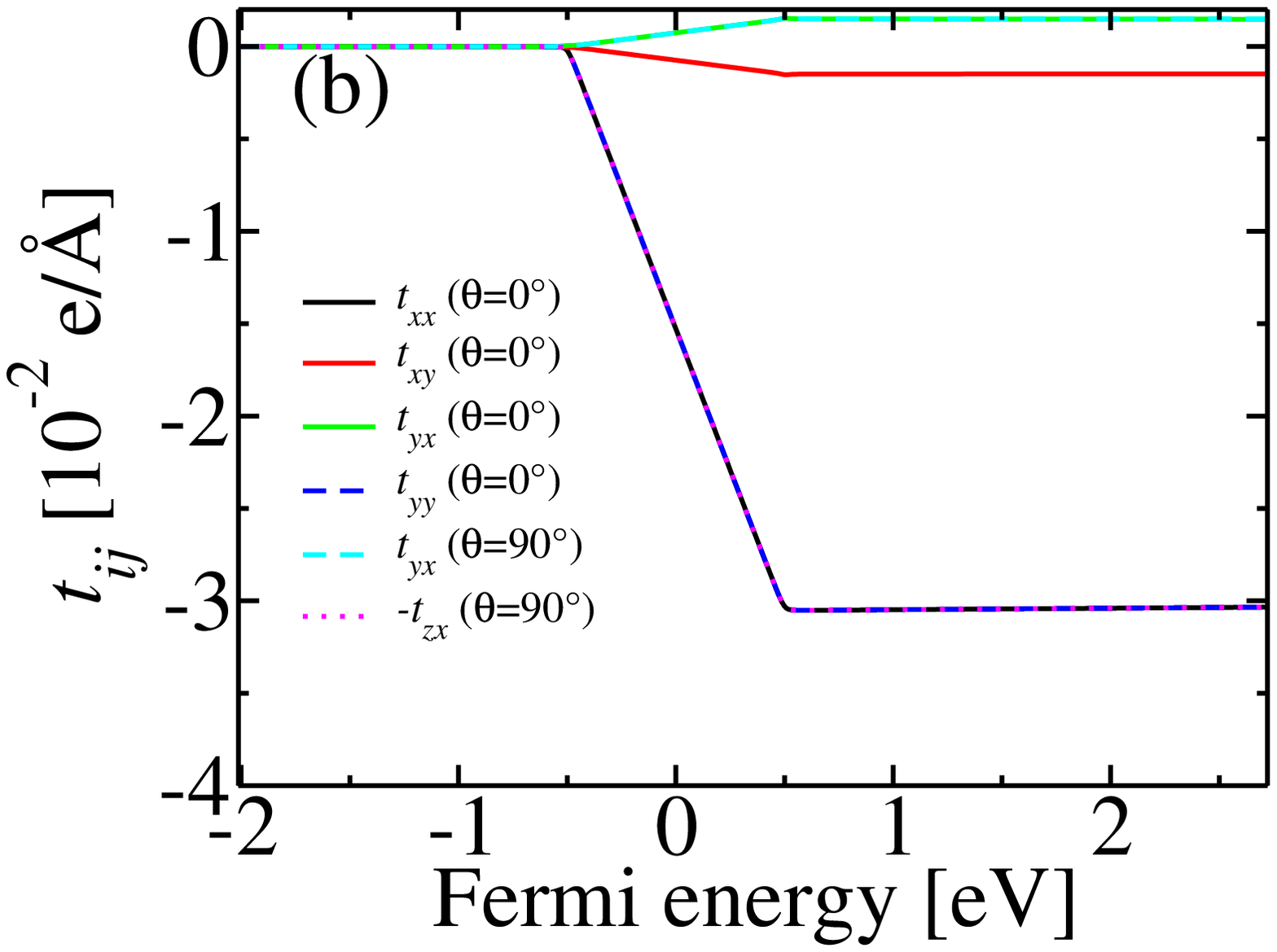}
\caption{\label{fig_alpha001ry_25meV}
Magnonic SOT (a) and
non-magnonic SOT (b) for $\alpha^{\rm R}=72$~meV~\AA\,
and $\Gamma=25$~meV.
}
\end{figure*}

\begin{figure*}
\includegraphics[width=0.49\linewidth]{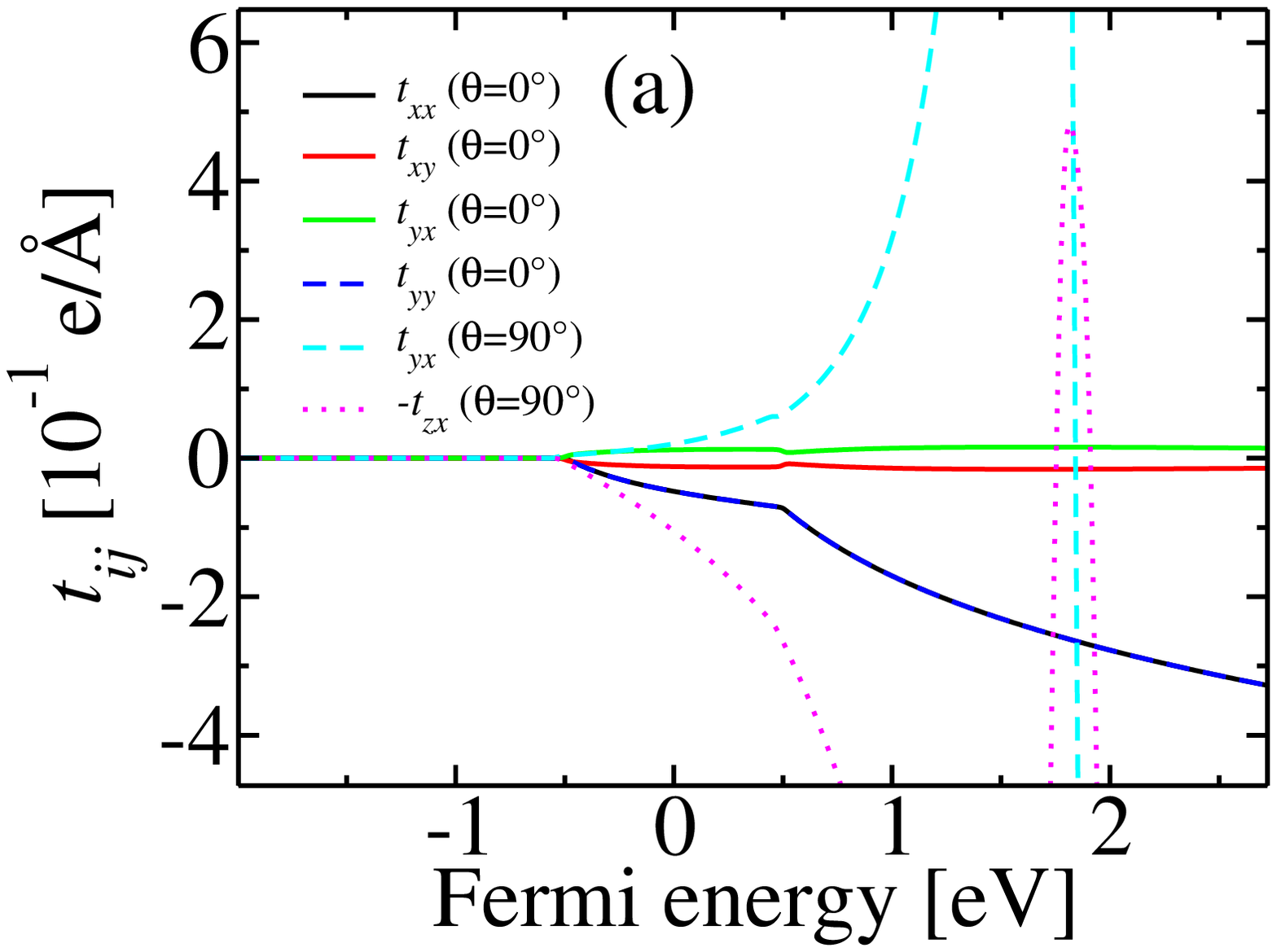}
\includegraphics[width=0.49\linewidth]{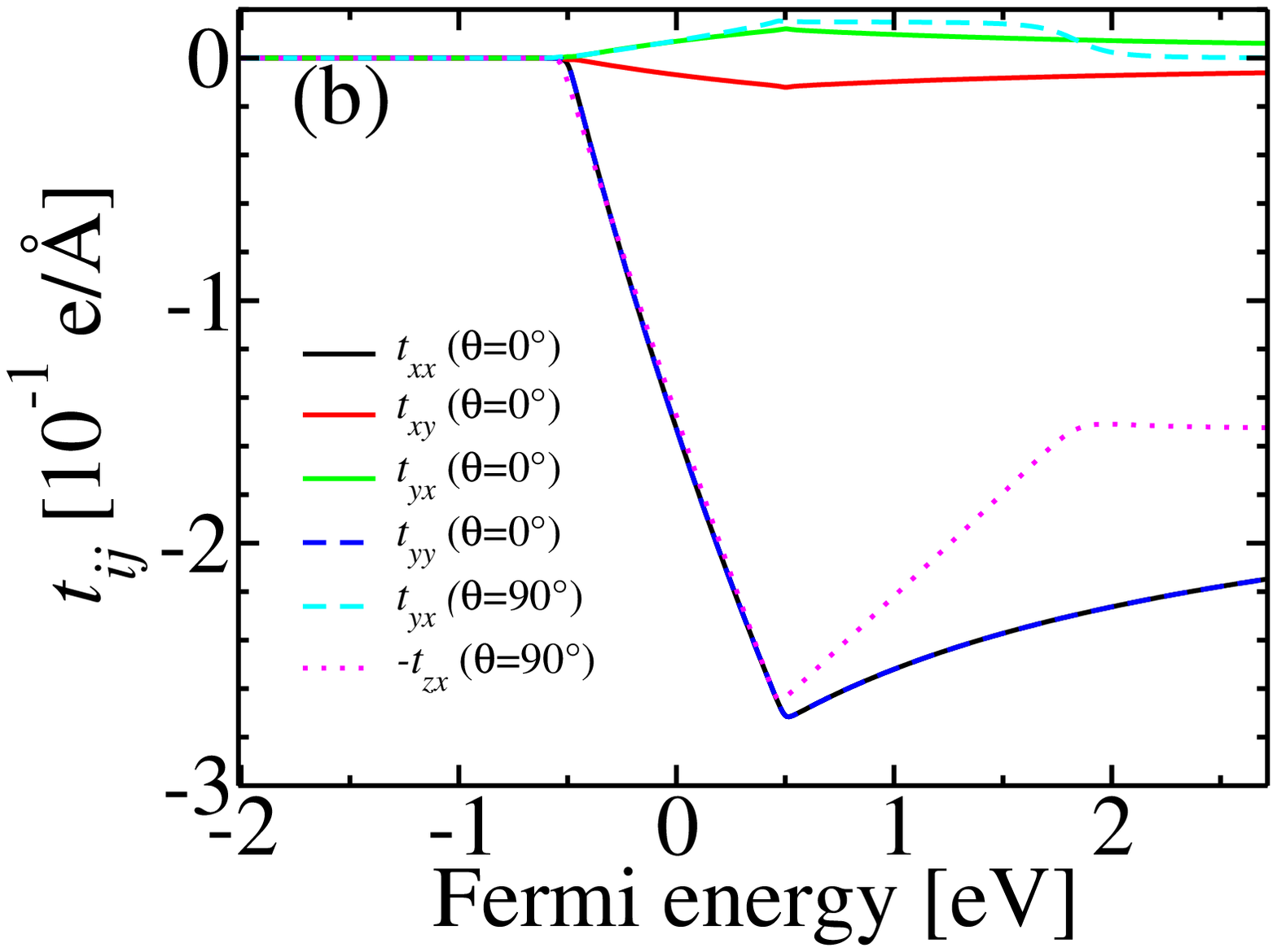}
\caption{\label{fig_alpha01ry_25meV}
Magnonic SOT (a) and
non-magnonic SOT (b) for $\alpha^{\rm R}=720$~meV~\AA\, 
and $\Gamma=25$~meV.
}
\end{figure*}

\begin{figure*}
\includegraphics[width=0.49\linewidth]{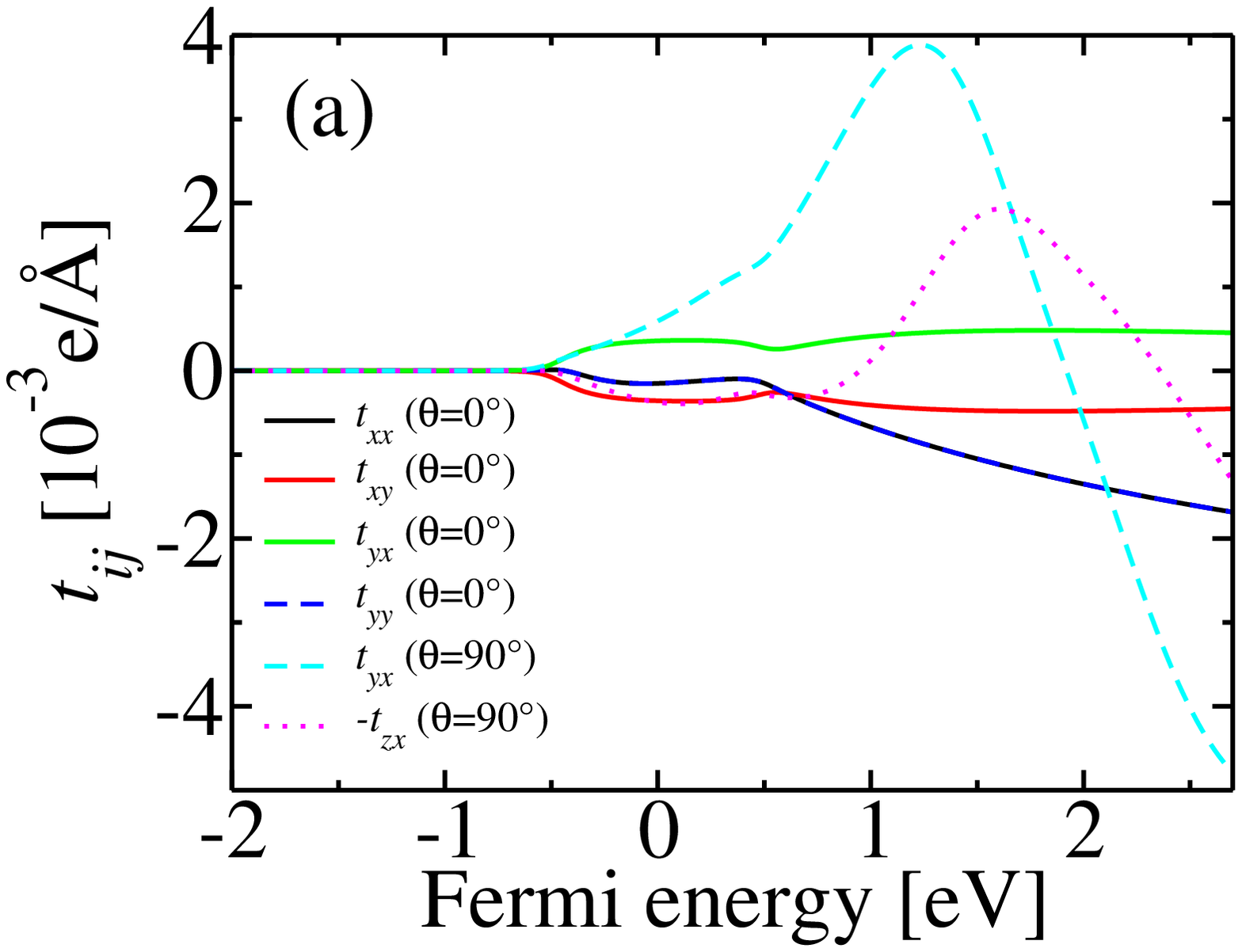}
\includegraphics[width=0.49\linewidth]{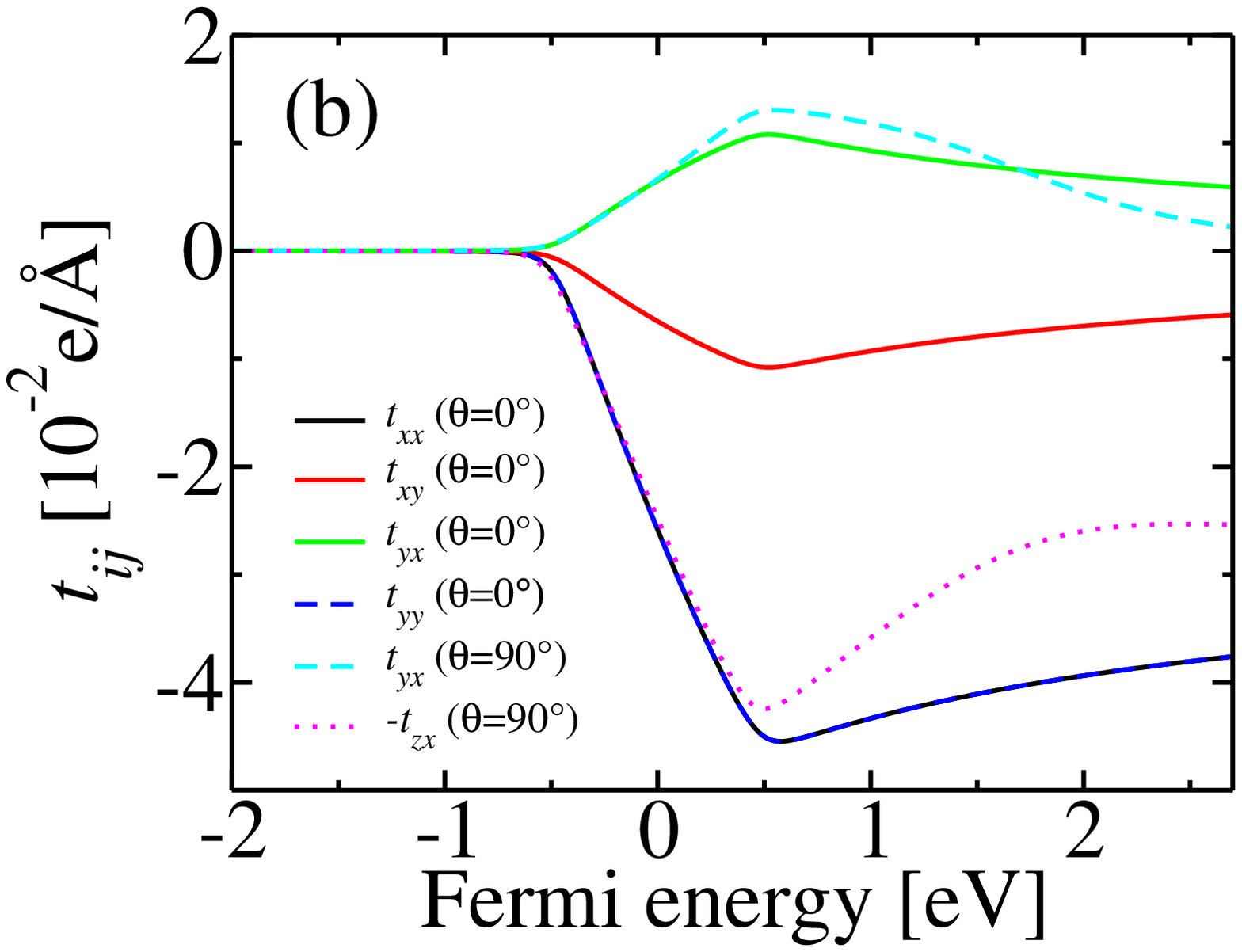}
\caption{\label{fig_alpha01ry}
Magnonic SOT (a) and
non-magnonic SOT (b) for $\alpha^{\rm R}=720$~meV~\AA\,
and $\Gamma=136$~meV.
}
\end{figure*}

In this section we study the magnonic SOT numerically  in the ferromagnetic
Rashba model~\cite{rashba_review}
\bege\label{eq_rashba_model}
H_{\vn{k}}=\frac{\hbar^2}{2m^{*}}k^2+
\alpha^{\rm R} (\vn{k}\times\hat{\vn{e}}_{z})\cdot\vn{\sigma}+
\frac{\Delta V}{2}\vn{\sigma}\cdot\hat{\vn{M}},
\ee
where $\alpha^{\rm R}$ is the Rashba parameter,
$\hat{\vn{M}}$ is the magnetization direction,
and $\Delta V$
is the exchange splitting.
We set the mass $m^{*}$ to the electron mass and the
exchange splitting to $\Delta V=1$~eV.
We use the expressions given in Sec.~\ref{sec_formalism_sot}
and in the Appendix~\ref{sec_appendix} to compute the magnonic SOT.
For the integrals $I^{(0,0)}A_{\rm mag}$, $I^{(2,0)}A_{\rm mag}$,
and $I^{(0,2)}A_{\rm mag}$ in these expressions we take
the values provided in Table~\ref{table_magnonic_integrals_twodim}
for the case of Co.
We introduce a broadening parameter $\Gamma$, which may be used to
model the effect of disorder, i.e., we use
\bege
\gret_{\Omega}=\hbar[\hbar\Omega-H_{\vn{k}}+i\Gamma]^{-1},
\ee
\bege
\gadv_{\Omega}=\hbar[\hbar\Omega-H_{\vn{k}}-i\Gamma]^{-1},
\ee
and $\gspec_{\Omega}=\gadv_{\Omega}-\gret_{\Omega}$ for the
Green's functions
in Eq.~\eqref{eq_g3_zeroth_order_magav},
Eq.~\eqref{eq_torque_20_one},
Eq.~\eqref{eq_app_t00_fermisea},
Eq.~\eqref{eq_torque_20_less2},
Eq.~\eqref{eq_torque_20_less3},
Eq.~\eqref{eq_torque_20_less4},
Eq.~\eqref{eq_app_secondq_torque_lesserone}, and
Eq.~\eqref{eq_app_secondq_torque_lessertwo}.
In order to compare the magnonic SOT to the non-magnonic one
we also compute the non-magnonic SOT according to
the equations in Ref.~\cite{ibcsoit}.

In Fig.~\ref{fig_alpha001ry_25meV} we show the SOTs when the
Rashba parameter and the broadening are $\alpha^{\rm R}=72$~meV~\AA\,
and $\Gamma=25$~meV, respectively.
The magnonic SOT shown in  Fig.~\ref{fig_alpha001ry_25meV}(a) is larger
than the
non-magnonic one shown in Fig.~\ref{fig_alpha001ry_25meV}(b) for this
choice of parameters.
In order to study the anisotropy of the SOT we show the torkances 
for $\theta=\phi=0$, i.e., $\hat{\vn{M}}$ along $\hat{\vn{e}}_{z}$,
and for $\theta=90^{\circ}$, $\phi=0^{\circ}$, i.e., $\hat{\vn{M}}$
along $\hat{\vn{e}}_{x}$.
We call the SOT anisotropic if $t_{xx}(\theta=0^{\circ})=-t_{zx}(\theta=90^{\circ})$ 
or $t_{yx}(\theta=0^{\circ})=t_{yx}(\theta=90^{\circ})$ are not satisfied.
While the magnonic SOT shows a small anisotropy, the anisotropy of the non-magnonic SOT
is invisible to the eye.

In Fig.~\ref{fig_alpha01ry_25meV} we show the SOTs when the
Rashba parameter and the broadening are $\alpha^{\rm R}=720$~meV~\AA\,
and $\Gamma=25$~meV, respectively.
Since the magnonic SOT is much larger for $\theta=90^{\circ}$ than it
is
for $\theta=0^{\circ}$, it is out of scale for several ranges of the
Fermi
energy in Fig.~\ref{fig_alpha01ry_25meV}(a). We show the full range
of the magnonic SOT at $\theta=90^{\circ}$ in Fig.~\ref{fig_zoomed}(a).
The non-magnonic SOT shown in Fig.~\ref{fig_alpha01ry_25meV}(b) is
rather
isotropic up to the Fermi energy 0.5~eV where its anisotropy starts to
become
significant. In contrast, for the magnonic SOT in Fig.~\ref{fig_alpha01ry_25meV}(a)
the relation $t_{yx}(\theta=0^{\circ})=t_{yx}(\theta=90^{\circ})$ is
satisfied approximately only up to the Fermi energy of 0~eV, where its anisotropy
starts to increase rapidly. The relation
$t_{xx}(\theta=0^{\circ})=-t_{zx}(\theta=90^{\circ})$
is satisfied approximately only for very small Fermi energies up to -0.3~eV.

In Fig.~\ref{fig_alpha01ry} we show the SOTs at the same Rashba
parameter $\alpha^{\rm R}=720$~meV~\AA, but at a larger broadening 
of $\Gamma=136$~meV. In agreement with the expectation~\cite{ibcsoit}
for the non-magnonic torque
we find that $t_{xy}\propto \Gamma^{0}$, 
$t_{yx}\propto \Gamma^{0}$, $t_{xx}\propto \Gamma^{-1}$, $t_{yy}\propto \Gamma^{-1}$,
and $t_{zx}\propto \Gamma^{-1}$ are approximately satisfied when we
compare Fig.~\ref{fig_alpha01ry}(b) and Fig.~\ref{fig_alpha01ry_25meV}(b).
In contrast, the magnonic SOT depends much stronger on $\Gamma$ and it 
is roughly one order of magnitude smaller than the non-magnonic one at
this value of the broadening of $\Gamma=136$~meV.

In Appendix~\ref{sec_appendix_plots} we provide the plots of the SOT
for
several additional choices of parameters, which confirms the trends
that
we discussed above using three examples. In general we find that the
magnonic
torque is sizable in comparison to the non-magnonic one if the
broadening parameter $\Gamma$ is small, i.e., when the disorder is
small. Additionally, we find that the anisotropy of the magnonic SOT
may
become gigantic if the Rashba parameter is large.

\section{Conclusions}
\label{sec_conclusions}
Using 3rd order perturbation theory within the framework of the Keldysh
nonequilibrium formalism we
derive suitable equations to assess the magnonic contributions to the SOT. 
In comparison to the purely electronic SOT, its magnonic counterpart
depends more strongly on the temperature.
We distinguish several contributions to the magnonic SOT, which 
depend differently on the spin-wave stiffness $\mathcal{A}$ 
and the temperature $T$. The dominating contribution
scales like $T^{2}/\mathcal{A}^{2}$,
which leads to a strong temperature-dependence of the magnonic 
contribution to the SOT, in agreement with experimental observations.
We compute the magnonic SOT 
in the ferromagnetic Rashba model.
It exhibits a strong anisotropy when the Rashba parameter is large
and it becomes larger than the non-magnonic SOT when the
quasiparticle broadening becomes small.
Since the magnonic SOT is sizable in comparison to its
purely electronic counterpart, magnons may therefore explain both the strong  
temperature dependence and the anisotropy of the SOT found in
some experiments.
\section*{Acknowledgments}
We gratefully acknowledge computing time on the supercomputers
of J\"ulich Supercomputing Center 
as well as funding by
Deutsche Forschungsgemeinschaft (DFG) through SPP 2137 
``Skyrmionics``, TRR 173 $-$ 268565370 (project A11), and DARPA TEE program through 
grant MIPR$\#$ HR0011831554 from DOI.

\bibliography{anismagnosot}

\begin{thebibliography}{28}%
\makeatletter
\providecommand \@ifxundefined [1]{%
 \@ifx{#1\undefined}
}%
\providecommand \@ifnum [1]{%
 \ifnum #1\expandafter \@firstoftwo
 \else \expandafter \@secondoftwo
 \fi
}%
\providecommand \@ifx [1]{%
 \ifx #1\expandafter \@firstoftwo
 \else \expandafter \@secondoftwo
 \fi
}%
\providecommand \natexlab [1]{#1}%
\providecommand \enquote  [1]{``#1''}%
\providecommand \bibnamefont  [1]{#1}%
\providecommand \bibfnamefont [1]{#1}%
\providecommand \citenamefont [1]{#1}%
\providecommand \href@noop [0]{\@secondoftwo}%
\providecommand \href [0]{\begingroup \@sanitize@url \@href}%
\providecommand \@href[1]{\@@startlink{#1}\@@href}%
\providecommand \@@href[1]{\endgroup#1\@@endlink}%
\providecommand \@sanitize@url [0]{\catcode `\\12\catcode `\$12\catcode
  `\&12\catcode `\#12\catcode `\^12\catcode `\_12\catcode `\%12\relax}%
\providecommand \@@startlink[1]{}%
\providecommand \@@endlink[0]{}%
\providecommand \url  [0]{\begingroup\@sanitize@url \@url }%
\providecommand \@url [1]{\endgroup\@href {#1}{\urlprefix }}%
\providecommand \urlprefix  [0]{URL }%
\providecommand \Eprint [0]{\href }%
\providecommand \doibase [0]{https://doi.org/}%
\providecommand \selectlanguage [0]{\@gobble}%
\providecommand \bibinfo  [0]{\@secondoftwo}%
\providecommand \bibfield  [0]{\@secondoftwo}%
\providecommand \translation [1]{[#1]}%
\providecommand \BibitemOpen [0]{}%
\providecommand \bibitemStop [0]{}%
\providecommand \bibitemNoStop [0]{.\EOS\space}%
\providecommand \EOS [0]{\spacefactor3000\relax}%
\providecommand \BibitemShut  [1]{\csname bibitem#1\endcsname}%
\let\auto@bib@innerbib\@empty
%</preamble>
\bibitem [{\citenamefont {Bhatti}\ \emph {et~al.}(2017)\citenamefont {Bhatti},
  \citenamefont {Sbiaa}, \citenamefont {Hirohata}, \citenamefont {Ohno},
  \citenamefont {Fukami},\ and\ \citenamefont
  {Piramanayagam}}]{mat_today_mram}%
  \BibitemOpen
  \bibfield  {author} {\bibinfo {author} {\bibfnamefont {S.}~\bibnamefont
  {Bhatti}}, \bibinfo {author} {\bibfnamefont {R.}~\bibnamefont {Sbiaa}},
  \bibinfo {author} {\bibfnamefont {A.}~\bibnamefont {Hirohata}}, \bibinfo
  {author} {\bibfnamefont {H.}~\bibnamefont {Ohno}}, \bibinfo {author}
  {\bibfnamefont {S.}~\bibnamefont {Fukami}},\ and\ \bibinfo {author}
  {\bibfnamefont {S.~N.}\ \bibnamefont {Piramanayagam}},\ }\bibfield  {title}
  {\bibinfo {title} {Spintronics based random access memory: a review},\ }\href
  {https://doi.org/10.1016/j.mattod.2017.07.007} {\bibfield  {journal}
  {\bibinfo  {journal} {MATERIALS TODAY}\ }\textbf {\bibinfo {volume} {20}},\
  \bibinfo {pages} {530} (\bibinfo {year} {2017})}\BibitemShut {NoStop}%
\bibitem [{\citenamefont {Manchon}\ \emph {et~al.}(2019)\citenamefont
  {Manchon}, \citenamefont {\ifmmode~\check{Z}\else \v{Z}\fi{}elezn\'y},
  \citenamefont {Miron}, \citenamefont {Jungwirth}, \citenamefont {Sinova},
  \citenamefont {Thiaville}, \citenamefont {Garello},\ and\ \citenamefont
  {Gambardella}}]{rmp_sot}%
  \BibitemOpen
  \bibfield  {author} {\bibinfo {author} {\bibfnamefont {A.}~\bibnamefont
  {Manchon}}, \bibinfo {author} {\bibfnamefont {J.}~\bibnamefont
  {\ifmmode~\check{Z}\else \v{Z}\fi{}elezn\'y}}, \bibinfo {author}
  {\bibfnamefont {I.~M.}\ \bibnamefont {Miron}}, \bibinfo {author}
  {\bibfnamefont {T.}~\bibnamefont {Jungwirth}}, \bibinfo {author}
  {\bibfnamefont {J.}~\bibnamefont {Sinova}}, \bibinfo {author} {\bibfnamefont
  {A.}~\bibnamefont {Thiaville}}, \bibinfo {author} {\bibfnamefont
  {K.}~\bibnamefont {Garello}},\ and\ \bibinfo {author} {\bibfnamefont
  {P.}~\bibnamefont {Gambardella}},\ }\bibfield  {title} {\bibinfo {title}
  {Current-induced spin-orbit torques in ferromagnetic and antiferromagnetic
  systems},\ }\href {https://doi.org/10.1103/RevModPhys.91.035004} {\bibfield
  {journal} {\bibinfo  {journal} {Rev. Mod. Phys.}\ }\textbf {\bibinfo {volume}
  {91}},\ \bibinfo {pages} {035004} (\bibinfo {year} {2019})}\BibitemShut
  {NoStop}%
\bibitem [{\citenamefont {Garello}\ \emph {et~al.}(2013)\citenamefont
  {Garello}, \citenamefont {Miron}, \citenamefont {Avci}, \citenamefont
  {Freimuth}, \citenamefont {Mokrousov}, \citenamefont {Bl\"ugel},
  \citenamefont {Auffret}, \citenamefont {Boulle}, \citenamefont {Gaudin},\
  and\ \citenamefont {Gambardella}}]{symmetry_spin_orbit_torques}%
  \BibitemOpen
  \bibfield  {author} {\bibinfo {author} {\bibfnamefont {K.}~\bibnamefont
  {Garello}}, \bibinfo {author} {\bibfnamefont {I.~M.}\ \bibnamefont {Miron}},
  \bibinfo {author} {\bibfnamefont {C.~O.}\ \bibnamefont {Avci}}, \bibinfo
  {author} {\bibfnamefont {F.}~\bibnamefont {Freimuth}}, \bibinfo {author}
  {\bibfnamefont {Y.}~\bibnamefont {Mokrousov}}, \bibinfo {author}
  {\bibfnamefont {S.}~\bibnamefont {Bl\"ugel}}, \bibinfo {author}
  {\bibfnamefont {S.}~\bibnamefont {Auffret}}, \bibinfo {author} {\bibfnamefont
  {O.}~\bibnamefont {Boulle}}, \bibinfo {author} {\bibfnamefont
  {G.}~\bibnamefont {Gaudin}},\ and\ \bibinfo {author} {\bibfnamefont
  {P.}~\bibnamefont {Gambardella}},\ }\bibfield  {title} {\bibinfo {title}
  {Symmetry and magnitude of spin-orbit torques in ferromagnetic
  heterostructures},\ }\href@noop {} {\bibfield  {journal} {\bibinfo  {journal}
  {Nature Nanotech.}\ }\textbf {\bibinfo {volume} {8}},\ \bibinfo {pages} {587}
  (\bibinfo {year} {2013})}\BibitemShut {NoStop}%
\bibitem [{\citenamefont {Hanke}\ \emph {et~al.}(2020)\citenamefont {Hanke},
  \citenamefont {Freimuth}, \citenamefont {Dup\'e}, \citenamefont {Sinova},
  \citenamefont {Kl\"aui},\ and\ \citenamefont
  {Mokrousov}}]{anisotropic_sot_hanke}%
  \BibitemOpen
  \bibfield  {author} {\bibinfo {author} {\bibfnamefont {J.-P.}\ \bibnamefont
  {Hanke}}, \bibinfo {author} {\bibfnamefont {F.}~\bibnamefont {Freimuth}},
  \bibinfo {author} {\bibfnamefont {B.}~\bibnamefont {Dup\'e}}, \bibinfo
  {author} {\bibfnamefont {J.}~\bibnamefont {Sinova}}, \bibinfo {author}
  {\bibfnamefont {M.}~\bibnamefont {Kl\"aui}},\ and\ \bibinfo {author}
  {\bibfnamefont {Y.}~\bibnamefont {Mokrousov}},\ }\bibfield  {title} {\bibinfo
  {title} {Engineering the dynamics of topological spin textures by anisotropic
  spin-orbit torques},\ }\href {https://doi.org/10.1103/PhysRevB.101.014428}
  {\bibfield  {journal} {\bibinfo  {journal} {Phys. Rev. B}\ }\textbf {\bibinfo
  {volume} {101}},\ \bibinfo {pages} {014428} (\bibinfo {year}
  {2020})}\BibitemShut {NoStop}%
\bibitem [{\citenamefont {Belashchenko}\ \emph {et~al.}(2019)\citenamefont
  {Belashchenko}, \citenamefont {Kovalev},\ and\ \citenamefont {van
  Schilfgaarde}}]{PhysRevMaterials.3.011401}%
  \BibitemOpen
  \bibfield  {author} {\bibinfo {author} {\bibfnamefont {K.~D.}\ \bibnamefont
  {Belashchenko}}, \bibinfo {author} {\bibfnamefont {A.~A.}\ \bibnamefont
  {Kovalev}},\ and\ \bibinfo {author} {\bibfnamefont {M.}~\bibnamefont {van
  Schilfgaarde}},\ }\bibfield  {title} {\bibinfo {title} {First-principles
  calculation of spin-orbit torque in a co/pt bilayer},\ }\href
  {https://doi.org/10.1103/PhysRevMaterials.3.011401} {\bibfield  {journal}
  {\bibinfo  {journal} {Phys. Rev. Materials}\ }\textbf {\bibinfo {volume}
  {3}},\ \bibinfo {pages} {011401(R)} (\bibinfo {year} {2019})}\BibitemShut
  {NoStop}%
\bibitem [{\citenamefont {Haney}\ \emph {et~al.}(2013)\citenamefont {Haney},
  \citenamefont {Lee}, \citenamefont {Lee}, \citenamefont {Manchon},\ and\
  \citenamefont {Stiles}}]{current_induced_torques_haney}%
  \BibitemOpen
  \bibfield  {author} {\bibinfo {author} {\bibfnamefont {P.~M.}\ \bibnamefont
  {Haney}}, \bibinfo {author} {\bibfnamefont {H.-W.}\ \bibnamefont {Lee}},
  \bibinfo {author} {\bibfnamefont {K.-J.}\ \bibnamefont {Lee}}, \bibinfo
  {author} {\bibfnamefont {A.}~\bibnamefont {Manchon}},\ and\ \bibinfo {author}
  {\bibfnamefont {M.~D.}\ \bibnamefont {Stiles}},\ }\bibfield  {title}
  {\bibinfo {title} {Current-induced torques and interfacial spin-orbit
  coupling},\ }\href@noop {} {\bibfield  {journal} {\bibinfo  {journal} {Phys.
  Rev. B}\ }\textbf {\bibinfo {volume} {88}},\ \bibinfo {pages} {214417}
  (\bibinfo {year} {2013})}\BibitemShut {NoStop}%
\bibitem [{\citenamefont {Freimuth}\ \emph
  {et~al.}(2014{\natexlab{a}})\citenamefont {Freimuth}, \citenamefont
  {Bl\"ugel},\ and\ \citenamefont {Mokrousov}}]{ibcsoit}%
  \BibitemOpen
  \bibfield  {author} {\bibinfo {author} {\bibfnamefont {F.}~\bibnamefont
  {Freimuth}}, \bibinfo {author} {\bibfnamefont {S.}~\bibnamefont {Bl\"ugel}},\
  and\ \bibinfo {author} {\bibfnamefont {Y.}~\bibnamefont {Mokrousov}},\
  }\bibfield  {title} {\bibinfo {title} {Spin-orbit torques in co/pt(111) and
  mn/w(001) magnetic bilayers from first principles},\ }\href@noop {}
  {\bibfield  {journal} {\bibinfo  {journal} {Phys. Rev. B}\ }\textbf {\bibinfo
  {volume} {90}},\ \bibinfo {pages} {174423} (\bibinfo {year}
  {2014}{\natexlab{a}})}\BibitemShut {NoStop}%
\bibitem [{\citenamefont {Ciccarelli}\ \emph {et~al.}(2016)\citenamefont
  {Ciccarelli}, \citenamefont {Anderson}, \citenamefont {Tshitoyan},
  \citenamefont {Ferguson}, \citenamefont {Gerhard}, \citenamefont {Gould},
  \citenamefont {Molenkamp}, \citenamefont {Gayles}, \citenamefont {Zelezny},
  \citenamefont {Smejkal}, \citenamefont {Yuan}, \citenamefont {Sinova},
  \citenamefont {Freimuth},\ and\ \citenamefont {Jungwirth}}]{sot_NiMnSb}%
  \BibitemOpen
  \bibfield  {author} {\bibinfo {author} {\bibfnamefont {C.}~\bibnamefont
  {Ciccarelli}}, \bibinfo {author} {\bibfnamefont {L.}~\bibnamefont
  {Anderson}}, \bibinfo {author} {\bibfnamefont {V.}~\bibnamefont {Tshitoyan}},
  \bibinfo {author} {\bibfnamefont {A.~J.}\ \bibnamefont {Ferguson}}, \bibinfo
  {author} {\bibfnamefont {F.}~\bibnamefont {Gerhard}}, \bibinfo {author}
  {\bibfnamefont {C.}~\bibnamefont {Gould}}, \bibinfo {author} {\bibfnamefont
  {L.~W.}\ \bibnamefont {Molenkamp}}, \bibinfo {author} {\bibfnamefont
  {J.}~\bibnamefont {Gayles}}, \bibinfo {author} {\bibfnamefont
  {J.}~\bibnamefont {Zelezny}}, \bibinfo {author} {\bibfnamefont
  {L.}~\bibnamefont {Smejkal}}, \bibinfo {author} {\bibfnamefont
  {Z.}~\bibnamefont {Yuan}}, \bibinfo {author} {\bibfnamefont {J.}~\bibnamefont
  {Sinova}}, \bibinfo {author} {\bibfnamefont {F.}~\bibnamefont {Freimuth}},\
  and\ \bibinfo {author} {\bibfnamefont {T.}~\bibnamefont {Jungwirth}},\
  }\bibfield  {title} {\bibinfo {title} {Room-temperature spin-orbit torque in
  nimnsb},\ }\href {https://doi.org/10.1038/NPHYS3772} {\bibfield  {journal}
  {\bibinfo  {journal} {NATURE PHYSICS}\ }\textbf {\bibinfo {volume} {12}},\
  \bibinfo {pages} {855} (\bibinfo {year} {2016})}\BibitemShut {NoStop}%
\bibitem [{\citenamefont {Qiu}\ \emph {et~al.}(2014)\citenamefont {Qiu},
  \citenamefont {Deorani}, \citenamefont {Narayanapillai}, \citenamefont {Lee},
  \citenamefont {Lee}, \citenamefont {Lee},\ and\ \citenamefont
  {Yang}}]{angular_and_temperature_dependence_Ta_CoFeB_MgO}%
  \BibitemOpen
  \bibfield  {author} {\bibinfo {author} {\bibfnamefont {X.}~\bibnamefont
  {Qiu}}, \bibinfo {author} {\bibfnamefont {P.}~\bibnamefont {Deorani}},
  \bibinfo {author} {\bibfnamefont {K.}~\bibnamefont {Narayanapillai}},
  \bibinfo {author} {\bibfnamefont {K.-S.}\ \bibnamefont {Lee}}, \bibinfo
  {author} {\bibfnamefont {K.-J.}\ \bibnamefont {Lee}}, \bibinfo {author}
  {\bibfnamefont {H.-W.}\ \bibnamefont {Lee}},\ and\ \bibinfo {author}
  {\bibfnamefont {H.}~\bibnamefont {Yang}},\ }\bibfield  {title} {\bibinfo
  {title} {Angular and temperature dependence of current induced spin-orbit
  effective fields in ta/cofeb/mgo nanowires},\ }\bibfield  {journal} {\bibinfo
   {journal} {SCIENTIFIC REPORTS}\ }\textbf {\bibinfo {volume} {4}},\ \href
  {https://doi.org/10.1038/srep04491} {10.1038/srep04491} (\bibinfo {year}
  {2014})\BibitemShut {NoStop}%
\bibitem [{\citenamefont {Kim}\ \emph {et~al.}(2014)\citenamefont {Kim},
  \citenamefont {Sinha}, \citenamefont {Mitani}, \citenamefont {Hayashi},
  \citenamefont {Takahashi}, \citenamefont {Maekawa}, \citenamefont
  {Yamanouchi},\ and\ \citenamefont {Ohno}}]{PhysRevB.89.174424}%
  \BibitemOpen
  \bibfield  {author} {\bibinfo {author} {\bibfnamefont {J.}~\bibnamefont
  {Kim}}, \bibinfo {author} {\bibfnamefont {J.}~\bibnamefont {Sinha}}, \bibinfo
  {author} {\bibfnamefont {S.}~\bibnamefont {Mitani}}, \bibinfo {author}
  {\bibfnamefont {M.}~\bibnamefont {Hayashi}}, \bibinfo {author} {\bibfnamefont
  {S.}~\bibnamefont {Takahashi}}, \bibinfo {author} {\bibfnamefont
  {S.}~\bibnamefont {Maekawa}}, \bibinfo {author} {\bibfnamefont
  {M.}~\bibnamefont {Yamanouchi}},\ and\ \bibinfo {author} {\bibfnamefont
  {H.}~\bibnamefont {Ohno}},\ }\bibfield  {title} {\bibinfo {title} {Anomalous
  temperature dependence of current-induced torques in
  $\text{CoFeB}/\text{MgO}$ heterostructures with ta-based underlayers},\
  }\href {https://doi.org/10.1103/PhysRevB.89.174424} {\bibfield  {journal}
  {\bibinfo  {journal} {Phys. Rev. B}\ }\textbf {\bibinfo {volume} {89}},\
  \bibinfo {pages} {174424} (\bibinfo {year} {2014})}\BibitemShut {NoStop}%
\bibitem [{\citenamefont {Ou}\ \emph {et~al.}(2016)\citenamefont {Ou},
  \citenamefont {Pai}, \citenamefont {Shi}, \citenamefont {Ralph},\ and\
  \citenamefont {Buhrman}}]{PhysRevB.94.140414}%
  \BibitemOpen
  \bibfield  {author} {\bibinfo {author} {\bibfnamefont {Y.}~\bibnamefont
  {Ou}}, \bibinfo {author} {\bibfnamefont {C.-F.}\ \bibnamefont {Pai}},
  \bibinfo {author} {\bibfnamefont {S.}~\bibnamefont {Shi}}, \bibinfo {author}
  {\bibfnamefont {D.~C.}\ \bibnamefont {Ralph}},\ and\ \bibinfo {author}
  {\bibfnamefont {R.~A.}\ \bibnamefont {Buhrman}},\ }\bibfield  {title}
  {\bibinfo {title} {Origin of fieldlike spin-orbit torques in heavy
  metal/ferromagnet/oxide thin film heterostructures},\ }\href
  {https://doi.org/10.1103/PhysRevB.94.140414} {\bibfield  {journal} {\bibinfo
  {journal} {Phys. Rev. B}\ }\textbf {\bibinfo {volume} {94}},\ \bibinfo
  {pages} {140414(R)} (\bibinfo {year} {2016})}\BibitemShut {NoStop}%
\bibitem [{\citenamefont {Li}\ \emph {et~al.}(2018)\citenamefont {Li},
  \citenamefont {Yun}, \citenamefont {Chen}, \citenamefont {Cui}, \citenamefont
  {Guo}, \citenamefont {Wu}, \citenamefont {Zuo}, \citenamefont {Yang},
  \citenamefont {Wang},\ and\ \citenamefont {Xi}}]{temperature_effects_PtCoC}%
  \BibitemOpen
  \bibfield  {author} {\bibinfo {author} {\bibfnamefont {D.}~\bibnamefont
  {Li}}, \bibinfo {author} {\bibfnamefont {J.}~\bibnamefont {Yun}}, \bibinfo
  {author} {\bibfnamefont {S.}~\bibnamefont {Chen}}, \bibinfo {author}
  {\bibfnamefont {B.}~\bibnamefont {Cui}}, \bibinfo {author} {\bibfnamefont
  {X.}~\bibnamefont {Guo}}, \bibinfo {author} {\bibfnamefont {K.}~\bibnamefont
  {Wu}}, \bibinfo {author} {\bibfnamefont {Y.}~\bibnamefont {Zuo}}, \bibinfo
  {author} {\bibfnamefont {D.}~\bibnamefont {Yang}}, \bibinfo {author}
  {\bibfnamefont {J.}~\bibnamefont {Wang}},\ and\ \bibinfo {author}
  {\bibfnamefont {L.}~\bibnamefont {Xi}},\ }\bibfield  {title} {\bibinfo
  {title} {Joule heating and temperature effects on current-induced
  magnetization switching in perpendicularly magnetized pt/co/c structures},\
  }\bibfield  {journal} {\bibinfo  {journal} {JOURNAL OF PHYSICS D-APPLIED
  PHYSICS}\ }\textbf {\bibinfo {volume} {51}},\ \href
  {https://doi.org/10.1088/1361-6463/aac7cc} {10.1088/1361-6463/aac7cc}
  (\bibinfo {year} {2018})\BibitemShut {NoStop}%
\bibitem [{\citenamefont {Mahfouzi}\ and\ \citenamefont
  {Kioussis}(2018)}]{PhysRevB.97.224426}%
  \BibitemOpen
  \bibfield  {author} {\bibinfo {author} {\bibfnamefont {F.}~\bibnamefont
  {Mahfouzi}}\ and\ \bibinfo {author} {\bibfnamefont {N.}~\bibnamefont
  {Kioussis}},\ }\bibfield  {title} {\bibinfo {title} {First-principles study
  of the angular dependence of the spin-orbit torque in pt/co and pd/co
  bilayers},\ }\href {https://doi.org/10.1103/PhysRevB.97.224426} {\bibfield
  {journal} {\bibinfo  {journal} {Phys. Rev. B}\ }\textbf {\bibinfo {volume}
  {97}},\ \bibinfo {pages} {224426} (\bibinfo {year} {2018})}\BibitemShut
  {NoStop}%
\bibitem [{\citenamefont {KASUYA}(1956)}]{resistance_ferromagnetic_kasuya}%
  \BibitemOpen
  \bibfield  {author} {\bibinfo {author} {\bibfnamefont {T.}~\bibnamefont
  {KASUYA}},\ }\bibfield  {title} {\bibinfo {title} {Electrical resistance of
  ferromagnetic metals},\ }\href {https://doi.org/10.1143/PTP.16.58} {\bibfield
   {journal} {\bibinfo  {journal} {PROGRESS OF THEORETICAL PHYSICS}\ }\textbf
  {\bibinfo {volume} {16}},\ \bibinfo {pages} {58} (\bibinfo {year}
  {1956})}\BibitemShut {NoStop}%
\bibitem [{\citenamefont {GOODINGS}(1963)}]{goodings}%
  \BibitemOpen
  \bibfield  {author} {\bibinfo {author} {\bibfnamefont {D.}~\bibnamefont
  {GOODINGS}},\ }\bibfield  {title} {\bibinfo {title} {Eletrical resistivity of
  ferromagnetic metals at low temperatures},\ }\href
  {https://doi.org/10.1103/PhysRev.132.542} {\bibfield  {journal} {\bibinfo
  {journal} {PHYSICAL REVIEW}\ }\textbf {\bibinfo {volume} {132}},\ \bibinfo
  {pages} {542} (\bibinfo {year} {1963})}\BibitemShut {NoStop}%
\bibitem [{\citenamefont {Misra}\ \emph {et~al.}(2009)\citenamefont {Misra},
  \citenamefont {Hebard}, \citenamefont {Muttalib},\ and\ \citenamefont
  {W\"olfle}}]{PhysRevB.79.140408}%
  \BibitemOpen
  \bibfield  {author} {\bibinfo {author} {\bibfnamefont {R.}~\bibnamefont
  {Misra}}, \bibinfo {author} {\bibfnamefont {A.~F.}\ \bibnamefont {Hebard}},
  \bibinfo {author} {\bibfnamefont {K.~A.}\ \bibnamefont {Muttalib}},\ and\
  \bibinfo {author} {\bibfnamefont {P.}~\bibnamefont {W\"olfle}},\ }\bibfield
  {title} {\bibinfo {title} {Spin-wave-mediated quantum corrections to the
  conductivity of thin ferromagnetic films of gadolinium},\ }\href
  {https://doi.org/10.1103/PhysRevB.79.140408} {\bibfield  {journal} {\bibinfo
  {journal} {Phys. Rev. B}\ }\textbf {\bibinfo {volume} {79}},\ \bibinfo
  {pages} {140408(R)} (\bibinfo {year} {2009})}\BibitemShut {NoStop}%
\bibitem [{\citenamefont {Danon}\ \emph {et~al.}(2014)\citenamefont {Danon},
  \citenamefont {Ricottone},\ and\ \citenamefont
  {Brouwer}}]{PhysRevB.90.024405}%
  \BibitemOpen
  \bibfield  {author} {\bibinfo {author} {\bibfnamefont {J.}~\bibnamefont
  {Danon}}, \bibinfo {author} {\bibfnamefont {A.}~\bibnamefont {Ricottone}},\
  and\ \bibinfo {author} {\bibfnamefont {P.~W.}\ \bibnamefont {Brouwer}},\
  }\bibfield  {title} {\bibinfo {title} {Spin-wave-induced correction to the
  conductivity of ferromagnets},\ }\href
  {https://doi.org/10.1103/PhysRevB.90.024405} {\bibfield  {journal} {\bibinfo
  {journal} {Phys. Rev. B}\ }\textbf {\bibinfo {volume} {90}},\ \bibinfo
  {pages} {024405} (\bibinfo {year} {2014})}\BibitemShut {NoStop}%
\bibitem [{\citenamefont {Kudrnovsk\'y}\ \emph {et~al.}(2012)\citenamefont
  {Kudrnovsk\'y}, \citenamefont {Drchal}, \citenamefont {Turek}, \citenamefont
  {Khmelevskyi}, \citenamefont {Glasbrenner},\ and\ \citenamefont
  {Belashchenko}}]{dlm_kudrnovsky_PhysRevB.86.144423}%
  \BibitemOpen
  \bibfield  {author} {\bibinfo {author} {\bibfnamefont {J.}~\bibnamefont
  {Kudrnovsk\'y}}, \bibinfo {author} {\bibfnamefont {V.}~\bibnamefont
  {Drchal}}, \bibinfo {author} {\bibfnamefont {I.}~\bibnamefont {Turek}},
  \bibinfo {author} {\bibfnamefont {S.}~\bibnamefont {Khmelevskyi}}, \bibinfo
  {author} {\bibfnamefont {J.~K.}\ \bibnamefont {Glasbrenner}},\ and\ \bibinfo
  {author} {\bibfnamefont {K.~D.}\ \bibnamefont {Belashchenko}},\ }\bibfield
  {title} {\bibinfo {title} {Spin-disorder resistivity of ferromagnetic metals
  from first principles: The disordered-local-moment approach},\ }\href
  {https://doi.org/10.1103/PhysRevB.86.144423} {\bibfield  {journal} {\bibinfo
  {journal} {Phys. Rev. B}\ }\textbf {\bibinfo {volume} {86}},\ \bibinfo
  {pages} {144423} (\bibinfo {year} {2012})}\BibitemShut {NoStop}%
\bibitem [{\citenamefont {Wysocki}\ \emph {et~al.}(2007)\citenamefont
  {Wysocki}, \citenamefont {Belashchenko}, \citenamefont {Velev},\ and\
  \citenamefont {van Schilfgaarde}}]{spin_disorder_landauer}%
  \BibitemOpen
  \bibfield  {author} {\bibinfo {author} {\bibfnamefont {A.~L.}\ \bibnamefont
  {Wysocki}}, \bibinfo {author} {\bibfnamefont {K.~D.}\ \bibnamefont
  {Belashchenko}}, \bibinfo {author} {\bibfnamefont {J.~P.}\ \bibnamefont
  {Velev}},\ and\ \bibinfo {author} {\bibfnamefont {M.}~\bibnamefont {van
  Schilfgaarde}},\ }\bibfield  {title} {\bibinfo {title} {Calculations of
  spin-disorder resistivity from first principles},\ }\href
  {https://doi.org/10.1063/1.2670472} {\bibfield  {journal} {\bibinfo
  {journal} {Journal of Applied Physics}\ }\textbf {\bibinfo {volume} {101}},\
  \bibinfo {pages} {09G506} (\bibinfo {year} {2007})}\BibitemShut {NoStop}%
\bibitem [{\citenamefont {Freimuth}\ \emph {et~al.}(2015)\citenamefont
  {Freimuth}, \citenamefont {Bl\"ugel},\ and\ \citenamefont
  {Mokrousov}}]{invsot}%
  \BibitemOpen
  \bibfield  {author} {\bibinfo {author} {\bibfnamefont {F.}~\bibnamefont
  {Freimuth}}, \bibinfo {author} {\bibfnamefont {S.}~\bibnamefont {Bl\"ugel}},\
  and\ \bibinfo {author} {\bibfnamefont {Y.}~\bibnamefont {Mokrousov}},\
  }\bibfield  {title} {\bibinfo {title} {Direct and inverse spin-orbit
  torques},\ }\href@noop {} {\bibfield  {journal} {\bibinfo  {journal} {Phys.
  Rev. B}\ }\textbf {\bibinfo {volume} {92}},\ \bibinfo {pages} {064415}
  (\bibinfo {year} {2015})}\BibitemShut {NoStop}%
\bibitem [{\citenamefont {Freimuth}\ \emph
  {et~al.}(2014{\natexlab{b}})\citenamefont {Freimuth}, \citenamefont
  {Bl\"ugel},\ and\ \citenamefont {Mokrousov}}]{mothedmisot}%
  \BibitemOpen
  \bibfield  {author} {\bibinfo {author} {\bibfnamefont {F.}~\bibnamefont
  {Freimuth}}, \bibinfo {author} {\bibfnamefont {S.}~\bibnamefont {Bl\"ugel}},\
  and\ \bibinfo {author} {\bibfnamefont {Y.}~\bibnamefont {Mokrousov}},\
  }\bibfield  {title} {\bibinfo {title} {Berry phase theory of
  dzyaloshinskii–moriya interaction and spin–orbit torques},\ }\href@noop
  {} {\bibfield  {journal} {\bibinfo  {journal} {Journal of physics: Condensed
  matter}\ }\textbf {\bibinfo {volume} {26}},\ \bibinfo {pages} {104202}
  (\bibinfo {year} {2014}{\natexlab{b}})}\BibitemShut {NoStop}%
\bibitem [{\citenamefont {Freimuth}\ \emph {et~al.}(2013)\citenamefont
  {Freimuth}, \citenamefont {Bamler}, \citenamefont {Mokrousov},\ and\
  \citenamefont {Rosch}}]{phase_space_berry}%
  \BibitemOpen
  \bibfield  {author} {\bibinfo {author} {\bibfnamefont {F.}~\bibnamefont
  {Freimuth}}, \bibinfo {author} {\bibfnamefont {R.}~\bibnamefont {Bamler}},
  \bibinfo {author} {\bibfnamefont {Y.}~\bibnamefont {Mokrousov}},\ and\
  \bibinfo {author} {\bibfnamefont {A.}~\bibnamefont {Rosch}},\ }\bibfield
  {title} {\bibinfo {title} {Phase-space berry phases in chiral magnets:
  Dzyaloshinskii-moriya interaction and the charge of skyrmions},\ }\href@noop
  {} {\bibfield  {journal} {\bibinfo  {journal} {Phys. Rev. B}\ }\textbf
  {\bibinfo {volume} {88}},\ \bibinfo {pages} {214409} (\bibinfo {year}
  {2013})}\BibitemShut {NoStop}%
\bibitem [{\citenamefont {Rammer}\ and\ \citenamefont
  {Smith}(1986)}]{rammer_smith}%
  \BibitemOpen
  \bibfield  {author} {\bibinfo {author} {\bibfnamefont {J.}~\bibnamefont
  {Rammer}}\ and\ \bibinfo {author} {\bibfnamefont {H.}~\bibnamefont {Smith}},\
  }\bibfield  {title} {\bibinfo {title} {Quantum field-theoretical methods in
  transport theory of metals},\ }\href
  {https://doi.org/10.1103/RevModPhys.58.323} {\bibfield  {journal} {\bibinfo
  {journal} {Rev. Mod. Phys.}\ }\textbf {\bibinfo {volume} {58}},\ \bibinfo
  {pages} {323} (\bibinfo {year} {1986})}\BibitemShut {NoStop}%
\bibitem [{\citenamefont {Loong}\ \emph {et~al.}(1984)\citenamefont {Loong},
  \citenamefont {Carpenter}, \citenamefont {Lynn}, \citenamefont {Robinson},\
  and\ \citenamefont {Mook}}]{magnetic_excitations_iron}%
  \BibitemOpen
  \bibfield  {author} {\bibinfo {author} {\bibfnamefont {C.}~\bibnamefont
  {Loong}}, \bibinfo {author} {\bibfnamefont {J.~M.}\ \bibnamefont
  {Carpenter}}, \bibinfo {author} {\bibfnamefont {J.~W.}\ \bibnamefont {Lynn}},
  \bibinfo {author} {\bibfnamefont {R.~A.}\ \bibnamefont {Robinson}},\ and\
  \bibinfo {author} {\bibfnamefont {H.~A.}\ \bibnamefont {Mook}},\ }\bibfield
  {title} {\bibinfo {title} {Neutron scattering study of the magnetic
  excitations in ferromagnetic iron at high energy transfers},\ }\href
  {https://doi.org/10.1063/1.333511} {\bibfield  {journal} {\bibinfo  {journal}
  {Journal of Applied Physics}\ }\textbf {\bibinfo {volume} {55}},\ \bibinfo
  {pages} {1895} (\bibinfo {year} {1984})}\BibitemShut {NoStop}%
\bibitem [{\citenamefont {Buczek}\ \emph {et~al.}(2011)\citenamefont {Buczek},
  \citenamefont {Ernst},\ and\ \citenamefont
  {Sandratskii}}]{magnons_iron_cobalt_nickel_sandratskii}%
  \BibitemOpen
  \bibfield  {author} {\bibinfo {author} {\bibfnamefont {P.}~\bibnamefont
  {Buczek}}, \bibinfo {author} {\bibfnamefont {A.}~\bibnamefont {Ernst}},\ and\
  \bibinfo {author} {\bibfnamefont {L.~M.}\ \bibnamefont {Sandratskii}},\
  }\bibfield  {title} {\bibinfo {title} {Different dimensionality trends in the
  landau damping of magnons in iron, cobalt, and nickel: Time-dependent density
  functional study},\ }\href {https://doi.org/10.1103/PhysRevB.84.174418}
  {\bibfield  {journal} {\bibinfo  {journal} {Phys. Rev. B}\ }\textbf {\bibinfo
  {volume} {84}},\ \bibinfo {pages} {174418} (\bibinfo {year}
  {2011})}\BibitemShut {NoStop}%
\bibitem [{\citenamefont {Mook}\ \emph {et~al.}(1969)\citenamefont {Mook},
  \citenamefont {Nicklow}, \citenamefont {Thompson},\ and\ \citenamefont
  {Wilkinson}}]{mook_spin_waves_Ni}%
  \BibitemOpen
  \bibfield  {author} {\bibinfo {author} {\bibfnamefont {H.~A.}\ \bibnamefont
  {Mook}}, \bibinfo {author} {\bibfnamefont {R.~M.}\ \bibnamefont {Nicklow}},
  \bibinfo {author} {\bibfnamefont {E.~D.}\ \bibnamefont {Thompson}},\ and\
  \bibinfo {author} {\bibfnamefont {M.~K.}\ \bibnamefont {Wilkinson}},\
  }\bibfield  {title} {\bibinfo {title} {Spin‐wave spectrum of nickel
  metal},\ }\href {https://doi.org/10.1063/1.1657713} {\bibfield  {journal}
  {\bibinfo  {journal} {Journal of Applied Physics}\ }\textbf {\bibinfo
  {volume} {40}},\ \bibinfo {pages} {1450} (\bibinfo {year}
  {1969})}\BibitemShut {NoStop}%
\bibitem [{\citenamefont {Ferriani}\ \emph {et~al.}(2008)\citenamefont
  {Ferriani}, \citenamefont {von Bergmann}, \citenamefont {Vedmedenko},
  \citenamefont {Heinze}, \citenamefont {Bode}, \citenamefont {Heide},
  \citenamefont {Bihlmayer}, \citenamefont {Bl\"ugel},\ and\ \citenamefont
  {Wiesendanger}}]{heide_dmi_mnw}%
  \BibitemOpen
  \bibfield  {author} {\bibinfo {author} {\bibfnamefont {P.}~\bibnamefont
  {Ferriani}}, \bibinfo {author} {\bibfnamefont {K.}~\bibnamefont {von
  Bergmann}}, \bibinfo {author} {\bibfnamefont {E.~Y.}\ \bibnamefont
  {Vedmedenko}}, \bibinfo {author} {\bibfnamefont {S.}~\bibnamefont {Heinze}},
  \bibinfo {author} {\bibfnamefont {M.}~\bibnamefont {Bode}}, \bibinfo {author}
  {\bibfnamefont {M.}~\bibnamefont {Heide}}, \bibinfo {author} {\bibfnamefont
  {G.}~\bibnamefont {Bihlmayer}}, \bibinfo {author} {\bibfnamefont
  {S.}~\bibnamefont {Bl\"ugel}},\ and\ \bibinfo {author} {\bibfnamefont
  {R.}~\bibnamefont {Wiesendanger}},\ }\href@noop {} {\bibfield  {journal}
  {\bibinfo  {journal} {Phys. Rev. Lett.}\ }\textbf {\bibinfo {volume} {101}},\
  \bibinfo {pages} {027201} (\bibinfo {year} {2008})}\BibitemShut {NoStop}%
\bibitem [{\citenamefont {Manchon}\ \emph {et~al.}(2015)\citenamefont
  {Manchon}, \citenamefont {Koo}, \citenamefont {Nitta}, \citenamefont
  {Frolov},\ and\ \citenamefont {Duine}}]{rashba_review}%
  \BibitemOpen
  \bibfield  {author} {\bibinfo {author} {\bibfnamefont {A.}~\bibnamefont
  {Manchon}}, \bibinfo {author} {\bibfnamefont {H.~C.}\ \bibnamefont {Koo}},
  \bibinfo {author} {\bibfnamefont {J.}~\bibnamefont {Nitta}}, \bibinfo
  {author} {\bibfnamefont {S.~M.}\ \bibnamefont {Frolov}},\ and\ \bibinfo
  {author} {\bibfnamefont {R.~A.}\ \bibnamefont {Duine}},\ }\bibfield  {title}
  {\bibinfo {title} {{N}ew perspectives for {R}ashba spin–orbit coupling},\
  }\href@noop {} {\bibfield  {journal} {\bibinfo  {journal} {Nature materials}\
  }\textbf {\bibinfo {volume} {14}},\ \bibinfo {pages} {871} (\bibinfo {year}
  {2015})}\BibitemShut {NoStop}%
\end{thebibliography}%
\onecolumngrid
\appendix

\section{Additional contributions}
\label{sec_appendix}

The torque from the lesser-two Green's function at the zeroth
order in $\omega_{\rm mag}$ and $\vn{q}$ is given by
\bege\label{eq_app_t00_fermisea}
\begin{aligned}
&\vn{T}^{\rm II, (0,0)}_{\rm mag}=
\frac{A_{\rm mag}  I^{(0,0)}(T)}{4\pi\hbar^3}
\int\rmd\Omega
\sum_{\gamma=x,y}
f(\hbar\Omega)
{\rm Tr}\Bigl\{
\vn{\mathcal{T}}\\
&\times\Bigl[  G^{\rm A}_{\Omega} \mathcal{T}_{\gamma} G^{\rm
  A}_{\Omega} \mathcal{T}_{\gamma} G^{\rm A}_{\Omega} \vn{v} G^{\rm
  A}_{\Omega} G^{\rm A}_{\Omega} +  G^{\rm A}_{\Omega}
\mathcal{T}_{\gamma} G^{\rm A}_{\Omega} \vn{v} G^{\rm A}_{\Omega}
\mathcal{T}_{\gamma} G^{\rm A}_{\Omega} G^{\rm A}_{\Omega} +G^{\rm A}_{\Omega} \mathcal{T}_{\gamma} G^{\rm A}_{\Omega} \vn{v} G^{\rm A}_{\Omega} G^{\rm A}_{\Omega} \mathcal{T}_{\gamma} G^{\rm A}_{\Omega} +  G^{\rm A}_{\Omega} \vn{v} G^{\rm A}_{\Omega} \mathcal{T}_{\gamma} G^{\rm A}_{\Omega} \mathcal{T}_{\gamma} G^{\rm A}_{\Omega} G^{\rm A}_{\Omega} \\
&+  G^{\rm A}_{\Omega} \vn{v} G^{\rm A}_{\Omega} \mathcal{T}_{\gamma} G^{\rm A}_{\Omega} G^{\rm A}_{\Omega} \mathcal{T}_{\gamma} G^{\rm A}_{\Omega}  + G^{\rm A}_{\Omega} \vn{v} G^{\rm A}_{\Omega} G^{\rm A}_{\Omega} \mathcal{T}_{\gamma} G^{\rm A}_{\Omega} \mathcal{T}_{\gamma} G^{\rm A}_{\Omega}  -  G^{\rm R}_{\Omega} \mathcal{T}_{\gamma} G^{\rm R}_{\Omega} \mathcal{T}_{\gamma} G^{\rm R}_{\Omega} \vn{v} G^{\rm R}_{\Omega} G^{\rm R}_{\Omega}  -  G^{\rm R}_{\Omega} \mathcal{T}_{\gamma} G^{\rm R}_{\Omega} \vn{v} G^{\rm R}_{\Omega} \mathcal{T}_{\gamma} G^{\rm R}_{\Omega} G^{\rm R}_{\Omega} \\
& -  G^{\rm R}_{\Omega} \mathcal{T}_{\gamma} G^{\rm R}_{\Omega} \vn{v} G^{\rm R}_{\Omega} G^{\rm R}_{\Omega} \mathcal{T}_{\gamma} G^{\rm R}_{\Omega}  -  G^{\rm R}_{\Omega} \vn{v} G^{\rm R}_{\Omega} \mathcal{T}_{\gamma} G^{\rm R}_{\Omega} \mathcal{T}_{\gamma} G^{\rm R}_{\Omega} G^{\rm R}_{\Omega}  -  G^{\rm R}_{\Omega} \vn{v} G^{\rm R}_{\Omega} \mathcal{T}_{\gamma}
G^{\rm R}_{\Omega} G^{\rm R}_{\Omega} \mathcal{T}_{\gamma} G^{\rm
  R}_{\Omega}  -  G^{\rm R}_{\Omega} \vn{v} G^{\rm R}_{\Omega} G^{\rm R}_{\Omega}
\mathcal{T}_{\gamma} G^{\rm R}_{\Omega} \mathcal{T}_{\gamma} G^{\rm
  R}_{\Omega}
\Bigr]
\cdot \vn{E}_{0}e
\Bigr\}
. \\
\end{aligned}
\ee

The torque from the lesser-two Green's function at the second order in
$\omega_{\rm mag}$ and at the zeroth order in $\vn{q}$ is given by
\bege\label{eq_torque_20_less2}
\begin{aligned}
&
\vn{T}^{\rm II, (2,0)}_{\rm mag}=
\frac{A_{\rm mag}  I^{(2,0)}(T)}{4\pi\hbar^5}
\int\rmd\Omega
\sum_{\gamma=x,y}
f(\hbar\Omega)
{\rm Tr}\Bigl\{
\vn{\mathcal{T}}\Bigl[ G^{\rm R}_{\Omega} \mathcal{T}_{\gamma} \frac{\partial^2}{\partial \Omega^2} G^{\rm R}_{\Omega} \mathcal{T}_{\gamma} G^{\rm R}_{\Omega} \vn{v} \frac{\partial}{\partial \Omega} G^{\rm R}_{\Omega}  +G^{\rm R}_{\Omega} \mathcal{T}_{\gamma} \frac{\partial^2}{\partial \Omega^2} G^{\rm R}_{\Omega} \vn{v} \frac{\partial}{\partial \Omega} G^{\rm R}_{\Omega} \mathcal{T}_{\gamma} G^{\rm R}_{\Omega} \\
& +G^{\rm R}_{\Omega} \mathcal{T}_{\gamma} \frac{\partial^2}{\partial \Omega^2} G^{\rm R}_{\Omega} \vn{v} G^{\rm R}_{\Omega} \mathcal{T}_{\gamma} \frac{\partial}{\partial \Omega} G^{\rm R}_{\Omega}  +G^{\rm R}_{\Omega} \mathcal{T}_{\gamma} G^{\rm R}_{\Omega} \vn{v} \frac{\partial^2}{\partial \Omega^2} G^{\rm R}_{\Omega} \mathcal{T}_{\gamma} \frac{\partial}{\partial \Omega} G^{\rm R}_{\Omega}  +G^{\rm R}_{\Omega} \mathcal{T}_{\gamma} G^{\rm R}_{\Omega} \vn{v}
\frac{\partial^{3}}{\partial \Omega^{3}} G^{\rm R}_{\Omega} \mathcal{T}_{\gamma} G^{\rm R}_{\Omega} \\
& +G^{\rm R}_{\Omega} \vn{v} \frac{\partial}{\partial \Omega} G^{\rm R}_{\Omega} \mathcal{T}_{\gamma} \frac{\partial^2}{\partial \Omega^2} G^{\rm R}_{\Omega} \mathcal{T}_{\gamma} G^{\rm R}_{\Omega}  +G^{\rm R}_{\Omega} \vn{v} G^{\rm R}_{\Omega} \mathcal{T}_{\gamma} \frac{\partial^2}{\partial \Omega^2} G^{\rm R}_{\Omega} \mathcal{T}_{\gamma} \frac{\partial}{\partial \Omega} G^{\rm R}_{\Omega}  +G^{\rm R}_{\Omega} \vn{v} G^{\rm R}_{\Omega} \mathcal{T}_{\gamma} \frac{\partial^{3}}{\partial \Omega^{3}} G^{\rm R}_{\Omega} \mathcal{T}_{\gamma} G^{\rm R}_{\Omega} \\
& - G^{\rm A}_{\Omega} \mathcal{T}_{\gamma} \frac{\partial^2}{\partial \Omega^2} G^{\rm A}_{\Omega} \mathcal{T}_{\gamma} G^{\rm A}_{\Omega} \vn{v} \frac{\partial}{\partial \Omega} G^{\rm A}_{\Omega}  - G^{\rm A}_{\Omega} \mathcal{T}_{\gamma} \frac{\partial^2}{\partial \Omega^2} G^{\rm A}_{\Omega} \vn{v} \frac{\partial}{\partial \Omega} G^{\rm A}_{\Omega} \mathcal{T}_{\gamma} G^{\rm A}_{\Omega}  - G^{\rm A}_{\Omega} \mathcal{T}_{\gamma} \frac{\partial^2}{\partial \Omega^2} G^{\rm A}_{\Omega} \vn{v} G^{\rm A}_{\Omega} \mathcal{T}_{\gamma} \frac{\partial}{\partial \Omega} G^{\rm A}_{\Omega} \\
& - G^{\rm A}_{\Omega} \mathcal{T}_{\gamma} G^{\rm A}_{\Omega} \vn{v} \frac{\partial^2}{\partial \Omega^2} G^{\rm A}_{\Omega} \mathcal{T}_{\gamma} \frac{\partial}{\partial \Omega} G^{\rm A}_{\Omega}  - G^{\rm A}_{\Omega} \mathcal{T}_{\gamma} G^{\rm A}_{\Omega} \vn{v} \frac{\partial^{3}}{\partial \Omega^{3}} G^{\rm A}_{\Omega} \mathcal{T}_{\gamma} G^{\rm A}_{\Omega}  - G^{\rm A}_{\Omega} \vn{v} \frac{\partial}{\partial \Omega} G^{\rm A}_{\Omega} \mathcal{T}_{\gamma} \frac{\partial^2}{\partial \Omega^2} G^{\rm A}_{\Omega} \mathcal{T}_{\gamma} G^{\rm A}_{\Omega} \\
& - G^{\rm A}_{\Omega} \vn{v} G^{\rm A}_{\Omega} \mathcal{T}_{\gamma} \frac{\partial^2}{\partial \Omega^2} G^{\rm A}_{\Omega} \mathcal{T}_{\gamma} \frac{\partial}{\partial \Omega} G^{\rm A}_{\Omega}  - G^{\rm A}_{\Omega} \vn{v} G^{\rm A}_{\Omega} \mathcal{T}_{\gamma} \frac{\partial^{3}}{\partial \Omega^{3}} G^{\rm A}_{\Omega} \mathcal{T}_{\gamma} G^{\rm A}_{\Omega}  - 2 G^{\rm A}_{\Omega} \mathcal{T}_{\gamma} \frac{\partial}{\partial \Omega} G^{\rm A}_{\Omega} \vn{v} \frac{\partial}{\partial \Omega} G^{\rm A}_{\Omega} \mathcal{T}_{\gamma} \frac{\partial}{\partial \Omega} G^{\rm A}_{\Omega} \\
& - 2 G^{\rm A}_{\Omega} \mathcal{T}_{\gamma} \frac{\partial}{\partial \Omega} G^{\rm A}_{\Omega} \vn{v} \frac{\partial^2}{\partial \Omega^2} G^{\rm A}_{\Omega} \mathcal{T}_{\gamma} G^{\rm A}_{\Omega}  +2 G^{\rm R}_{\Omega} \mathcal{T}_{\gamma} \frac{\partial}{\partial \Omega} G^{\rm R}_{\Omega} \vn{v} \frac{\partial}{\partial \Omega} G^{\rm R}_{\Omega} \mathcal{T}_{\gamma} \frac{\partial}{\partial \Omega} G^{\rm R}_{\Omega}  +2 G^{\rm R}_{\Omega} \mathcal{T}_{\gamma} \frac{\partial}{\partial \Omega} G^{\rm
  R}_{\Omega} \vn{v} \frac{\partial^2}{\partial \Omega^2} G^{\rm R}_{\Omega}
\mathcal{T}_{\gamma} G^{\rm R}_{\Omega} \Bigr]\Bigr\}\cdot \vn{E}_{0} e.\\
\end{aligned}
\ee
Additionally, there are the following torques from the
lesser-three and lesser-four Green's functions:
\bege\label{eq_torque_20_less3}
\begin{aligned}
&
\vn{T}^{\rm III, (2,0)}_{\rm mag}=-
\frac{A_{\rm mag}  I^{(2,0)}(T)}{4\pi\hbar^3}
\int\rmd\Omega
\sum_{\gamma=x,y}
f''(\hbar\Omega)
{\rm Tr}\Bigl\{
\vn{\mathcal{T}}\Bigl[ G^{\rm R}_{\Omega} \mathcal{T}_{\gamma} G^{\rm S}_{\Omega} \mathcal{T}_{\gamma} G^{\rm A}_{\Omega} \vn{v} \frac{\partial}{\partial \Omega} G^{\rm A}_{\Omega} \\
& + G^{\rm R}_{\Omega} \mathcal{T}_{\gamma} G^{\rm S}_{\Omega} \vn{v} \frac{\partial}{\partial \Omega} G^{\rm A}_{\Omega} \mathcal{T}_{\gamma} G^{\rm A}_{\Omega}  + G^{\rm R}_{\Omega} \mathcal{T}_{\gamma} G^{\rm S}_{\Omega} \vn{v} G^{\rm A}_{\Omega} \mathcal{T}_{\gamma} \frac{\partial}{\partial \Omega} G^{\rm A}_{\Omega}  + G^{\rm R}_{\Omega} \mathcal{T}_{\gamma} G^{\rm R}_{\Omega} \vn{v} G^{\rm S}_{\Omega} \mathcal{T}_{\gamma} \frac{\partial}{\partial \Omega} G^{\rm A}_{\Omega} \\
& + G^{\rm R}_{\Omega} \vn{v} \frac{\partial}{\partial \Omega} G^{\rm R}_{\Omega} \mathcal{T}_{\gamma} G^{\rm S}_{\Omega} \mathcal{T}_{\gamma} G^{\rm A}_{\Omega}  + G^{\rm R}_{\Omega} \vn{v} G^{\rm R}_{\Omega} \mathcal{T}_{\gamma} G^{\rm S}_{\Omega} \mathcal{T}_{\gamma} \frac{\partial}{\partial \Omega} G^{\rm A}_{\Omega}  + 3 G^{\rm R}_{\Omega} \mathcal{T}_{\gamma} G^{\rm R}_{\Omega} \vn{v} \frac{\partial}{\partial \Omega} G^{\rm S}_{\Omega} \mathcal{T}_{\gamma} G^{\rm A}_{\Omega} \\
& + 3 G^{\rm R}_{\Omega} \vn{v} G^{\rm R}_{\Omega} \mathcal{T}_{\gamma} \frac{\partial}{\partial \Omega} G^{\rm S}_{\Omega} \mathcal{T}_{\gamma} G^{\rm A}_{\Omega}  + 2 G^{\rm R}_{\Omega} \mathcal{T}_{\gamma} \frac{\partial}{\partial \Omega} G^{\rm
  R}_{\Omega} \vn{v} G^{\rm S}_{\Omega} \mathcal{T}_{\gamma} G^{\rm
  A}_{\Omega} \Bigr]\Bigr\}\cdot \vn{E}_{0}e\\
\end{aligned}
\ee
and
\bege\label{eq_torque_20_less4}
\begin{aligned}
&
\vn{T}^{\rm IV, (2,0)}_{\rm mag}=-
\frac{A_{\rm mag}  I^{(2,0)}(T)}{4\pi\hbar^2}
\int\rmd\Omega
\sum_{\gamma=x,y}
f'''(\hbar\Omega)
{\rm Tr}\Bigl\{
\vn{\mathcal{T}}\\
&\times\Bigl[  G^{\rm R}_{\Omega} \mathcal{T}_{\gamma} G^{\rm R}_{\Omega} \vn{v} G^{\rm S}_{\Omega} \mathcal{T}_{\gamma} G^{\rm A}_{\Omega} + G^{\rm R}_{\Omega} \vn{v} G^{\rm R}_{\Omega} \mathcal{T}_{\gamma}
G^{\rm S}_{\Omega}  \mathcal{T}_{\gamma} G^{\rm A}_{\Omega}\Bigr] \Bigr\}\cdot \vn{E}_{0}e.\\
\end{aligned}
\ee

The torque from the lesser-one Green's function at the second order in $q$ and zeroth order in $\omega_{\rm mag}$ is
given by:
\bege\label{eq_app_secondq_torque_lesserone}
\begin{aligned}
&\vn{T}^{\rm I, (0,2)}_{\rm mag}=-
\sum_{ij}\frac{eA_{\rm mag}  I^{(0,2)}_{ij}(T)}{4\pi\hbar^2}
\int\rmd\Omega
\sum_{\gamma=x,y}
{\rm Tr}\Bigl\{
\vn{\mathcal{T}} f'(\hbar\Omega)\gret_{\Omega} \vn{E}_{0}\cdot\Bigl[
\vn{v}  
\gspec_{\Omega} 
\mathcal{T}_{\gamma} 
\frac{\partial^2 \gadv_{\Omega,\vn{q} }}{\partial q_i \partial q_j}
\mathcal{T}_{\gamma}
\gadv_{\Omega }+
\vn{v} 
\gret_{\Omega} 
\mathcal{T}_{\gamma} 
\frac{\partial^2 \gspec_{\Omega,\vn{q} }}{\partial q_i \partial q_j}
\mathcal{T}_{\gamma}
\gadv_{\Omega }\\
&+
\vn{v} 
\gret_{\Omega} 
\mathcal{T}_{\gamma} 
\frac{\partial^2 \gret_{\Omega,\vn{q} }}{\partial q_i \partial q_j}
\mathcal{T}_{\gamma}
\gspec_{\Omega }+\mathcal{T}_{\gamma}\frac{\partial^2
\gret_{\Omega,\vn{q}} 
}{\partial q_i \partial q_j }
\vn{v}
\gspec_{\Omega} 
\mathcal{T}_{\gamma}
\gadv_{\Omega}+\mathcal{T}_{\gamma}
\frac{\partial^2
\gret_{\Omega,\vn{q}} 
}{\partial q_i \partial q_j }
\vn{v}
\gret_{\Omega} 
\mathcal{T}_{\gamma}
\gspec_{\Omega} 
+\mathcal{T}_{\gamma}
\frac{\partial
\gret_{\Omega,\vn{q}} 
}{\partial q_i   }
\vn{v}
\frac{\partial
\gspec_{\Omega,\vn{q}} 
}{\partial q_j }
\mathcal{T}_{\gamma}
\gadv_{\Omega} \\
&+\mathcal{T}_{\gamma}
\frac{\partial
\gret_{\Omega,\vn{q}} 
}{\partial q_i}
\vn{v}
\frac{\partial
\gret_{\Omega,\vn{q}} 
}{\partial q_j}
\mathcal{T}_{\gamma}
\gspec_{\Omega} 
+\mathcal{T}_{\gamma}
\frac{\partial
\gret_{\Omega,\vn{q}} 
}{\partial q_j}
\vn{v}
\frac{\partial
\gspec_{\Omega,\vn{q}} 
}{\partial q_i  }
\mathcal{T}_{\gamma}
\gadv_{\Omega}+\mathcal{T}_{\gamma}
\frac{\partial
\gret_{\Omega,\vn{q}} 
}{\partial q_j}
\vn{v}
\frac{\partial
\gret_{\Omega,\vn{q}} 
}{\partial q_i}
\mathcal{T}_{\gamma}
\gspec_{\Omega}\\
&+\mathcal{T}_{\gamma}
\gret_{\Omega} 
\vn{v}
\frac{\partial^2
\gspec_{\Omega,\vn{q}} 
}{\partial q_i  \partial q_j  }
\mathcal{T}_{\gamma}
\gadv_{\Omega}+\mathcal{T}_{\gamma}
\gret_{\Omega} 
\vn{v}
\frac{\partial^2
\gret_{\Omega,\vn{q}} 
}{\partial q_i  \partial q_j}
\mathcal{T}_{\gamma}
\gspec_{\Omega}
+\mathcal{T}_{\gamma}\frac{\partial
\gret_{\Omega,\vn{q}} 
}{\partial q_i  }
\frac{\partial
\vn{v}}{\partial q_j}
\gspec_{\Omega} 
\mathcal{T}_{\gamma}
\gadv_{\Omega}\\
&+\mathcal{T}_{\gamma}\frac{\partial
\gret_{\Omega,\vn{q}} 
}{\partial q_j  }
\frac{\partial
\vn{v}}{\partial q_i}
\gspec_{\Omega} 
\mathcal{T}_{\gamma}
\gadv_{\Omega}+
\mathcal{T}_{\gamma}\gret_{\Omega} 
\frac{\partial
\vn{v}}{\partial q_j}
\frac{\partial
\gspec_{\Omega,\vn{q}} 
}{\partial q_i}
\mathcal{T}_{\gamma}
\gadv_{\Omega}
+
\mathcal{T}_{\gamma}\gret_{\Omega} 
\frac{\partial
\vn{v}}{\partial q_i}
\frac{\partial
\gspec_{\Omega,\vn{q}} 
}{\partial q_j}
\mathcal{T}_{\gamma}
\gadv_{\Omega}
+\mathcal{T}_{\gamma}\frac{\partial
\gret_{\Omega,\vn{q}} 
}{\partial q_i  }
\frac{\partial
\vn{v}}{\partial q_j}
\gret_{\Omega} 
\mathcal{T}_{\gamma}
\gspec_{\Omega}\\
&+\mathcal{T}_{\gamma}\frac{\partial
\gret_{\Omega,\vn{q}} 
}{\partial q_j  }
\frac{\partial
\vn{v}}{\partial q_i}
\gret_{\Omega} 
\mathcal{T}_{\gamma}
\gspec_{\Omega}+
\mathcal{T}_{\gamma}\gret_{\Omega} 
\frac{\partial
\vn{v}}{\partial q_j}
\frac{\partial
\gret_{\Omega,\vn{q}} 
}{\partial q_i}
\mathcal{T}_{\gamma}
\gspec_{\Omega}
+
\mathcal{T}_{\gamma}\gret_{\Omega} 
\frac{\partial
\vn{v}}{\partial q_i}
\frac{\partial
\gret_{\Omega,\vn{q}} 
}{\partial q_j}
\mathcal{T}_{\gamma}
\gspec_{\Omega}+
\mathcal{T}_{\gamma}
\frac{\partial^2
\gret_{\Omega,\vn{q}} 
}{\partial q_i  \partial q_j}
\mathcal{T}_{\gamma}
\gret_{\Omega} 
\vn{v}
\gspec_{\Omega}\Bigr]\Bigr\}.
\end{aligned}
\ee
Additionally, we obtain the following Fermi sea contribution:
\bege\label{eq_app_secondq_torque_lessertwo}
\begin{aligned}
&\vn{T}^{\rm II, (0,2)}_{\rm mag}=\sum_{ij}
\frac{A_{\rm mag}  I^{(0,2)}_{ij}(T)}{4\pi\hbar^3}
\int\rmd\Omega
\sum_{\gamma=x,y}
{\rm Tr}\Bigl\{
\vn{\mathcal{T}} f(\hbar\Omega) 
\Bigl[ G^{\rm R}_{\Omega} \vn{v} \frac{\partial}{\partial \Omega} G^{\rm R}_{\Omega} \mathcal{T}_{\gamma} \left. \frac{\partial^{2}}{\partial q_i \partial q_j} G^{\rm R}_{\Omega} \right|_{\substack{ q=0 }} \mathcal{T}_{\gamma} G^{\rm R}_{\Omega} \\
&  +G^{\rm R}_{\Omega} \vn{v} G^{\rm R}_{\Omega} \mathcal{T}_{\gamma} \left. \frac{\partial^{2}}{\partial q_i \partial q_j} G^{\rm R}_{\Omega} \right|_{\substack{ q=0 }} \mathcal{T}_{\gamma} \frac{\partial}{\partial \Omega} G^{\rm R}_{\Omega}  +G^{\rm R}_{\Omega} \vn{v} G^{\rm R}_{\Omega} \mathcal{T}_{\gamma} \left. \frac{\partial^{3}}{\partial q_i \partial q_j\partial \Omega} G^{\rm R}_{\Omega} \right|_{\substack{ q=0 }} \mathcal{T}_{\gamma} G^{\rm R}_{\Omega} \\
& -  G^{\rm A}_{\Omega} \vn{v} \frac{\partial}{\partial \Omega} G^{\rm A}_{\Omega} \mathcal{T}_{\gamma} \left. \frac{\partial^{2}}{\partial q_i \partial q_j} G^{\rm A}_{\Omega} \right|_{\substack{ q=0 }} \mathcal{T}_{\gamma} G^{\rm A}_{\Omega}  -  G^{\rm A}_{\Omega} \vn{v} G^{\rm A}_{\Omega} \mathcal{T}_{\gamma} \left. \frac{\partial^{2}}{\partial q_i \partial q_j} G^{\rm A}_{\Omega} \right|_{\substack{ q=0 }} \mathcal{T}_{\gamma} \frac{\partial}{\partial \Omega} G^{\rm A}_{\Omega} \\
& -  G^{\rm A}_{\Omega} \vn{v} G^{\rm A}_{\Omega}
\mathcal{T}_{\gamma} \left. \frac{\partial^{3}}{\partial q_i \partial q_j\partial \Omega} G^{\rm A}_{\Omega} \right|_{\substack{ q=0
  }} \mathcal{T}_{\gamma} G^{\rm A}_{\Omega}  + G^{\rm R}_{\Omega} \mathcal{T}_{\gamma} \left. \frac{\partial^{2}}{\partial q_i \partial q_j} G^{\rm R}_{\Omega} \right|_{\substack{ q=0 }} \vn{v} \frac{\partial}{\partial \Omega} G^{\rm R}_{\Omega}{\left(e,0 \right)} \mathcal{T}_{\gamma} G^{\rm R}_{\Omega} \\
& + G^{\rm R}_{\Omega} \mathcal{T}_{\gamma} \left. \frac{\partial^{2}}{\partial q_i \partial q_j} G^{\rm R}_{\Omega} \right|_{\substack{ q=0 }} \vn{v} G^{\rm R}_{\Omega}{\left(e,0 \right)} \mathcal{T}_{\gamma} \frac{\partial}{\partial \Omega} G^{\rm R}_{\Omega}  + G^{\rm R}_{\Omega} \mathcal{T}_{\gamma} G^{\rm R}_{\Omega}{\left(e,0 \right)} \vn{v} \left. \frac{\partial^{2}}{\partial q_i \partial q_j} G^{\rm R}_{\Omega} \right|_{\substack{ q=0 }} \mathcal{T}_{\gamma} \frac{\partial}{\partial \Omega} G^{\rm R}_{\Omega} \\
& + G^{\rm R}_{\Omega} \mathcal{T}_{\gamma} G^{\rm R}_{\Omega}{\left(e,0 \right)} \vn{v} \left. \frac{\partial^{3}}{\partial q_i \partial q_j\partial \Omega} G^{\rm R}_{\Omega} \right|_{\substack{ q=0 }} \mathcal{T}_{\gamma} G^{\rm R}_{\Omega}  -  G^{\rm A}_{\Omega} \mathcal{T}_{\gamma} \left. \frac{\partial^{2}}{\partial q_i \partial q_j} G^{\rm A}_{\Omega} \right|_{\substack{ q=0 }} \vn{v} \frac{\partial}{\partial \Omega} G^{\rm A}_{\Omega}{\left(e,0 \right)} \mathcal{T}_{\gamma} G^{\rm A}_{\Omega} \\
& -  G^{\rm A}_{\Omega} \mathcal{T}_{\gamma} \left. \frac{\partial^{2}}{\partial q_i \partial q_j} G^{\rm A}_{\Omega} \right|_{\substack{ q=0 }} \vn{v} G^{\rm A}_{\Omega}{\left(e,0 \right)} \mathcal{T}_{\gamma} \frac{\partial}{\partial \Omega} G^{\rm A}_{\Omega}  -  G^{\rm A}_{\Omega} \mathcal{T}_{\gamma} G^{\rm A}_{\Omega}{\left(e,0 \right)} \vn{v} \left. \frac{\partial^{2}}{\partial q_i \partial q_j} G^{\rm A}_{\Omega} \right|_{\substack{ q=0 }} \mathcal{T}_{\gamma} \frac{\partial}{\partial \Omega} G^{\rm A}_{\Omega} \\
& -  G^{\rm A}_{\Omega} \mathcal{T}_{\gamma} G^{\rm A}_{\Omega}{\left(e,0 \right)} \vn{v} \left. \frac{\partial^{3}}{\partial q_i \partial q_j\partial \Omega} G^{\rm A}_{\Omega} \right|_{\substack{ q=0 }} \mathcal{T}_{\gamma} G^{\rm A}_{\Omega}  - 2  G^{\rm A}_{\Omega} \mathcal{T}_{\gamma} \left. \frac{\partial}{\partial q_i} G^{\rm A}_{\Omega} \right|_{\substack{ q=0 }} \vn{v} \left. \frac{\partial}{\partial q_j} G^{\rm A}_{\Omega} \right|_{\substack{ q=0 }} \mathcal{T}_{\gamma} \frac{\partial}{\partial \Omega} G^{\rm A}_{\Omega} \\
& - 2  G^{\rm A}_{\Omega} \mathcal{T}_{\gamma} \left. \frac{\partial}{\partial q_i} G^{\rm A}_{\Omega} \right|_{\substack{ q=0 }} \vn{v} \left. \frac{\partial^{2}}{\partial q_j\partial \Omega} G^{\rm A}_{\Omega} \right|_{\substack{ q=0 }} \mathcal{T}_{\gamma} G^{\rm A}_{\Omega} +2  G^{\rm R}_{\Omega} \mathcal{T}_{\gamma} \left. \frac{\partial}{\partial q_i} G^{\rm R}_{\Omega} \right|_{\substack{ q=0 }} \vn{v} \left. \frac{\partial}{\partial q_j} G^{\rm R}_{\Omega} \right|_{\substack{ q=0 }} \mathcal{T}_{\gamma} \frac{\partial}{\partial \Omega} G^{\rm R}_{\Omega} \\
& +2  G^{\rm R}_{\Omega} \mathcal{T}_{\gamma} \left. \frac{\partial}{\partial q_i} G^{\rm R}_{\Omega} \right|_{\substack{ q=0 }} \vn{v} \left. \frac{\partial^{2}}{\partial q_j\partial \Omega} G^{\rm R}_{\Omega} \right|_{\substack{ q=0 }} \mathcal{T}_{\gamma} G^{\rm R}_{\Omega}  + G^{\rm R}_{\Omega} \mathcal{T}_{\gamma} \left. \frac{\partial^{2}}{\partial q_i \partial q_j} G^{\rm R}_{\Omega} \right|_{\substack{ q=0 }} \mathcal{T}_{\gamma} G^{\rm R}_{\Omega} \vn{v} \frac{\partial}{\partial \Omega} G^{\rm R}_{\Omega} \\
& -  G^{\rm A}_{\Omega} \mathcal{T}_{\gamma}
\left. \frac{\partial^{2}}{\partial q_i \partial q_j} G^{\rm
    A}_{\Omega} \right|_{\substack{ q=0 }}
\mathcal{T}_{\gamma} G^{\rm A}_{\Omega} \vn{v}
\frac{\partial}{\partial \Omega}  G^{\rm
  A}_{\Omega} \Bigr ]+2 G^{\rm R}_{\Omega} \mathcal{T}_{\gamma} \left. \frac{\partial}{\partial q_i} G^{\rm R}_{\Omega} \right|_{\substack{ q=0 }} \frac{\partial \vn{v}}{\partial q_j} \frac{\partial}{\partial \Omega} G^{\rm R}_{\Omega} \mathcal{T}_{\gamma} G^{\rm R}_{\Omega} \\
& +2 G^{\rm R}_{\Omega} \mathcal{T}_{\gamma} \left. \frac{\partial}{\partial q_i} G^{\rm R}_{\Omega} \right|_{\substack{ q=0 }} \frac{\partial \vn{v}}{\partial q_j} G^{\rm R}_{\Omega} \mathcal{T}_{\gamma} \frac{\partial}{\partial \Omega} G^{\rm R}_{\Omega}  +2 G^{\rm R}_{\Omega} \mathcal{T}_{\gamma} G^{\rm R}_{\Omega} \frac{\partial \vn{v}}{\partial q_j} \left. \frac{\partial}{\partial q_i} G^{\rm R}_{\Omega} \right|_{\substack{ q=0 }} \mathcal{T}_{\gamma} \frac{\partial}{\partial \Omega} G^{\rm R}_{\Omega} \\
& +2 G^{\rm R}_{\Omega} \mathcal{T}_{\gamma} G^{\rm R}_{\Omega} \frac{\partial \vn{v}}{\partial q_j} \left. \frac{\partial^{2}}{\partial q_i\partial \Omega} G^{\rm R}_{\Omega} \right|_{\substack{ q=0 }} \mathcal{T}_{\gamma} G^{\rm R}_{\Omega} \\
& - 2 G^{\rm A}_{\Omega} \mathcal{T}_{\gamma} \left. \frac{\partial}{\partial q_i} G^{\rm A}_{\Omega} \right|_{\substack{ q=0 }} \frac{\partial \vn{v}}{\partial q_j} \frac{\partial}{\partial \Omega} G^{\rm A}_{\Omega} \mathcal{T}_{\gamma} G^{\rm A}_{\Omega}  - 2 G^{\rm A}_{\Omega} \mathcal{T}_{\gamma} \left. \frac{\partial}{\partial q_i} G^{\rm A}_{\Omega} \right|_{\substack{ q=0 }} \frac{\partial \vn{v}}{\partial q_j} G^{\rm A}_{\Omega} \mathcal{T}_{\gamma} \frac{\partial}{\partial \Omega} G^{\rm A}_{\Omega} \\
& - 2 G^{\rm A}_{\Omega} \mathcal{T}_{\gamma} G^{\rm A}_{\Omega}
\frac{\partial \vn{v}}{\partial q_j} \left. \frac{\partial}{\partial
    q_i} G^{\rm A}_{\Omega} \right|_{\substack{ q=0 }}
\mathcal{T}_{\gamma} \frac{\partial}{\partial \Omega} G^{\rm
  A}_{\Omega}  - 2 G^{\rm A}_{\Omega} \mathcal{T}_{\gamma} G^{\rm
  A}_{\Omega} \frac{\partial \vn{v}}{\partial q_j}
\left. \frac{\partial^{2}}{\partial q_i\partial \Omega} G^{\rm
    A}_{\Omega} \right|_{\substack{ q=0 }} \mathcal{T}_{\gamma} G^{\rm
  A}_{\Omega} \Bigr]\cdot\vn{E}_{0}e \Bigr\}.\\
\end{aligned}
\ee

\section{Additional plots of the SOT}
\label{sec_appendix_plots}
\begin{figure*}
\includegraphics[width=0.49\linewidth]{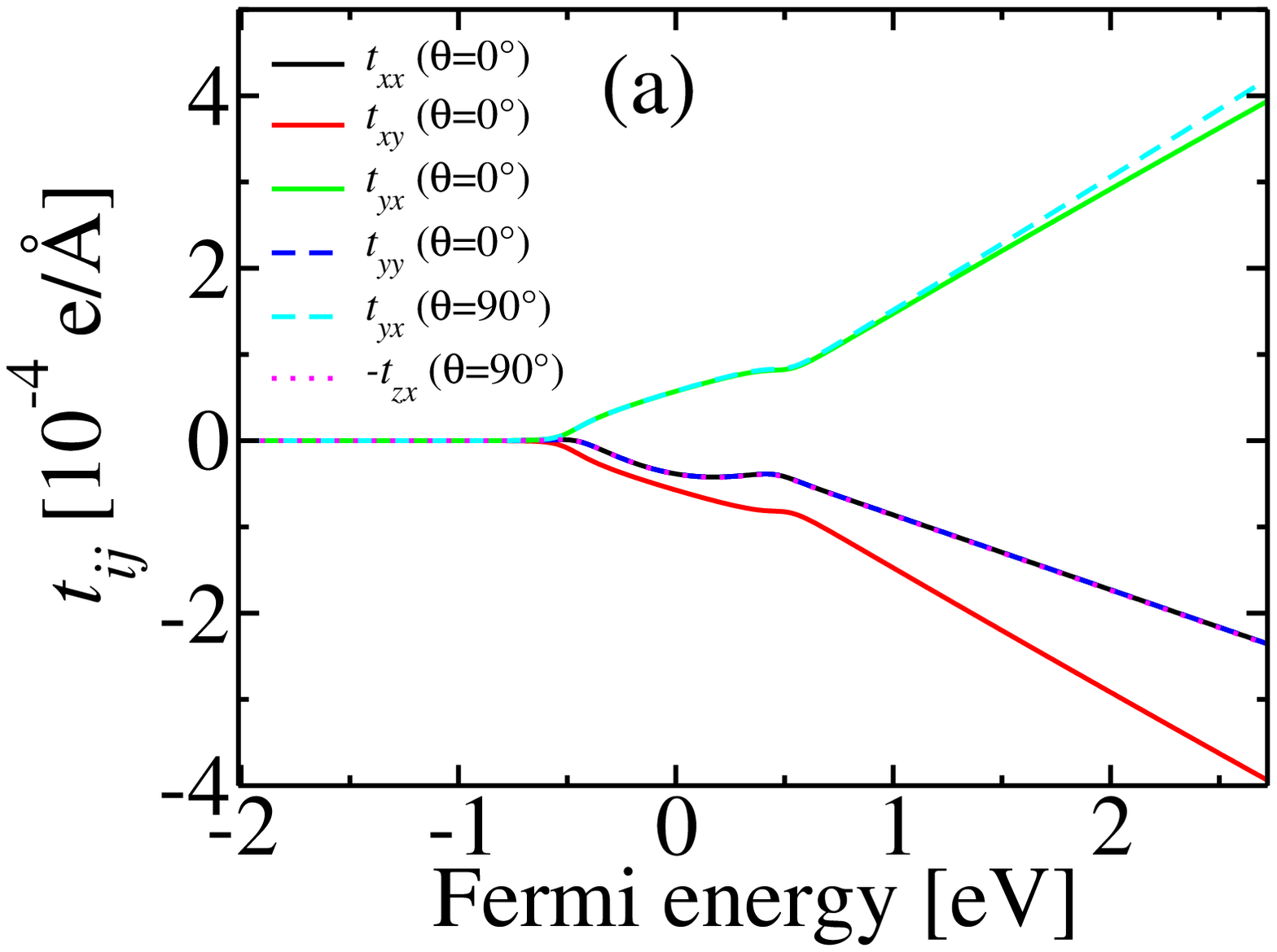}
\includegraphics[width=0.49\linewidth]{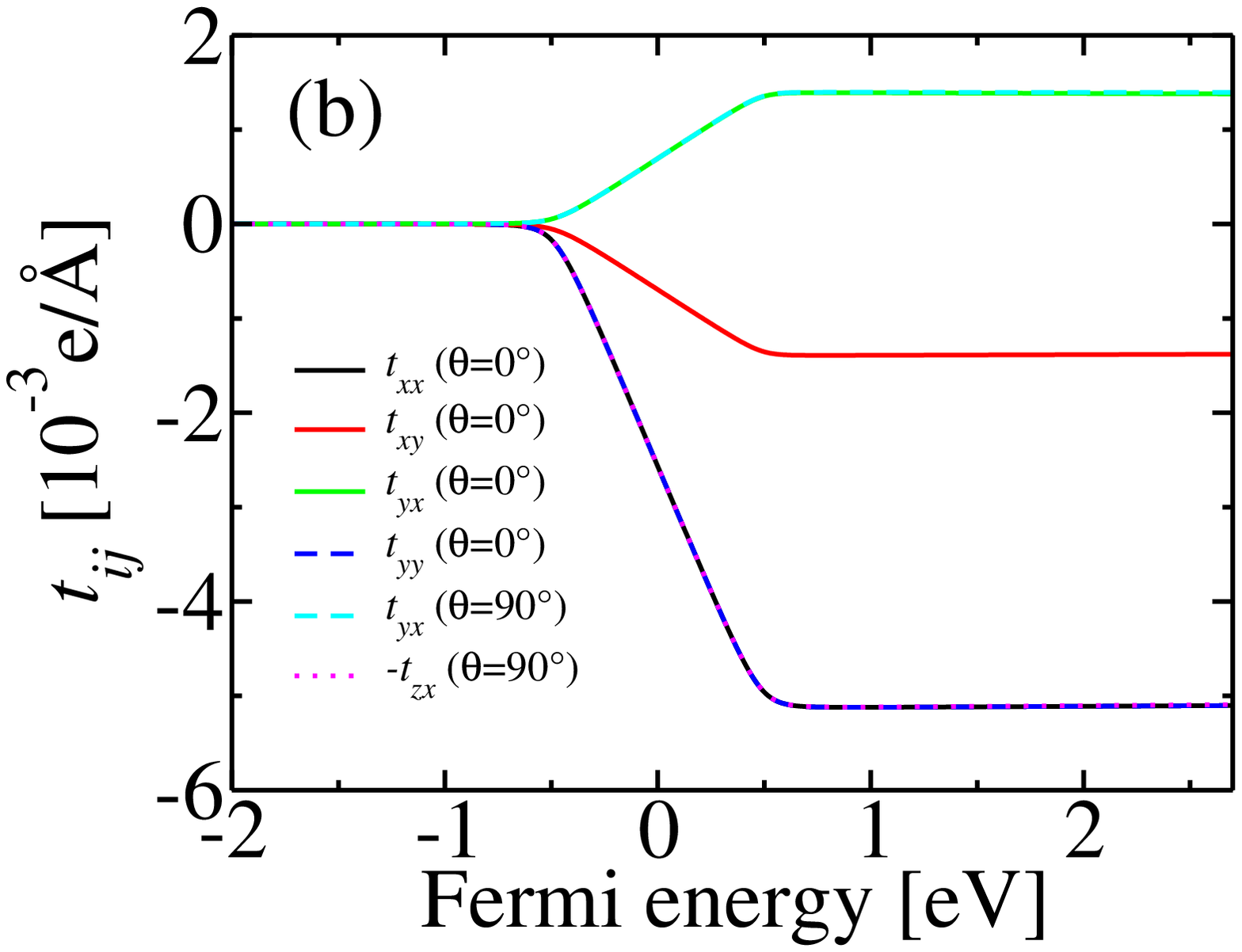}
\caption{\label{fig_alpha001ry}
Magnonic SOT (a) and
non-magnonic SOT (b) for $\alpha^{\rm R}=72$~meV~\AA\, and $\Gamma=136$~meV.
}
\end{figure*}

\begin{figure*}
\includegraphics[width=0.49\linewidth]{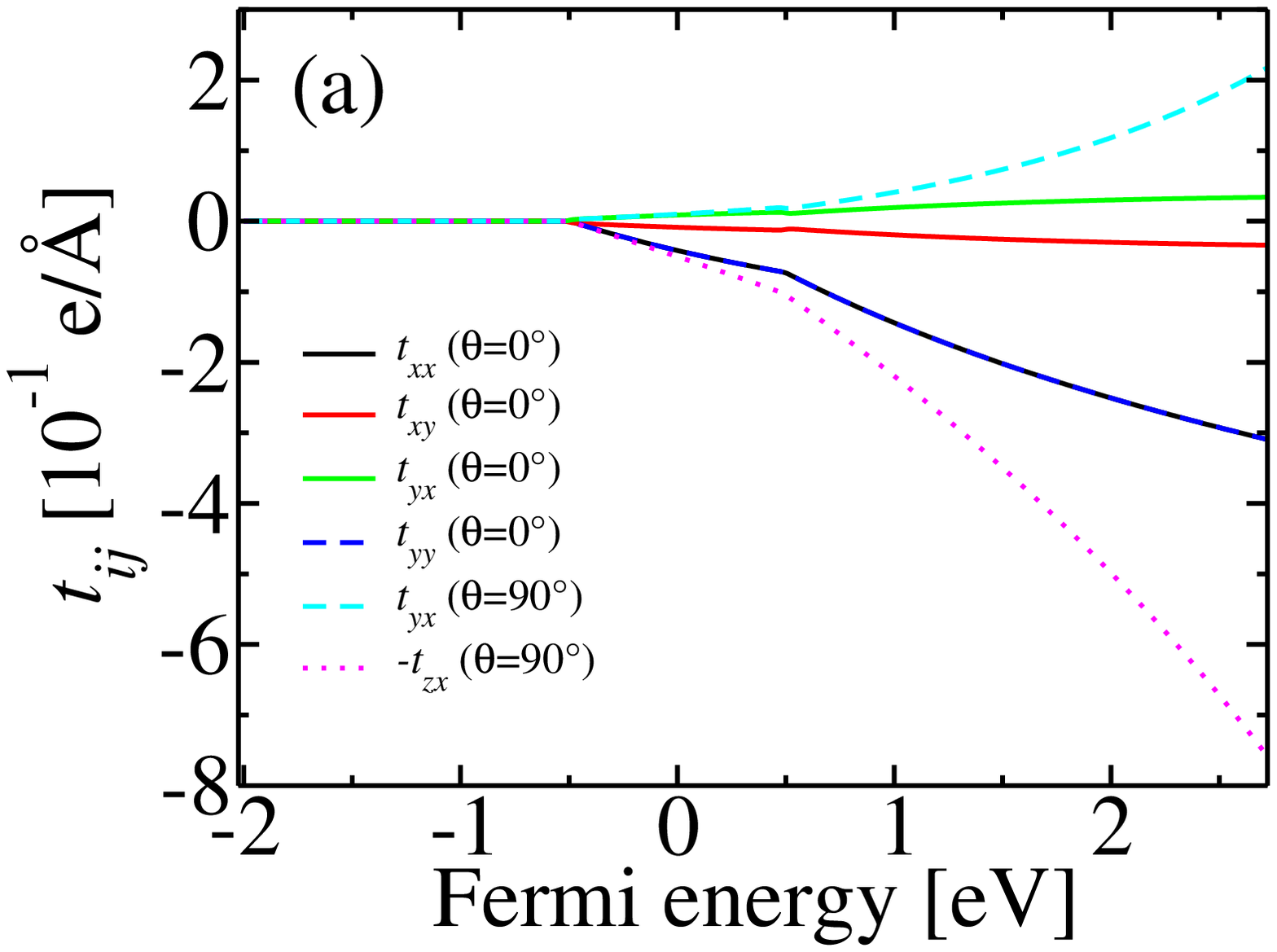}
\includegraphics[width=0.49\linewidth]{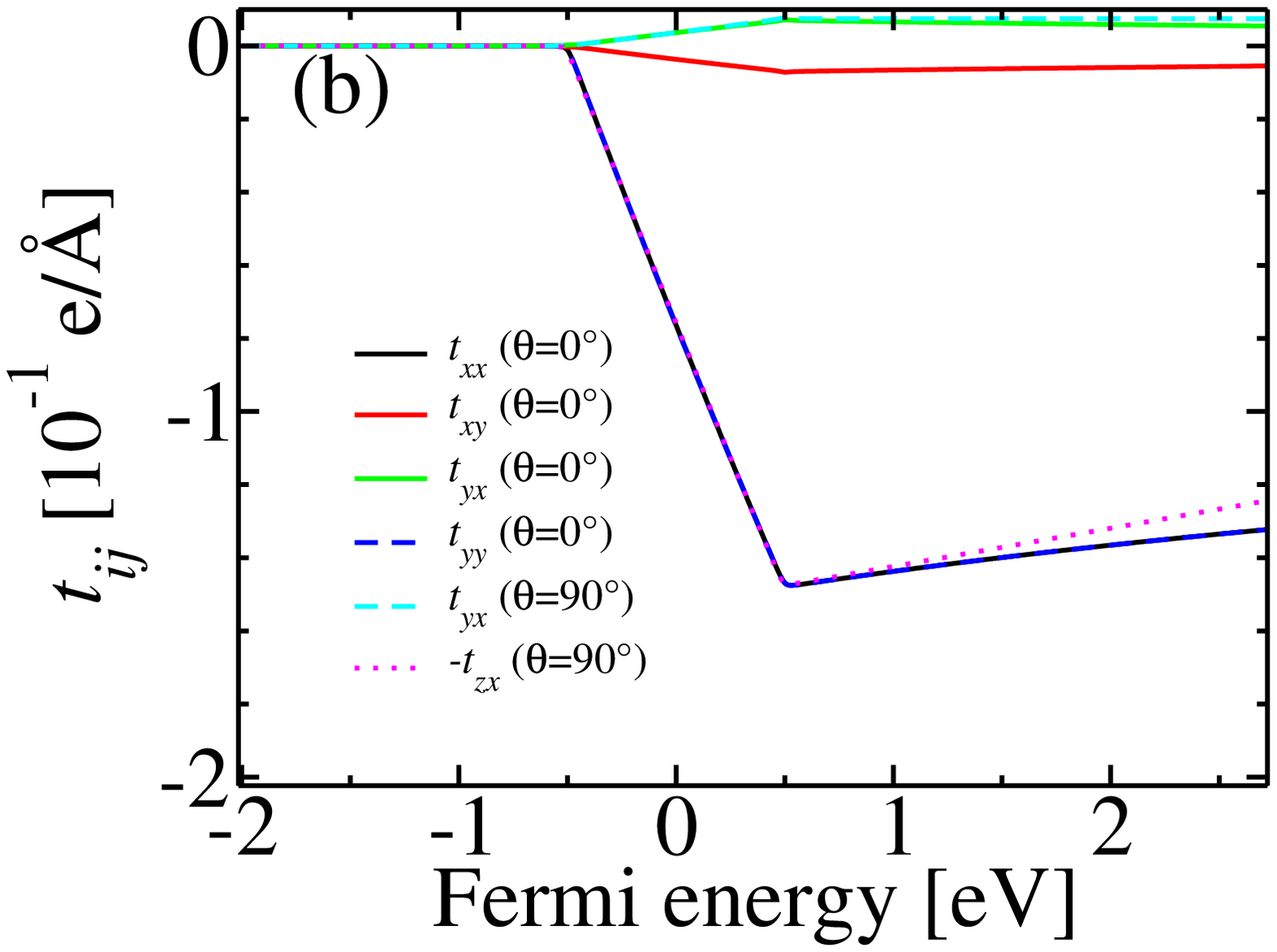}
\caption{\label{fig_alpha005ry_25meV}
Magnonic SOT (a) and
non-magnonic SOT (b) for $\alpha^{\rm R}=360$~meV~\AA\,
and $\Gamma=25$~meV.
}
\end{figure*}

\begin{figure*}
\includegraphics[width=0.49\linewidth]{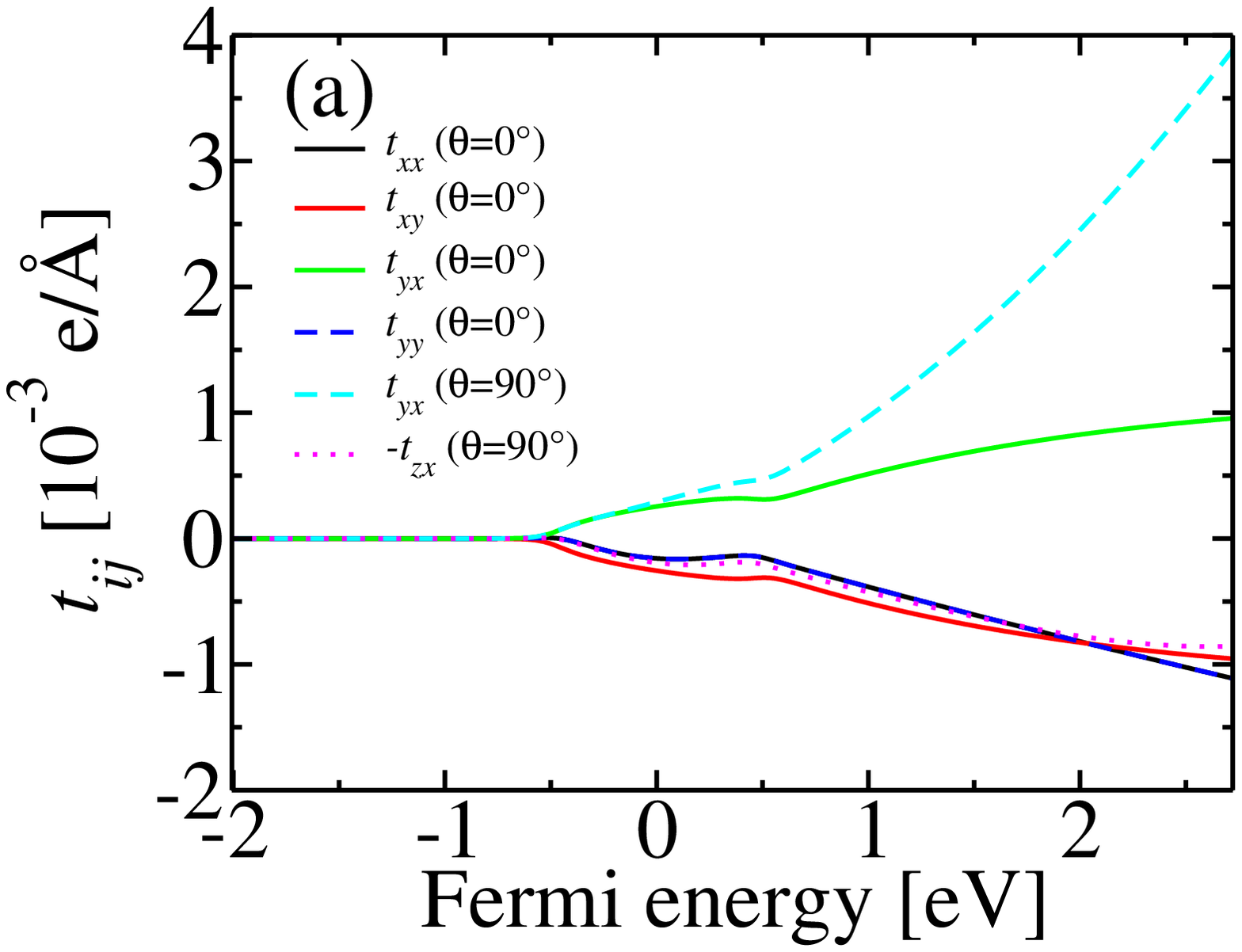}
\includegraphics[width=0.49\linewidth]{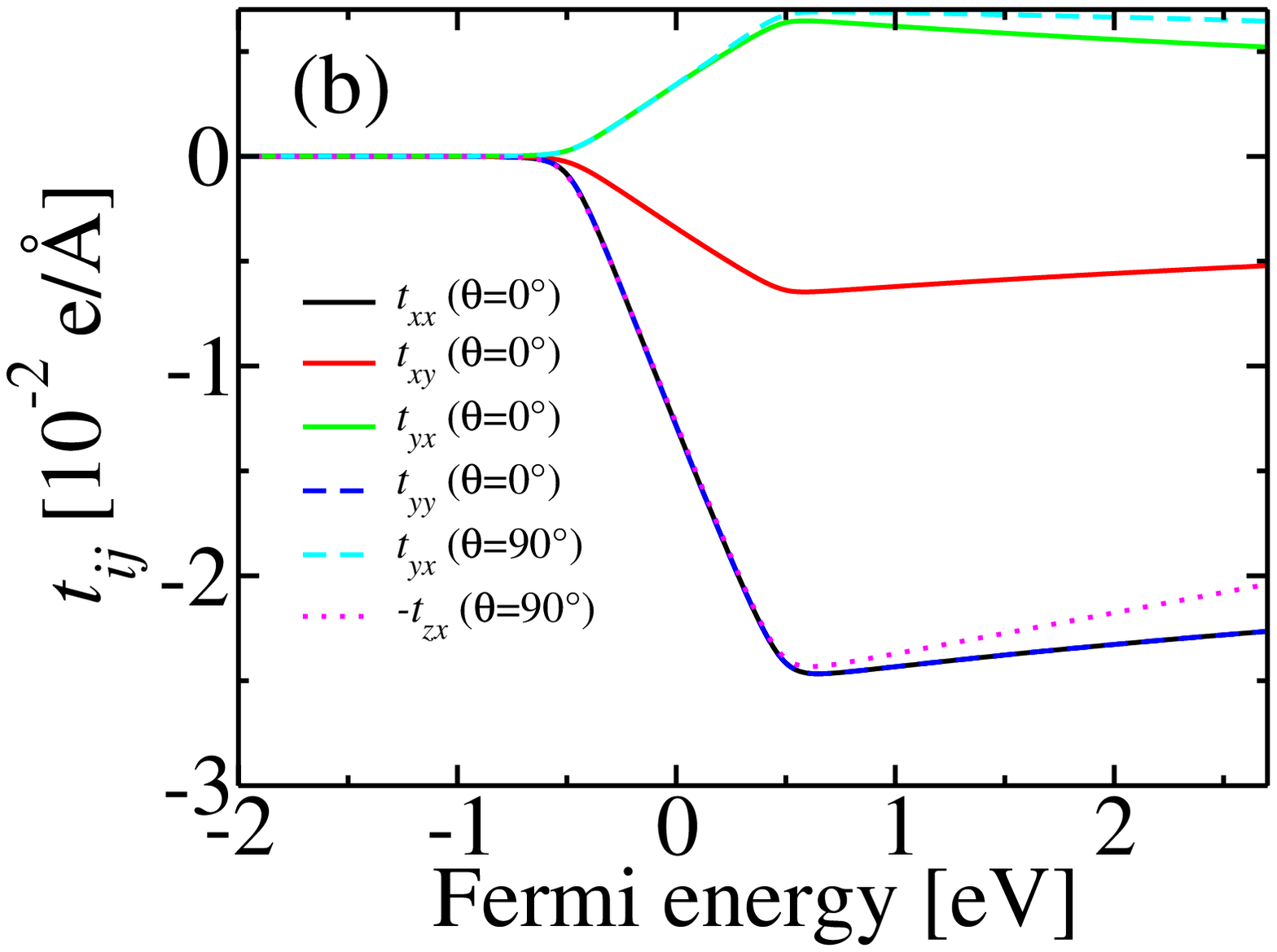}
\caption{\label{fig_alpha005ry}
Magnonic SOT (a) and
non-magnonic SOT (b) for $\alpha^{\rm R}=360$~meV~\AA\,
and $\Gamma=136$~meV.
}
\end{figure*}

\begin{figure*}
\includegraphics[width=0.49\linewidth]{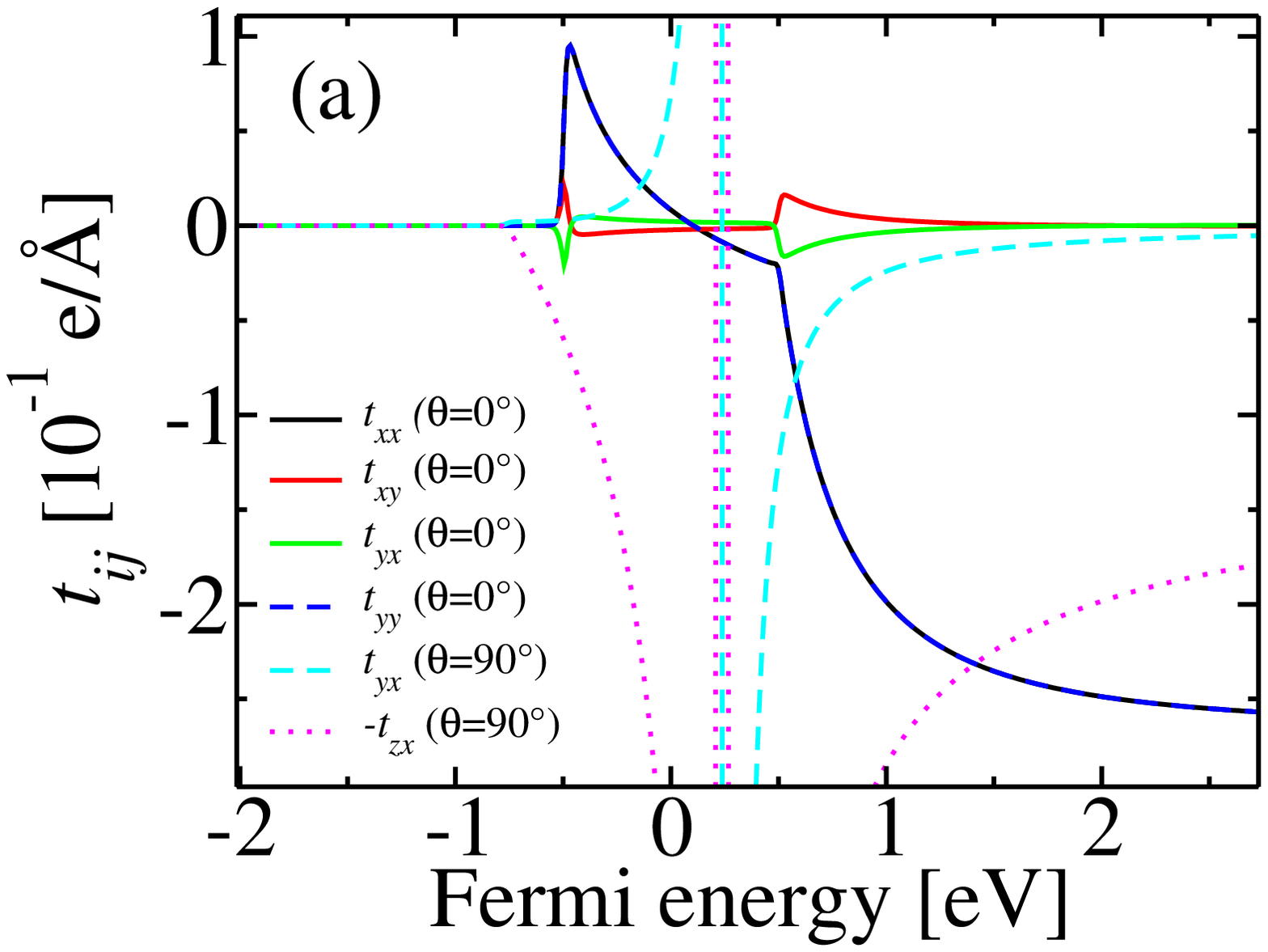}
\includegraphics[width=0.49\linewidth]{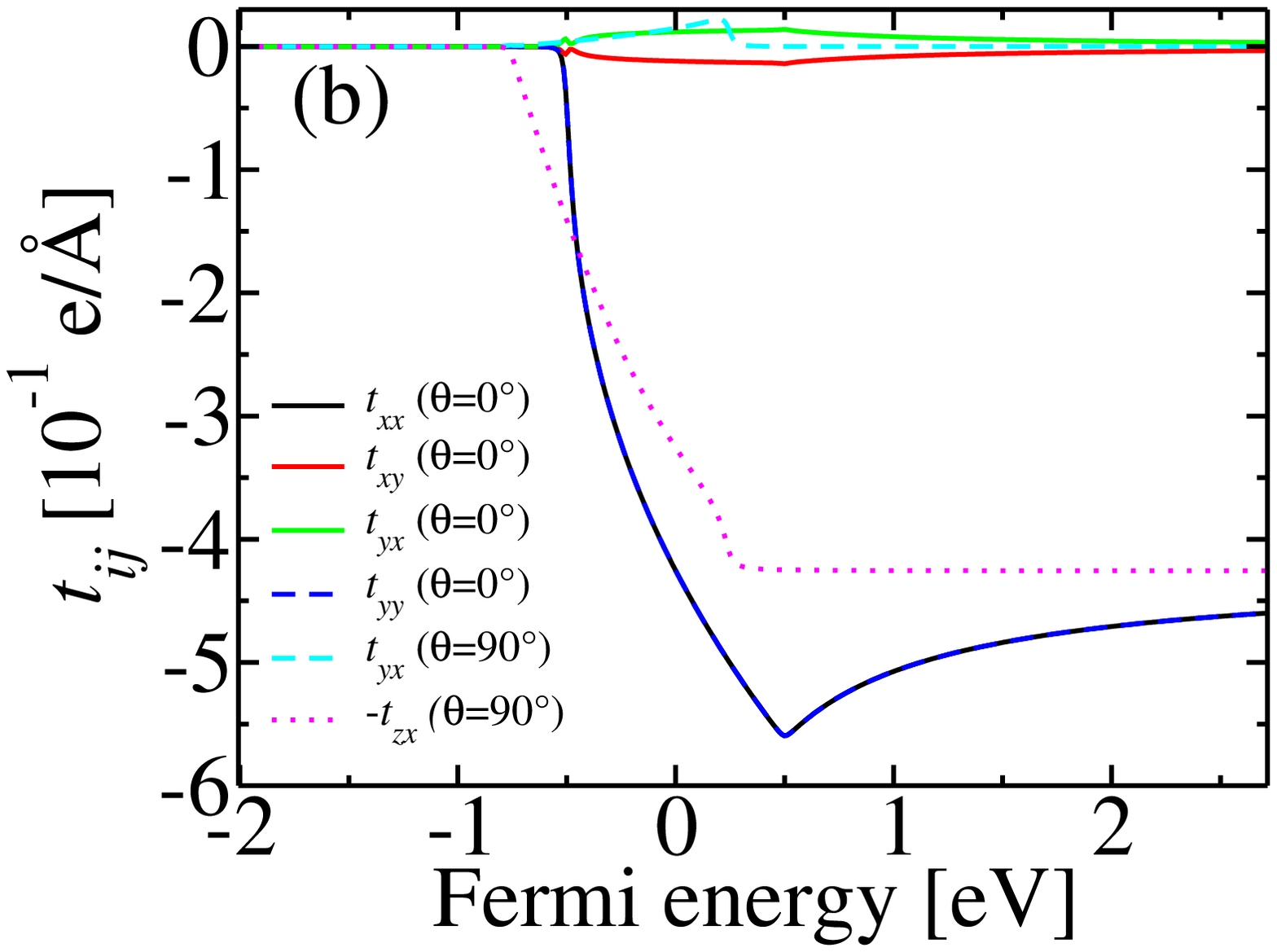}
\caption{\label{fig_bigalpha_25meV}
Magnonic SOT (a) and
non-magnonic SOT (b) for $\alpha^{\rm R}=2$~eV~\AA\,
and $\Gamma=25$~meV.
}
\end{figure*}

\begin{figure*}
\includegraphics[width=0.49\linewidth]{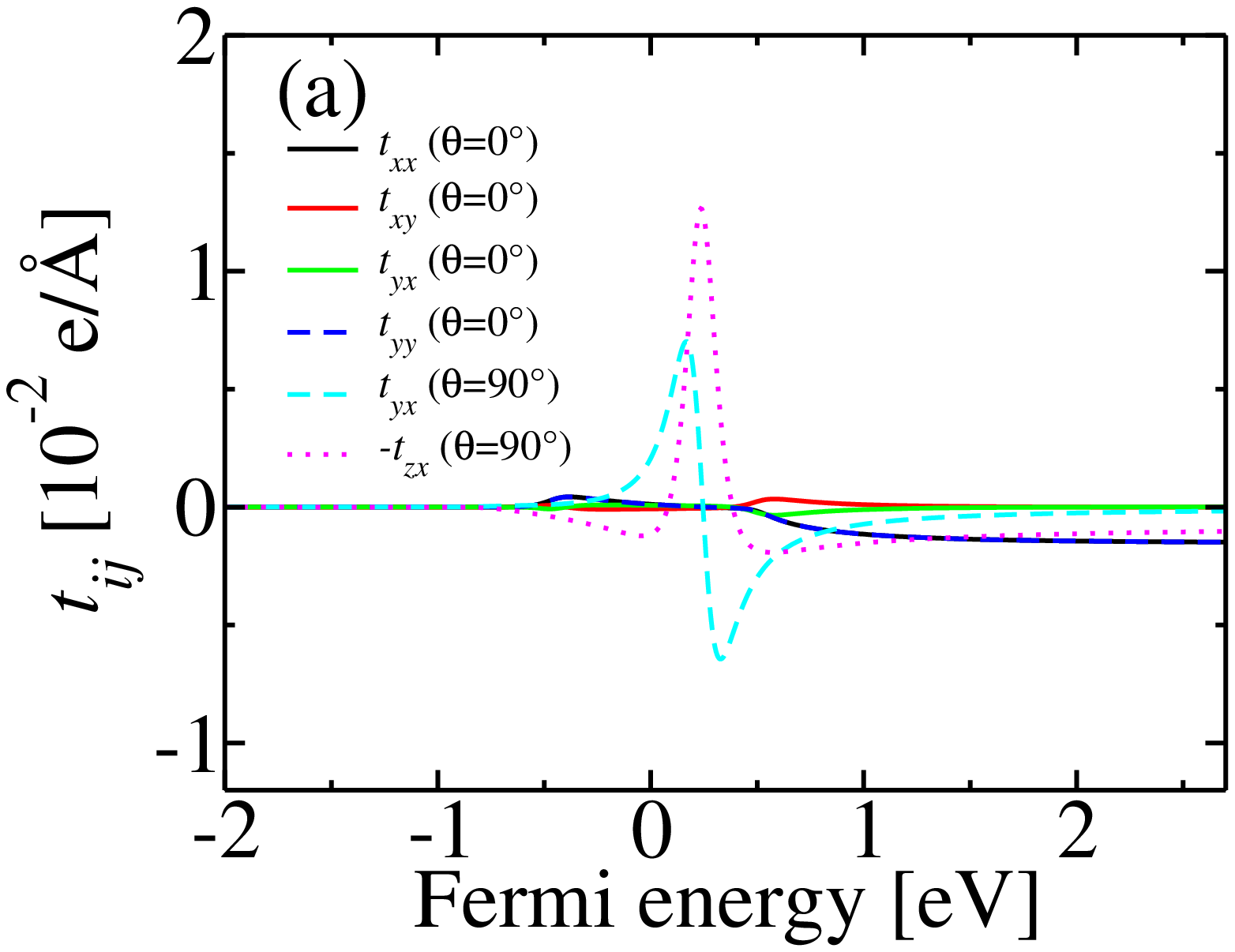}
\includegraphics[width=0.49\linewidth]{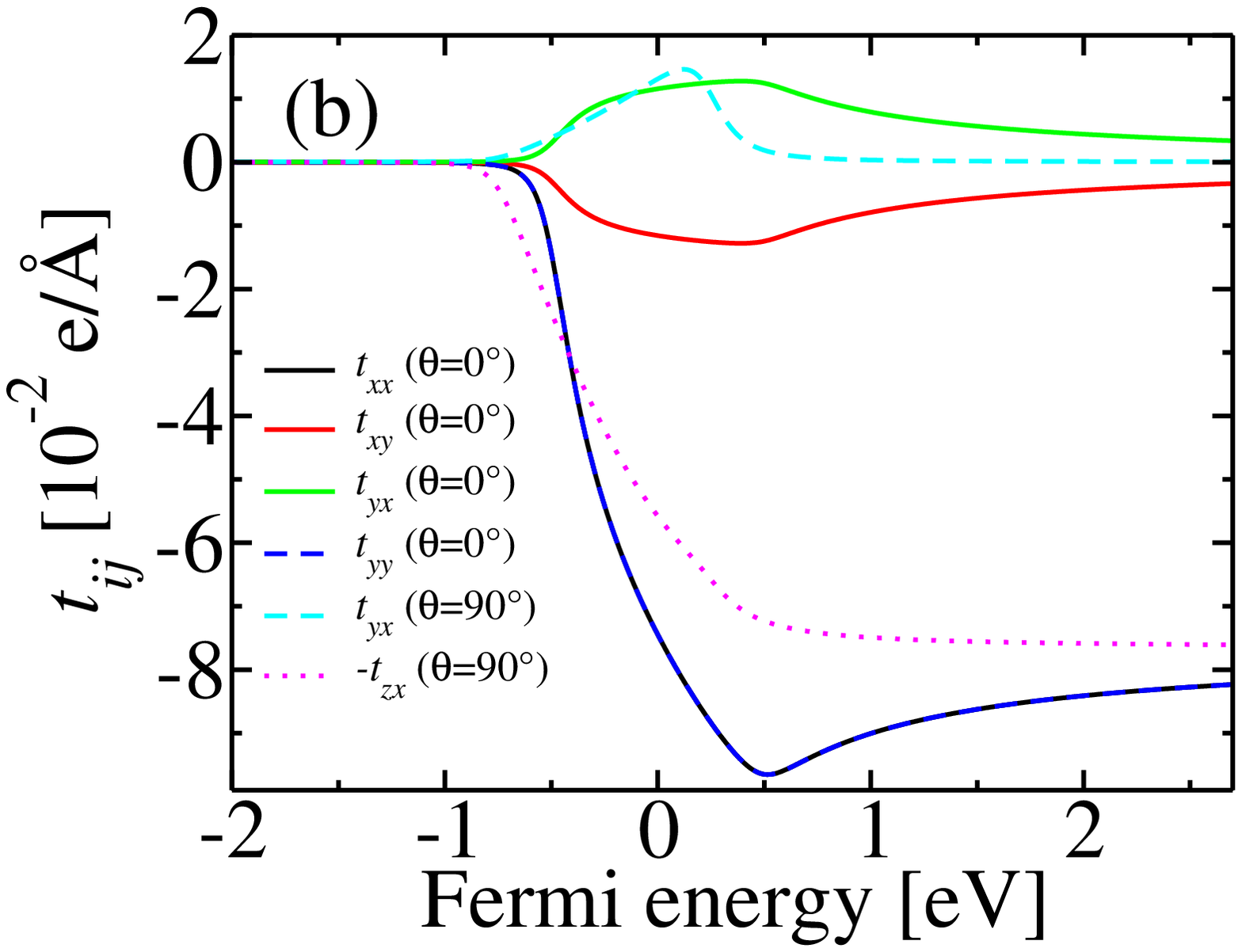}
\caption{\label{fig_bigalpha}
Magnonic SOT (a) and
non-magnonic SOT (b) for $\alpha^{\rm R}=2$~eV~\AA\,
and $\Gamma=136$~meV.
}
\end{figure*}

\begin{figure*}
\includegraphics[width=0.49\linewidth]{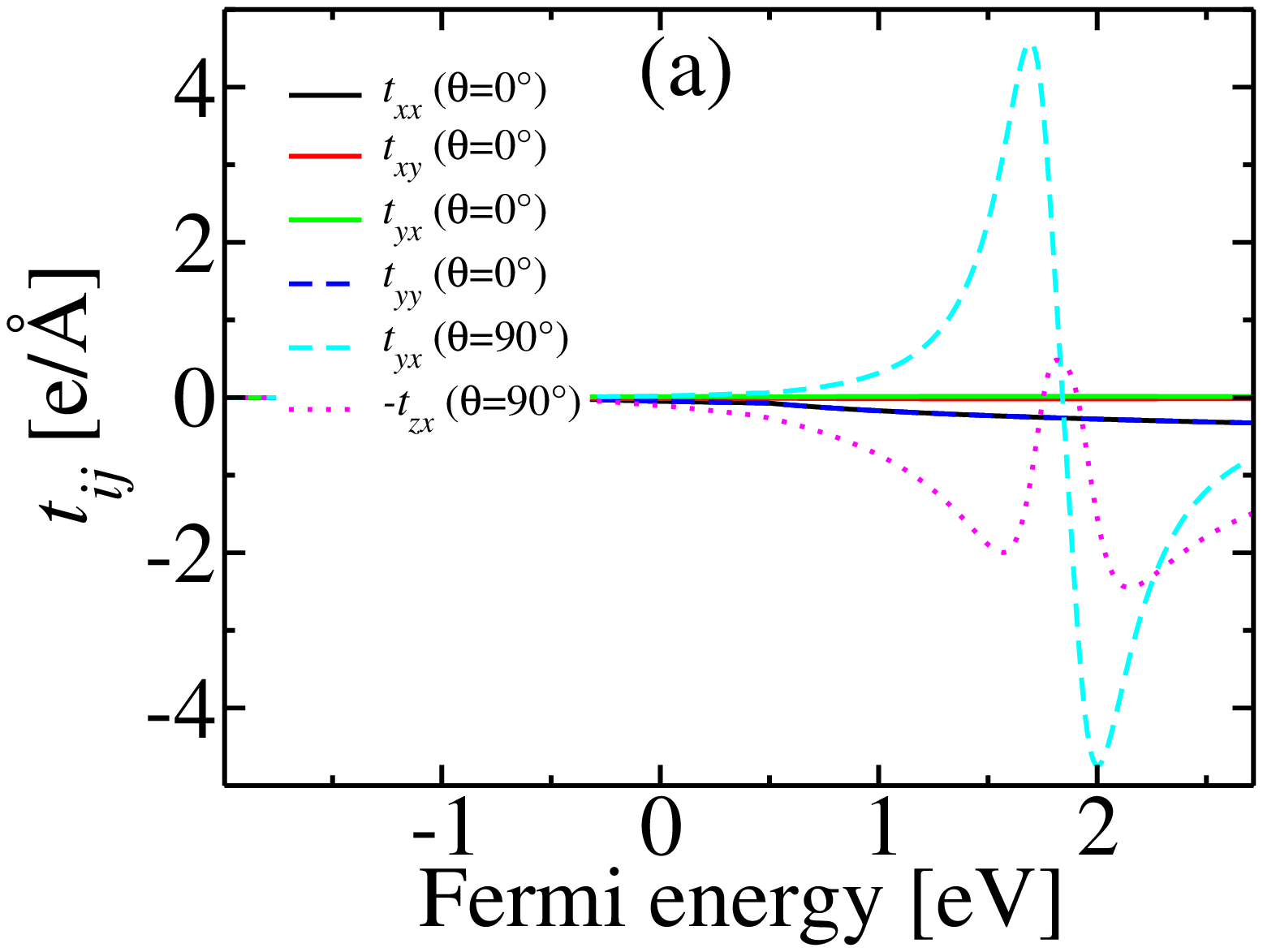}
\includegraphics[width=0.49\linewidth]{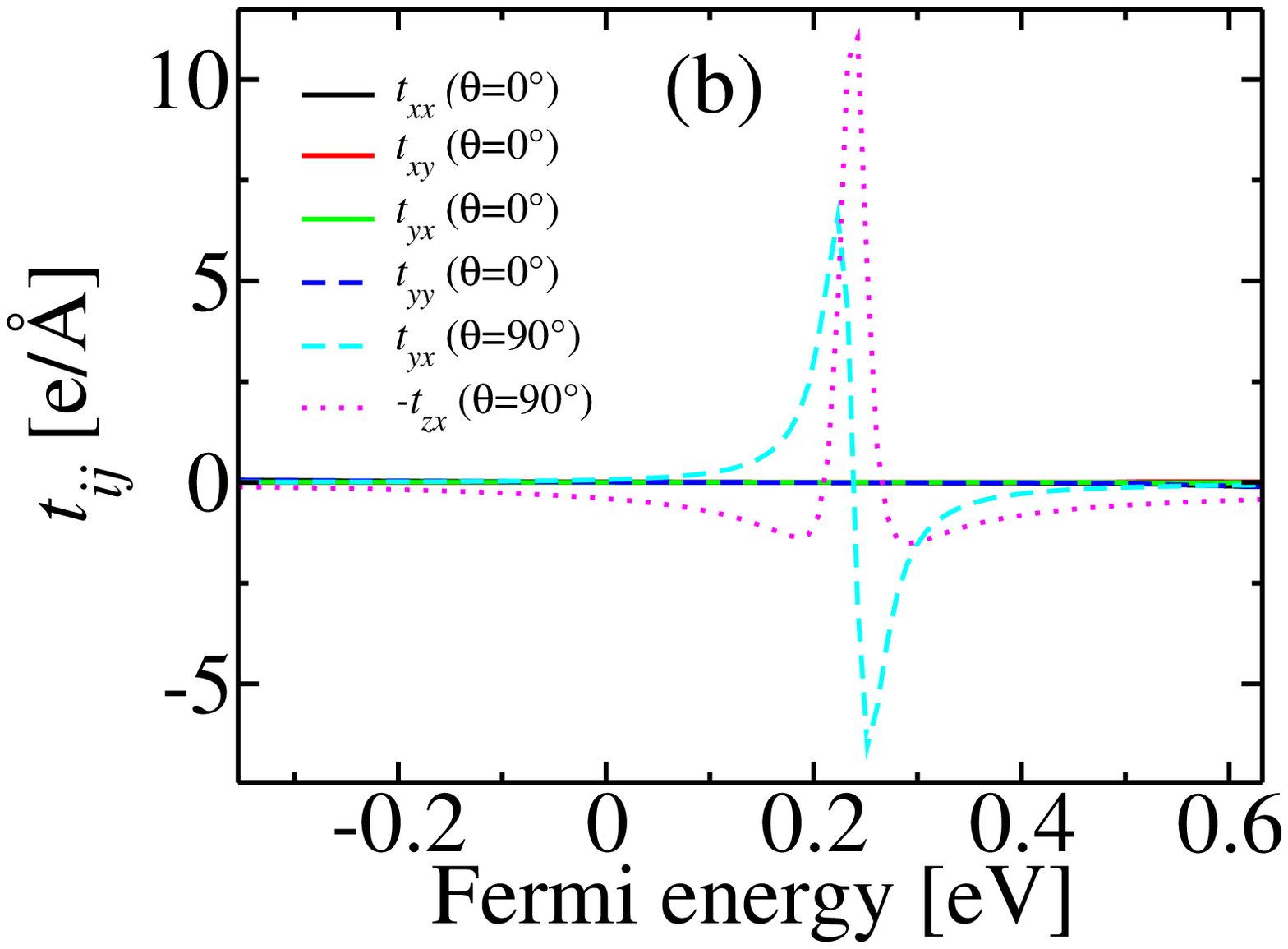}
\caption{\label{fig_zoomed}
Magnonic SOT for $\alpha^{\rm R}=720$~meV~\AA\, 
and $\Gamma=25$~meV (a) and 
for $\alpha^{\rm R}=2$~eV~\AA\, 
and $\Gamma=25$~meV (b).
}
\end{figure*}

In this appendix we provide additional plots of the SOT for different
Rashba and broadening parameters. 
In Fig.~\ref{fig_alpha001ry} we show the SOT for
the Rashba and broadening parameters of 
$\alpha^{\rm R}=72$~meV~\AA\, and $\Gamma=136$~meV, respectively.
In contrast to Fig.~\ref{fig_alpha001ry_25meV} the magnonic SOT is
roughly an order of magnitude smaller than the non-magnonic one
due to the larger broadening.
In Fig.~\ref{fig_alpha005ry_25meV} we show the SOTs for
the Rashba and broadening parameters of 
$\alpha^{\rm R}=360$~meV~\AA\, and $\Gamma=25$~meV, respectively.
The anisotropy of the magnonic SOT is much larger than the one of the
non-magnonic SOT due to the Rashba parameter, which is larger than in
Fig.~\ref{fig_alpha001ry_25meV} and in Fig.~\ref{fig_alpha001ry}.
In Fig.~\ref{fig_alpha005ry} we show the SOTs at the same Rashba
parameter,
but with a larger broadening of $\Gamma=136$~meV. In contrast to
Fig.~\ref{fig_alpha005ry_25meV},
where the magnonic SOT is larger than the non-magnonic one, the
magnonic
SOT is smaller here due to the larger broadening.
In Fig.~\ref{fig_bigalpha_25meV} we show the SOT for the
Rashba and broadening parameters of 
$\alpha^{\rm R}=2$~eV~\AA\, and $\Gamma=25$~meV, respectively.
The anisotropy of the magnonic SOT is gigantic at this large value of
the
Rashba parameter, and it is much larger than the anisotropy of the
non-magnonic SOT. When the broadening is increased to 136~meV
the magnonic SOT for $\theta=90^{\circ}$ is still sizable in
comparison
to the non-magnonic SOT for Fermi energies around 0, while it is
suppressed
otherwise, as shown in Fig.~\ref{fig_bigalpha}.
In Fig.~\ref{fig_zoomed}(a) we replot Fig.~\ref{fig_alpha01ry_25meV}
with a different scale of the vertical axis in order to show the full
range of the magnonic SOT for $\theta=90^{\circ}$.
Similarly, we replot Fig.~\ref{fig_bigalpha_25meV}(a) in Fig.~\ref{fig_zoomed}(b).

\end{document}